\documentclass[fleqn,usenatbib]{mnras}
\usepackage{amsmath}
\usepackage{mathptmx}
\usepackage[all]{hypcap}
\usepackage{txfonts}
\usepackage{soul}
\usepackage{listings}
\usepackage[T1]{fontenc}
\usepackage{ae,aecompl}
\usepackage{graphicx}
\usepackage{xcolor}
\usepackage{amssymb}
\usepackage{geometry}
\usepackage[normalem]{ulem}
\usepackage{longtable}
\usepackage{multirow}
\usepackage{multicol}
\usepackage{appendix}
\usepackage{tablefootnote}
\newcommand{\nn}{\nonumber}

\usepackage{siunitx}

\def\aj{AJ}
\def\apj{ApJ}
\def\apjl{ApJL}
\def\apjs{ApJS}
\def\jcap{JCAP}
\def\mnras{MNRAS}
\def\aap{A\&A}

\def\nat{Nature}
\def\apjs{ApJS}
\def\apjl{ApJ Letters}

\title[ Grand-design and Flocculent Spirals with DCNN]{Identification of Grand-design and Flocculent Spirals from SDSS using Deep Convolutional Neural Network}

\author[Sarkar, S., Narayanan , G., Banerjee, A., Prakash, P.] {Suman Sarkar$^{1}$, Ganesh Narayanan$^{1}$, Arunima Banerjee $^{1}$\thanks{arunima@iisertirupati.ac.in}, Prem Prakash $^{1}$ \\$^1$ Department of Physics, Indian Institute of Science Education and Research ( IISER ) Tirupati, Tirupati - 517507, India}

\date{}

\pubyear{2020}

\begin{document}
\label{firstpage}
\pagerange{\pageref{firstpage}--\pageref{lastpage}}
\maketitle

\begin{abstract}
Spiral galaxies can be classified into the {\it Grand-designs} and {\it Flocculents} based on the nature of their spiral arms. The {\it Grand-designs} exhibit almost continuous and high contrast spiral arms and are believed to be  driven by stationary density waves, while the {\it Flocculents} have patchy and low-contrast spiral features and are primarily stochastic in origin. We train a Deep Convolutional neural network (DCNN) model to classify spirals into {\it Grand-designs} and {\it Flocculents}, with a testing accuracy of $\mathrm{97.2\%}$. We then use the above model for classifying $\mathrm{1,354}$ spirals from the SDSS. Out of these, $\mathrm{721}$ were identified as {\it Flocculents}, and the rest as {\it Grand-designs}. Interestingly, we find  the mean asymptotic rotational velocities of our newly classified  {\it Grand-designs} and  {\it Flocculents} are  $218 \pm 86 \mathrm{\;Km\; s^{-1}}$ and $146 \pm 67 \mathrm{\;Km\; s^{-1}}$ respectively, indicating that the {\it Grand-designs} are mostly the high-mass and the {\it Flocculents} the intermediate-mass spirals. This is further corroborated by the observation that the mean morphological indices of the {\it Grand-designs} and {\it Flocculents} are $2.6 \pm 1.8$ and $4.7 \pm 1.9$ respectively, implying that the {\it Flocculents} primarily consist of a late-type galaxy population in contrast to the {\it Grand-designs}. Finally, an almost equal fraction of bars $\sim$ 0.3  in both the classes of spiral galaxies reveals that the presence of a bar component does not regulate the type of spiral arm hosted by a galaxy.  Our results may have important implications for formation and evolution of spiral arms in galaxies.

\end{abstract}


\section{Introduction} 

The magnificent spiral arms constitute the most notable characteristic of disc galaxies, and distinguish them from other astronomical objects. However, not all disc galaxies have spiral arms, and those which have discernible spiral features could be of different types depending on their origin. A {\it {\it {\it Grand-design}}} 
spiral galaxy has almost continuous and well-defined  spiral arms. A {\it {\it Flocculent}} spiral galaxy, on the other hand, has patchy or fragmentary spiral arms which are not quite well-defined. See, for example, \citet{elmegreen82}. Spiral arms play a crucial role in galaxy formation and evolution as well. In addition to being the primary sites of star formation in galaxies, they help transfer angular momentum from the inner to the outer galaxy. If the spiral features were material arms, then the differential rotation of the galactic disc would have wound them up into tightly-coiled spirals in a few rotations. The theoretically predicted pitch angle of such a material arm in a typical galaxy like the Milky Way is $\sim$ 0.14$^{\circ}$. However, observational evidence indicates that the pitch-angle varies between 10 - 15$^{\circ}$, thus ruling out the possibility that the spiral features are material arms \citep{Binney2008}.

The alternative model is that of density waves, wherein the constituent stars change with time, unlike in a material arm.  The local motion of the disc stars may be approximated as radial perturbations on a circular orbit known as the epicyclic theory. Since the frequency of the radial oscillation is higher than that of the azimuthal oscillation, each orbit undergoes precession at a frequency determined by the radial and azimuthal frequencies. If a family of nearby, concentric epicyclic orbits has similar precession rates, it will constitute a pattern (or figure of rotation) with a well-defined pattern (or precession) speed. Further, if the position angles of the ellipses increase continuously, the geometrical pattern traced out will be close to that of a spiral which will rotate slowly with the above pattern speed. However, the self-gravity of the spiral features is neglected in the above kinematical model. In a more realistic dynamical model, the formation of spiral arms in galactic discs is triggered by disc dynamical instabilities (See, for example, \citet{Binney2008}). \citet{ toomre1964gravitational} analytically studied the response of a 1-component, self-gravitating galactic disc, differentially-rotating and assumed to be infinitesimally thin, to local axi-symmetric perturbations and introduced the Toomre $Q$ parameter such that for a stable disc $Q >$  1 . The formalism was later generalised to more realistic 2-component or multi-component discs and/or with finite disc thickness \citep{Bertin1988,Jog1996,Romero2011}. Similarly, the study of the response of a 1-component, sheared disc to local, non-axisymmetric perturbations was pioneered by \cite{Goldreich1965} and \cite{Julian1966}.

A density wave can be constructed by the coherent excitation of epicyclic motions in an axisymmetric disk, and may also be a leading wave with respect to the galacto-centric direction. However, the differential rotation of the galactic disc will eventually shear that density pattern, and its own amplitude will evolve as a result. Such waves can get amplified very strongly as the shear sweeps them around. As the epicycle and the perturbation rotate in the same direction, so stars stay in the perturbation longer than they would under other conditions. Discs are stabilized by random motion on small scales, and by rotation on large scales. However, the stabilizing effects of random motion can be temporarily suppressed under the above special circumstance. This is particularly important in the case when the gaseous equivalent  to Toomre's Q is as low as unity and the azimuthal wavelength matches the critical value. The resultant form of the density wave ensures that individual stars remain within the overdense region for a substantial fraction of the epicyclic period $\kappa$. If this happens, small perturbations of the disk can be swing amplified. \cite{Toomre1981} coined the term swing amplification for this phenomenon, since the growth is maximum as a wave swings from a leading to a trailing position. \cite{Jog1992}, for example, generalized the study to a 2-component system with application to the Galaxy. Swing amplification thus also explains the preference of trailing spirals in galaxies. 

\begin{figure*}
\resizebox{18 cm}{!}{\rotatebox{0}{\includegraphics{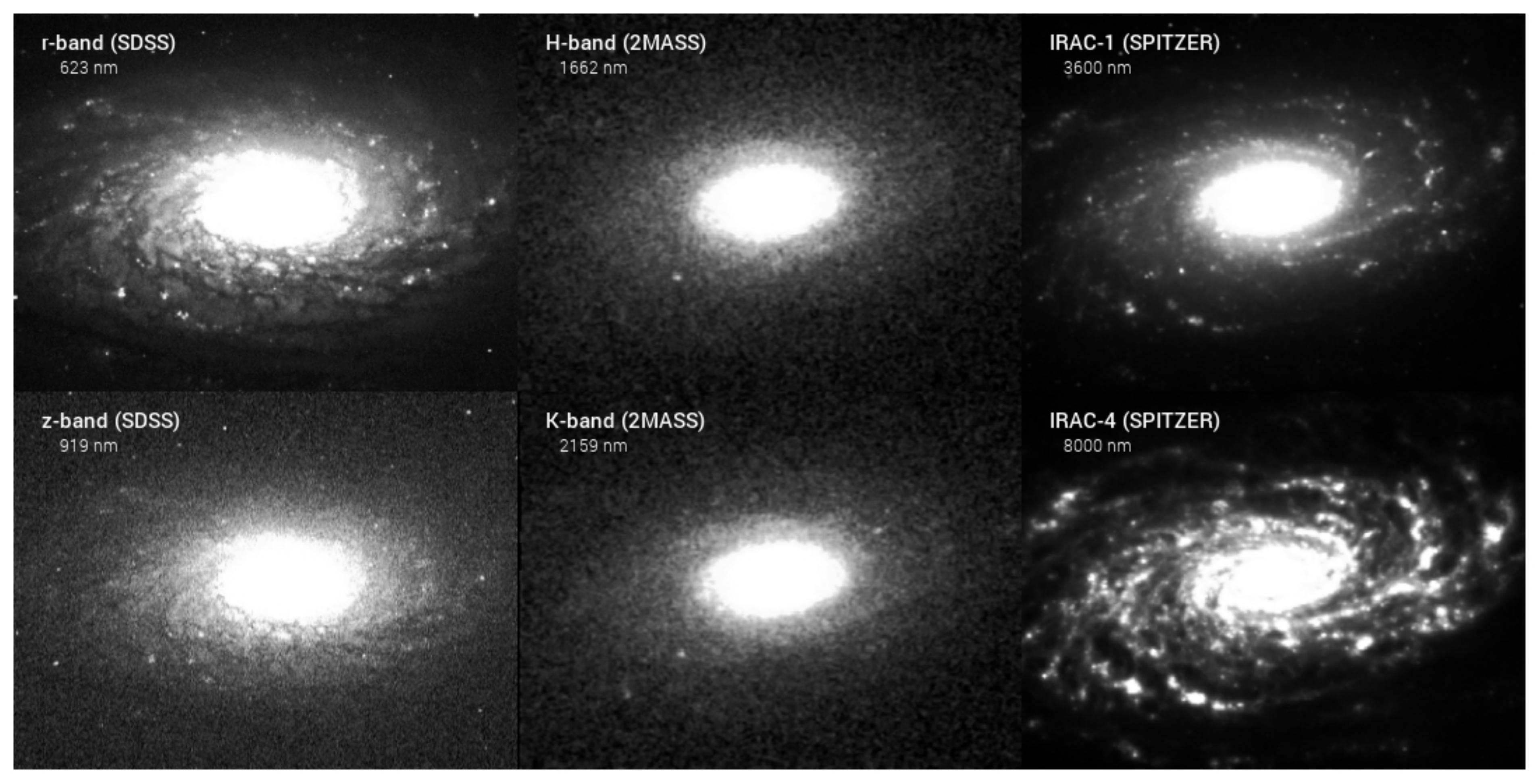}}}
\label{fig:ngc5055}
\caption{Galaxy {\it NGC 5055} observed in different optical, near-IR and IR bands. Images taken from InfraRed Science Archive (IRSA) \protect\footnotemark }
\end{figure*}

While {\it Flocculent} spirals are understood to be formed by the swing amplification of random local instabilities \citep{elmegreen11}, the Density Wave Theory forms the basic analytical framework for understanding the dynamics of the {\it Grand-design}s \citep{lin66}. The essence of density-wave theory is  that the external non-axisymmetric perturbation is one of the normal modes of oscillation of the galactic disc. For such a normal mode, the gravitational field associated with the perturbation equals the field required to elicit the non-axisymmetric response. Further, for a standing spiral wave mode rotating rigidly around the disc, a feedback mechanism needs to be generated such that the wave returns to its site of origin and a cycle is established. Theoretical studies have indicated that such reflections may happen at the Lindblad resonances where the angular velocity of the pattern with respect to that of the disc is commensurate with the radial frequency. For a two-armed {\it Grand-design} spiral, for instance, the relative velocity has to be half of the epicyclic frequency. Finally, if a single mode dominates over other modes, the spiral structure thus triggered will be a quasi-stationary one i.e., it would retain its fixed shape by rotating rigidly with a fixed pattern speed, thus forming a {\it Grand-design} spiral. The more irregular systems, where the pattern is multi-armed or filamentary or patchy like the {\it Flocculents} are understood to be formed by the superposition of different modes \citep{shu82,bertin14}.

The formation and evolution of spiral structures in galaxies is still not very well understood when we combine theoretical models with observational studies and numerical simulations. \citet{1998Seigar} found that the {\it Grand Designs} are associated with presence of a bar component. Besides \cite{Bittner2017}, who studied the correlation between the properties of bars and spiral arms in galaxies using Spitzer $S^4G$ survey, found that existence of bulges are necessary in  {\it Grand-design} galaxies if we believe that they are driven by standing spiral wave modes. Both of these are in compliance with the Density Wave theory because a central hot component with high Toomre Q would reflect the density wave moving inward, and thus help in the sustenance of a standing wave pattern of a  {\it Grand-design} spiral. {\it Flocculents}, on the other hand, are usually marked by a stochastic star formation, propagating outwards as a shock wave due to stellar winds \citep{Gerola1978, Thornley1997}. However, there are several open questions. Firstly, the numerical simulation of a stationary spiral pattern in is still an open problem. Simulations indicate that a spiral pattern can also be obtained from a superposition of the transient modes \citep{Sellwood_2014}. But transient spirals may heat up the stellar disc, rendering it stable against the presence of a spiral arm, and consequently the spiral pattern may disappear over time \citep{1984Sellwood}. \cite{Sellwood2011} shows long-lived spiral modes are not reproducible in simulations.  \cite{SellwoodCarlberg2014} show continuously changing recurrent transient spiral patterns arise in an isolated, unbarred disk galaxy models in N-body simulations.  Secondly, according to the manifold theory, the quadrupolar potential of a bar should trigger a spiral density wave \citep{athan09,athan09b}. As follow-ups of the above, a host of theoretical studies have indicated that the dynamical properties of the bar and the spiral arm in a galaxy should be correlated \citep{schwarz84, salo10, filistov12, diaz19, garma21, mondal21}. Interestingly, observational studies, do not always comply with theoretical predictions. In fact, the distribution of pitch-angles of barred and unbarred spiral galaxies is found to be roughly the same \citep{voglis06} and no correlation could be found between the bar-induced gravitational torque and the pitch angle \citep{diaz19}. Besides, the effect of the bar pattern speed on the strength of the spiral arms also appears to be insignificant. \cite{schwarz84} as well as \cite{buta09} report that although, in general, strong spirals are expected to be associated with strong bars, two of the strongly barred galaxies in their sample (NGC 7513 and UGC 10862)  showed weak spiral amplitudes. Further, \cite{romanishin85} studied the difference in star formation properties in  {\it Grand-designs} and in {\it Flocculents} and found that the  {\it Grand-designs} were bluer than {\it Flocculents} having the same atomic hydrogen mass, thus indicating different star formation histories of the two populations. Finally, low surface brightness galaxies mostly host {\it Flocculent} spiral arms \citep{McGaugh1995,Schombert2011A}, the reason behind which is not quite well explored.  Therefore, it is useful to carry out the classification of a large sample of spiral galaxies into  {\it Grand-designs}  and {\it Flocculents}, possibly spanning a wider range of the parameter space, to impose stringent constraints on the formation and evolution models of galactic spiral arms.

Depending on their nature and appearance of the spiral arms, \citet{elmegreen82} introduced an arm-class based, 12-point classification scheme for the spiral galaxies (\S 2.1). Arm classes 1-4 were considered as {\it Flocculents} and the rest, up to arm-class 12, as {\it {\it Grand-design}}s. Based on the same, \citet{elmegreen87} determined the arm classes of $762$ galaxies in $B$-band. Using the $3.6 \; \mu m$ images of $46$ galaxies from the SPITZER ($S^4G$) survey, \cite{elmegreen11} observed that the arm-interarm contrast decreases from  {\it Grand-designs} to {\it Flocculents}. In fact, it was observed that {\it Flocculents} had arm-contrasts ranging between 0.7 and 1.4, and  {\it Grand-designs} above 1.4. Further, the {\it Multiple -arms} type was introduced as a class intermediate between {\it Grand-designs} and {\it Flocculents}. \citet{buta15} used this technique to find the arm classes of a combined set of $1,114$ galaxies from $S^4G$. 

The conventional approach of morphological classification is based on visual inspection. Popular citizen science projects like \textit{Galaxy zoo} \citep{lintott11, willett13} are known for performing morphological typecasting  of galaxies visually. Depending on the vote fraction, the galaxies are assigned different morphological tags. However, this process of manual identification is time-consuming, and is not applicable to huge data sets. When it comes to efficient as well as fast morphological classification of a large sample, artificial intelligence is the preferred option. Machine learning approaches are viable in situations where sample completeness is more important than classification accuracy. Machine learning and artificial neural networks \citep{samuel59, fukushima75} have been the cornerstones of high-end model construction since a decade or so. In recent years, there has been a remarkable increase in the use of supervised, unsupervised, and transfer learning techniques in different areas of research. In the field of astronomy, machine learning has proven to be quite effective in resolving problems such as distinguishing stars from galaxies \citep{odewahn92, weir95, kim17}, classification of type-Ia supernovae \citep{moller16, hossein20}, gravitational lens detection  \citep{jacobs17, schaefer18, lanusse18}, finding optical transients, radio sources and exoplanets \citep{cabrera17,aniyan17,visser21}. Convolutional Neural Network (DCNN) is the next generation of the feed forward Artificial Neural Networks (ANN) \citep{mcculloch43a} which takes into account the back propagation of errors as well \citep{werbos75,rumel86,lecun90}. DCNNs are administered mainly to image recognition and classification problems, where the feature detection is automated and unsupervised. It has been successfully used to classify galaxy morphology in a variety of image datasets from various optical and infrared sky surveys \citep{dieleman15,huertas15, abraham18, goddard20, cheng21}. It has also been used to model the dynamics of interacting galaxies and mergers in various redshift surveys and cosmological simulations \citep{ackermann18, bottrell19, prakash20, bickley21}.\\

In this paper, we develop a Deep Convolutional Neural network (DCNN) model to classify spiral galaxies into  {\it Grand-designs} and {\it Flocculents}. We use the {\it Grand-designs} and {\it Flocculents} classified by \citet{buta15} as our training, validation and testing set. We then use this model to classify the SDSS DR17 images of $1,354$ spiral galaxies, hitherto unclassified but identified as spiral galaxies by Galaxy zoo2, into the two afore-mentioned classes. Our results will help identify a large number of spiral galaxies of the above two categories, shedding light on their relative abundance, and distribution of their physical properties, their local environments as well as their correlation with other disc components like bars and bulges, which may indirectly impose tight constraints on the dynamical models of their formation and evolution. The rest of the paper is organised as follows. In \S 2, we present the training and testing of the model, in \S 3 the new classifications done using the DCNN model thus developed, and finally the conclusions in \S 4.
% footnote for figure {fig:ngc5055} 
\footnotetext{https://irsa.ipac.caltech.edu/irsaviewer/}

\section{Training and Testing of the DCNN model}
\subsection{ Data for training and testing }
For training and testing our DCNN model, we need a sample of images of spiral galaxies from the literature, which have already been classified into  
{\it Grand-designs} and {\it Flocculents}. As noted before, \citet{elmegreen82} introduced a 12-point classification scheme for identification of  
{\it Grand-design} and {\it Flocculent} galaxies. Depending on the nature and appearance, the spiral arms were categorized into 12 arm-classes; 
Arm-Class 1: chaotic, fragmented and asymmetric arms, Arm-Class 2: Fragmented spiral arm pieces with no regular pattern, Arm-Class 3: Fragmented 
arms uniformly distributed around the galactic center, Arm-Class 4: Only one prominent arm, otherwise fragmented arm, Arm Class 5: Irregular broken 
arms in outer part, two short symmetric arms in inner part, Arm Class 6: Two symmetric inner arms, feathery ring-like outer structure, Arm-Class 7: 
Two symmetric, long outer arms, feathery ring-like outer structure, Arm-Class 8: Tightly-wrapped, ring-like arms, Arm-Class 9: Two symmetric inner arms, 
multiple long and continuous outer arms and so on, Arm-Class 10: Two long extended arms and a prominent bar, Arm-Class 11: Two prominent arms, one of 
them extending to the outskirts, and finally, Arm-Class 12: Two long, symmetric arms dominating the optical disc. Arm classes 1-4 were considered as 
{\it Flocculents} and the rest, up to arm-class 12, as {\it {\it Grand-design}}s. Based on the same, \citet{elmegreen87} determined the arm classes of 
$762$ galaxies in $B$-Band. Later on, a modified classification was employed by \cite{elmegreen11}, where galaxies were classified depending on their 
arm-inter-arm contrast in three different classes. The arm-inter-arm contrast of a spiral galaxy is defined as 
$$ A(R) = 2.5 \; \log\,\left(\frac{2I_{\rm{arm}}}{I_{\rm{interarm1}}+I_{\rm{interarm2}}}\right),$$ 

\noindent where $I_{\rm{arm}}$ average intensity of peak of azimuthal intensity distribution and $I_{\rm{interarm}}$ corresponds the intensity of adjacent 
inter-arm regions. \citet{buta15} use the technique introduced by \citet{elmegreen11} and update the arm classes of $1,114$ galaxies from $S^4G$ into the three aforementioned classes. We use the images of the {\it Grand-designs} and the {\it Flocculents} as classified in \citet{buta15} as our training and testing samples. It is important to note here that \citet{elmegreen11} also indicate that the spiral arms which appear to be {\it Flocculent} in optical mostly remain {\it Flocculent} in the 
mid-IR as well. Only a few {\it Flocculents} show weak two arm structures in mid-IR ($3.6 \; \mu m$), which are absent in optical bands. In fact, it was observed 
that only 2 of the 13 {\it Flocculent} galaxies had different arm-classes in the optical and in the mid-IR bands. However, similar to the $B$-band images, 
the $S^4G$ images mostly trace the star-forming regions,  as the heated dust mainly emits in the mid-infrared. Therefore, we checked for possible variation of the arm-class as compared to the NIR bands as well. For example, \autoref{fig:ngc5055} shows NGC 5055 in different optical, near-IR and IR bands. It is an {\it ``optically flocculent''} galaxy that in near-IR is found to have a {\it Grand-design} like feature in the inner region \citep{thornley96}. We find that despite the arm contrast of the galaxy being variable in different bands, the characteristic fleecy nature 
is noticed both in the IR and optical bands. Hence, the classification of spiral arms done in the mid-IR bands are mostly applicable to the optical bands as well.

\begin{figure*}
\resizebox{8.5 cm}{!}{\rotatebox{0}{\includegraphics{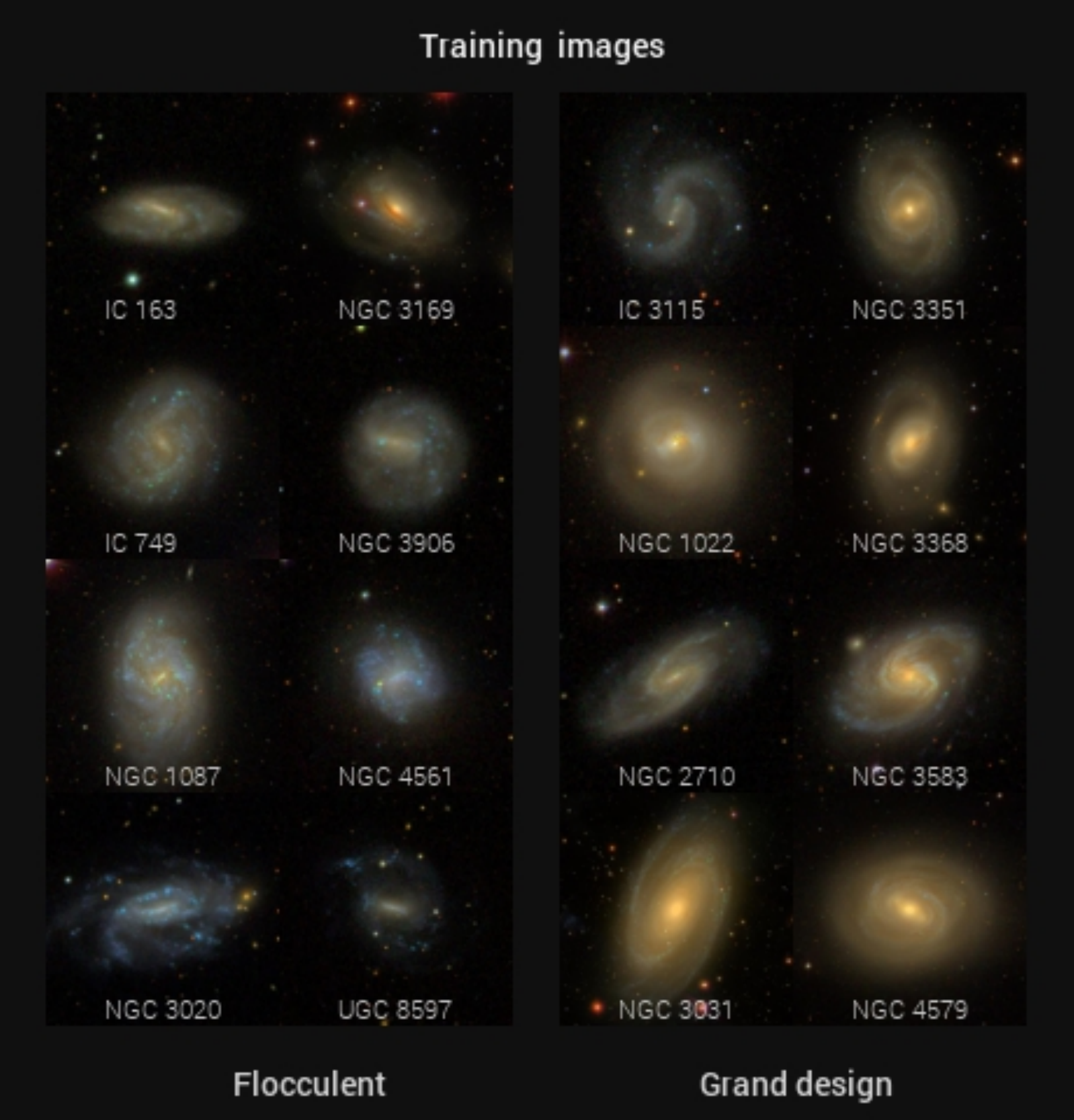}}} \hspace{0.6 cm}
\resizebox{8.5 cm}{!}{\rotatebox{0}{\includegraphics{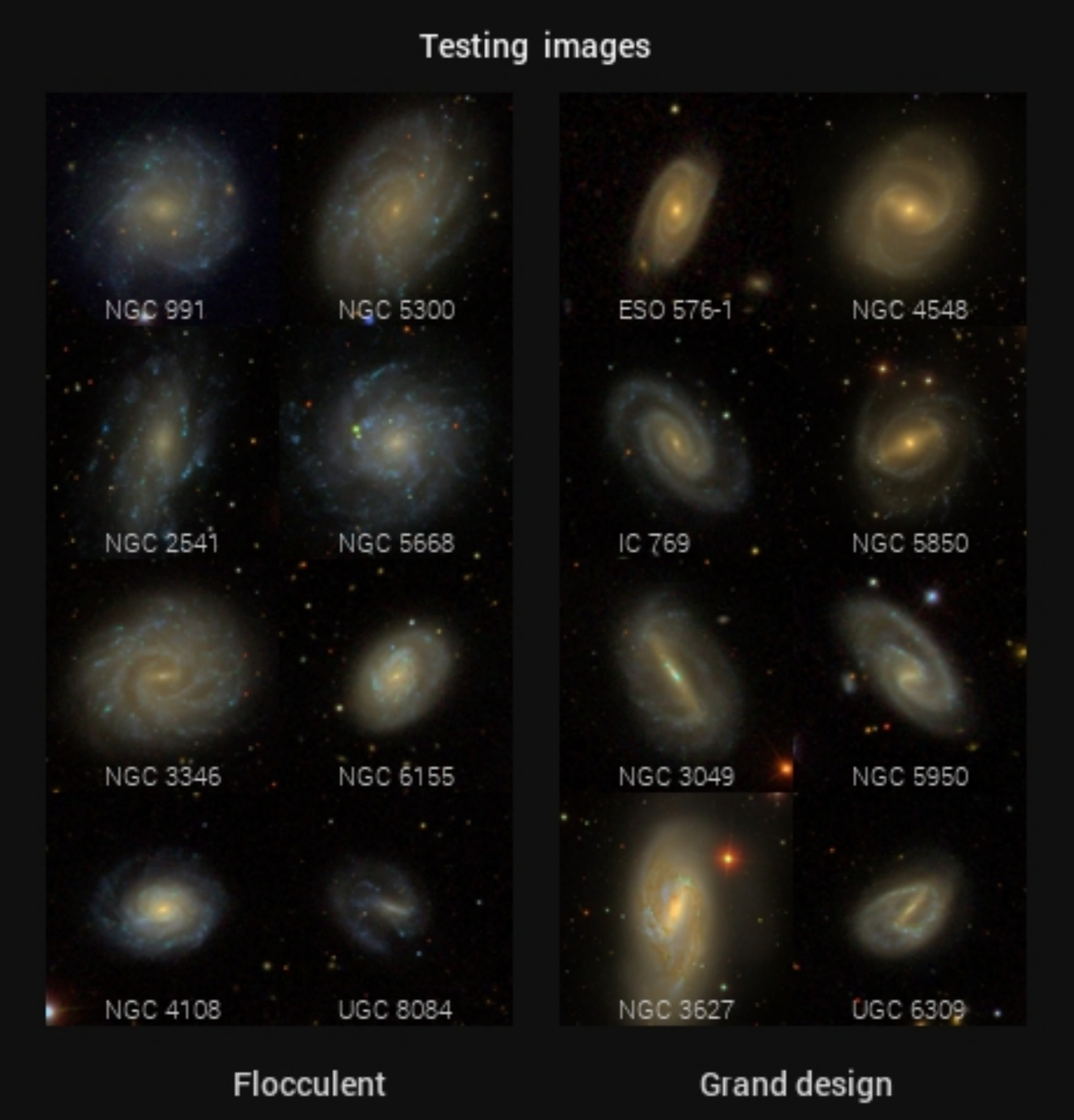}}}
\label{fig:traintest}
\caption{[Left] A subset of the sample of {\it Flocculent} and  {\it Grand-design} galaxies used for training and validation of our DCNN. [Right] A subset of the sample of {\it Flocculent} and  {\it Grand-design} galaxies used for testing our DCNN }
\end{figure*}

In the last two decades, SDSS \citep{york00} successfully mapped one third of the celestial sky, having one of the largest sky coverages. SDSS offers high 
resolution optical pictures of nearby galaxies with a seeing width (full width half maxima in the point spread function) of $\sim \SI{1.3}{\arcsecond}$ and sky 
brightness sensitivity of $\sim 20.8 \mathrm{\; mag/arcsec^{2}}$ in $r$-band \citep{fukugita96,gunn98,ross11}. We attempt to obtain the image cut-outs of $123$ 
{\it Grand-designs} and $321$ {\it Flocculent} spirals from SDSS DR17 image server, out of the $201$ {\it Grand-designs} and $553$ {\it Flocculent} galaxies 
classified by \cite{buta15}. Images outside SDSS footprint could not be retrieved. Also, we do not consider the {\it Multiple-arm} spirals, which were also part 
of the above classification. From the remaining, images with very low contrast or very high interference from nearby sources are discarded. We visually inspect 
and neglect the images where the galaxy (in optical band) shows discrepancy over the morphological tag provided by \citet{buta15} (in mid-IR). Finally, we are 
left with $360$ images, among which $270$ are {\it Flocculents} and 90 {\it Grand-designs}. We randomly select 288 (80\%) of the images for training of the DCNN 
model, and the rest of the 72 (20\%) images are kept for testing its performance. In \autoref{tab:traintest}, we list all the galaxies from SDSS DR17 which 
constitute our training, validation and testing sample. \autoref{fig:traintest} [Left Panel] represents a sample of images from the training set and \autoref{fig:traintest} [Right Panel] a sample of images from the testing set.

\small{
\begin{table*}
\caption{Training, Validation and Testing sets of  {\it Grand-design} (G) and {\it Flocculent} (F) galaxies for our DCNN model}
\label{tab:traintest}
\begin{tabular}{cccccccccccc}
\hline 
\multicolumn{12}{c}{Training \& Validation set}\\
\hline
NGC	157	    &	G	&	NGC	3627	&	G	&	NGC	4496A	&	F	&	NGC	5713	&	F	&	UGC	5238	&	F	&	UGC	9936	&	F	\\
NGC	298	    &	F	&	NGC	3629	&	F	&	NGC	4502	&	F	&	NGC	5774	&	F	&	UGC	5391	&	F	&	UGC	9951	&	F	\\
NGC	337	    &	F	&	NGC	3659	&	F	&	NGC	4517	&	F	&	NGC	5789	&	F	&	UGC	5478	&	F	&	UGC	10020	&	F	\\
NGC	337A    &	F	&	NGC	3664	&	F	&	NGC	4519	&	F	&	NGC	5916	&	G	&	UGC	5522	&	F	&	UGC	10041	&	F	\\
NGC	428	    &	F	&	NGC	3689	&	G	&	NGC	4525	&	F	&	NGC	5957	&	G	&	UGC	5676	&	F	&	UGC	10310	&	F	\\
NGC	450	    &	F	&	NGC	3718	&	G	&	NGC	4561	&	F	&	NGC	5958	&	F	&	UGC	5695	&	F	&	UGC	10437	&	F	\\
NGC	672	    &	F	&	NGC	3755	&	F	&	NGC	4571	&	F	&	NGC	5962	&	F	&	UGC	5707	&	F	&	UGC	10445	&	F	\\
NGC	755	    &	F	&	NGC	3782	&	F	&	NGC	4592	&	F	&	NGC	6015	&	F	&	UGC	5740	&	F	&	UGC	10791	&	F	\\
NGC	941	    &	F	&	NGC	3795A	&	F	&	NGC	4595	&	F	&	NGC	6106	&	F	&	UGC	5832	&	F	&	UGC	10854	&	F	\\
NGC	1022	&	G	&	NGC	3846A	&	F	&	NGC	4618	&	F	&	NGC	6207	&	F	&	UGC	5934	&	F	&	UGC	12151	&	F	\\
NGC	1051	&	F	&	NGC	3850	&	F	&	NGC	4625	&	F	&	NGC	6239	&	F	&	UGC	5976	&	F	&	UGC	12681	&	F	\\
NGC	1068	&	G	&	NGC	3876	&	F	&	NGC	4632	&	F	&	NGC	6267	&	F	&	UGC	5989	&	F	&	UGC	12682	&	F	\\
NGC	1084	&	F	&	NGC	3906	&	F	&	NGC	4635	&	F	&	NGC	6395	&	F	&	UGC	6162	&	F	&	UGC	 2443	&	G	\\
NGC	1087	&	F	&	NGC	3913	&	F	&	NGC	4658	&	F	&	NGC	7416	&	G	&	UGC	6249	&	F	&	UGC	 6903	&	G	\\
NGC	1299	&	F	&	NGC	3930	&	F	&	NGC	4668	&	F	&	NGC	7479	&	G	&	UGC	6320	&	F	&	UGC	 8658	&	G	\\
NGC	2543	&	G	&	NGC	3949	&	F	&	NGC	4765	&	F	&	NGC	7497	&	F	&	UGC	6345	&	F	&	UGC	 9858	&	G	\\
NGC	2681	&	G	&	NGC	3985	&	F	&	NGC	4900	&	F	&	NGC	7625	&	F	&	UGC	6534	&	F	&	PGC	7900	&	F	\\
NGC	2684	&	F	&	NGC	4020	&	F	&	NGC	4904	&	F	&	NGC	 4165	&	G	&	UGC	6816	&	F	&	PGC	8295	&	F	\\
NGC	2701	&	F	&	NGC	4032	&	F	&	NGC	4961	&	F	&	NGC	 4258	&	G	&	UGC	6818	&	F	&	PGC	35705	&	F	\\
NGC	2710	&	G	&	NGC	4037	&	F	&	NGC	5002	&	F	&	NGC	 4260	&	G	&	UGC	6849	&	F	&	PGC	43020	&	F	\\
NGC	2735	&	G	&	NGC	4049	&	F	&	NGC	5068	&	F	&	NGC	 4314	&	G	&	UGC	6879	&	F	&	PGC	49521	&	F	\\
NGC	2743	&	F	&	NGC	4080	&	F	&	NGC	5117	&	F	&	NGC	 4321	&	G	&	UGC	6900	&	F	&	PGC	66559	&	F	\\
NGC	2780	&	G	&	NGC	4100	&	G	&	NGC	5147	&	F	&	NGC	 4378	&	G	&	UGC	6930	&	F	&	PGC	68771	&	F	\\
NGC	2854	&	G	&	NGC	4108B	&	F	&	NGC	5194	&	G	&	NGC	 4412	&	G	&	UGC	7009	&	F	&	PGC	69293	&	F	\\
NGC	2856	&	G	&	NGC	4116	&	F	&	NGC	5204	&	F	&	NGC	 4413	&	G	&	UGC	7133	&	F	&	PGC	72252	&	F	\\
NGC	2964	&	G	&	NGC	4136	&	F	&	NGC	5205	&	G	&	NGC	 4450	&	G	&	UGC	7271	&	F	&	PGC	 11248	&	G	\\
NGC	3020	&	F	&	NGC	4141	&	F	&	NGC	5289	&	G	&	NGC	 4531	&	G	&	UGC	7699	&	F	&	IC	163	    &	F	\\
NGC	3023	&	F	&	NGC	4142	&	F	&	NGC	5339	&	G	&	NGC	 4579	&	G	&	UGC	7700	&	F	&	IC	167	    &   G	\\
NGC	3031	&	G	&	NGC	4178	&	F	&	NGC	5347	&	G	&	NGC	 4725	&	G	&	UGC	8041	&	F	&	IC	718	    &	F	\\
NGC	3057	&	F	&	NGC	4180	&	G	&	NGC	5371	&	G	&	NGC	 4795	&	G	&	UGC	8042	&	F	&	IC	749	    &	F	\\
NGC	3153	&	F	&	NGC	4192	&	G	&	NGC	5383	&	F	&	NGC	 5116	&	G	&	UGC	8053	&	F	&	IC	776	    &	F	\\
NGC	3169	&	F	&	NGC	4204	&	F	&	NGC	5443	&	G	&	NGC	 5248	&	G	&	UGC	8153	&	F	&	IC	797	    &	F	\\
NGC	3177	&	G	&	NGC	4238	&	F	&	NGC	5448	&	G	&	NGC	 6181	&	G	&	UGC	8282	&	F	&	IC	800	    &	F	\\
NGC	3185	&	G	&	NGC	4276	&	F	&	NGC	5474	&	F	&	NGC	 7328	&	G	&	UGC	8385	&	F	&	IC	851	    &	F	\\
NGC	3206	&	F	&	NGC	4288	&	F	&	NGC	5486	&	F	&	NGC	 7463	&	G	&	UGC	8449	&	F	&	IC	1014	&	F	\\
NGC	3225	&	F	&	NGC	4294	&	F	&	NGC	5566	&	G	&	NGC	 7714	&	G	&	UGC	8516	&	F	&	IC	1066	&	F	\\
NGC	3227	&	G	&	NGC	4299	&	F	&	NGC	5569	&	F	&	NGC	 7743	&	G	&	UGC	8588	&	F	&	IC	1125	&	F	\\
NGC	3287	&	F	&	NGC	4303A	&	F	&	NGC	5577	&	F	&	NGC	 7798	&	G	&	UGC	8597	&	F	&	IC	1151	&	G	\\
NGC	3299	&	F	&	NGC	4351	&	F	&	NGC	5587	&	G	&	NGC	 7817	&	G	&	UGC	8733	&	F	&	IC	1251	&	F	\\
NGC	3351	&	G	&	NGC	4353	&	F	&	NGC	5600	&	F	&	UGC	1014	&	F	&	UGC	8851	&	F	&	IC	2604	&	F	\\
NGC	3368	&	G	&	NGC	4376	&	F	&	NGC	5624	&	F	&	UGC	1551	&	F	&	UGC	8877	&	F	&	IC	3102	&	G	\\
NGC	3381	&	F	&	NGC	4384	&	F	&	NGC	5630	&	F	&	UGC	1862	&	F	&	UGC	8909	&	F	&	IC	3115	&	G	\\
NGC	3455	&	F	&	NGC	4385	&	F	&	NGC	5661	&	G	&	UGC	2081	&	F	&	UGC	8995	&	F	&	IC	3259	&	F	\\
NGC	3488	&	F	&	NGC	4390	&	F	&	NGC	5667	&	F	&	UGC	4499	&	F	&	UGC	9215	&	F	&	IC	3391	&	F	\\
NGC	3504	&	G	&	NGC	4409	&	F	&	NGC	5669	&	F	&	UGC	4543	&	F	&	UGC	9299	&	F	&	IC	3476	&	F	\\
NGC	3549	&	G	&	NGC	4416	&	F	&	NGC	5678	&	F	&	UGC	4787	&	F	&	UGC	9601	&	F	&	IC	3517	&	F	\\
NGC	3583	&	G	&	NGC	4470	&	F	&	NGC	5693	&	F	&	UGC	4871	&	F	&	UGC	9663	&	F	&	IC	3742	&	F	\\
NGC	3589	&	F	&	NGC	4490	&	F	&	NGC	5708	&	F	&	UGC	5114	&	F	&	UGC	9875	&	F	&	ESO	539-7	&	F	\\
\hline
\multicolumn{12}{c}{ Testing set } \\
\hline
NGC	718 	&	G	&	NGC	3389	&	F	&	NGC	4534	&	F	&	NGC	5798	&	F	&	UGC	6309	&	G	&	UGC	 9569	&	F	\\
NGC	991	    &	F	&	NGC	3433	&	G	&	NGC	4548	&	G	&	NGC	5850	&	G	&	UGC	6575	&	F	&	UGC	 9661	&	F	\\
NGC	2537	&	F	&	NGC	3445	&	F	&	NGC	4569	&	G	&	NGC	5899	&	G	&	UGC	6713	&	F	&	UGC	10290	&	F	\\
NGC	2541	&	F	&	NGC	3507	&	G	&	NGC	4597	&	F	&	NGC	5915	&	F	&	UGC	7239	&	F	&	UGC	12707	&	F	\\
NGC	2552	&	F	&	NGC	3623	&	G	&	NGC	4630	&	F	&	NGC	5949	&	F	&	UGC	 7590	&	F	&	UGC	12856	&	F	\\
NGC	2919	&	G	&	NGC	3626	&	G	&	NGC	4633	&	F	&	NGC	5950	&	G	&	UGC	 7690	&	F	&	PGC	6667	&	F	\\
NGC	3049	&	G	&	NGC	3794	&	F	&	NGC	5300	&	F	&	NGC	6155	&	F	&	UGC	 8056	&	F	&	PGC	42868	&	F	\\
NGC	3213	&	F	&	NGC	3982	&	G	&	NGC	5334	&	F	&	NGC	6255	&	F	&	UGC	 8084	&	F	&	PGC	66242	&	F	\\
NGC	3246	&	F	&	NGC	4108	&	F	&	NGC	5346	&	F	&	UGC	1195	&	F	&	UGC	 8489	&	F	&	IC	750	    &	G	\\
NGC	3264	&	F	&	NGC	4120	&	F	&	NGC	5585	&	F	&	UGC	1547	&	F	&	UGC	 8892	&	F	&	IC	769	    &	G	\\
NGC	3274	&	F	&	NGC	4234	&	F	&	NGC	5645	&	F	&	UGC	4621	&	G	&	UGC	 9274	&	F	&	ESO	576-1	&	G	\\
NGC	3346	&	F	&	NGC	4498	&	F	&	NGC	5668	&	F	&	UGC	5172	&	F	&	UGC	 9310	&	F	&	ESO	576-32	&	F	\\
\hline
\end{tabular}
\end{table*}
}

\subsection{Preprocessing and augmentation}
 A number of different augmentation techniques 
are tried and tested. The process described here results in the maximum testing accuracy that we achieve. There are 4 steps to the processing of images for 
training. In the first step, we rotate the images in multiples of $10 ^\circ$, which leads to an $36-fold$ augmentation. Next, all the rotated images are flipped 
horizontally and vertically, generating an $3-fold$ augmentation for all the rotated images. In the third step, we perform another $3-fold$ augmentation by varying 
the sharpness of the images. Finally, the brightness and contrast of all the images are enhanced by $10\%$ before resizing them to $256 \times 256$ pixel. The number 
of augmented images remains unchanged in this final step. For each of the input image, we get $324$ augmented images after processing.  The training-validation set 
of $288$ images are augmented into $93312$ images, $20\%$ of which are randomly chosen for validation, and the rest for training. A summary of the number of original 
images and augmented images used for training and validation is presented in \autoref{tab:train_test_num}.

\begin{table}
	\centering
	\caption{Sample size for training and testing}
	\label{tab:train_test_num}
	\begin{tabular}{lcc} 
	    \hline
     &  Before augmentation & After augmentation  \\ 
		  \hline
        Training    & 230  & 74520 \\
        Validation  &  58  & 18792 \\
        Testing     & \multicolumn{2}{c}{72}   \\
\hline
	\end{tabular}
\end{table}

\subsection{Design of the network}
Deep Convolutional Neural Networks (DCNN)s were initially used for modelling 2-dimensional data to address problems like automated identification of handwritten characters \citep{lecun90,lecun98}. Later, it was popularized in all forms of image classifications. Convolution and pooling are primary steps in the feature extraction process of a DCNN, where the features in an input image are collected as a set of matrices and mapped into a normalised number representing an element in a categorical distribution. The hidden layers between the input and the output are sequentially connected through neurons and activated through activation functions. These neurons transport the weighted information to adjacent layers. A series of training and validation exercises are performed to estimate the best weights that model the input data for a set of images with predetermined classification. This process includes numerous iterations of forward-pass and back-propagation. One cycle of repeated training and validation with the entire training data is termed an {\it epoch}. The model weights are continuously updated after each epoch to reach the optimum value. The trained model with the best weight is then used to classify undetermined categories of images. We have implemented the DCNN model architecture based on the {\it AlexNet} architecture \citep{krizhevsky12}. The full network architecture is shown in \autoref{tab:model}.

\noindent The convolution operation is one of the most important operations in a DCNN, convolving the input matrix with the filter matrix of given dimensions. Their effective contribution is evaluated by employing an activation function, and the output feature maps are obtained by sliding the filter over the entire input matrix. Using a kernel of size $K$ with a stride of $S$ and padding $P$ on either side, an input image of $h \times w$ pixels turns into a $h^{\prime} \times w^{\prime}$ feature map after the convolution, where the height($h^{\prime}$) and width ($w^{\prime}$) evolves as
\begin{eqnarray}
\label{eq:lshape}
h^{\prime} = \frac{h - K + 2P}{S} + 1  \hspace{20 px} \mathrm{and} \hspace{20 px }
w^{\prime} = \frac{w - K + 2P}{S} + 1 .
\end{eqnarray}
 
\noindent The convolved output matrix $X^{\prime}$ after the convolution of input matrix $X$ is given as
\begin{eqnarray}
\label{eq:conv}
 X^{\prime}_{l,m}=  & ( W \ast X ) _ {l,m} & = \mathrm{{\cal F}_A} \left( \,  \sum_{i=-\rho}^{\rho} \sum_{j=-\rho}^{\rho} \; W_{i,j} \; X_{l+i,m+j} \, + \, b \, \right).
\end{eqnarray}
\noindent Here $\rho=\frac{K-1}{2}$, for a kernel of size $K\times K$. The pre-computed weight vector $W$ and the bias parameter $b$ are validated and updated at the end of each iteration. ${\cal F}_{A}$ is the activation function that determines whether a neuron should be taken into account or not. Here, we use the {\it Rectified Linear Unit (ReLU)} \citep{jarrett09,nair10} as the activation function. The function ${\cal F}_{A} (x) = max \left( 0,x \right)$ nullifies the negative values and retains the positive ones as they are. \\

\noindent The next {operation} is the {\it max-pooling}, where a downsized output layer is produced by retaining the maximum values from each specified sized kernel.
This process drastically reduces the computation cost by keeping only the sharpest features in the convoluted map. The output of the max-pooled layer is obtained as
\begin{eqnarray}
\label{eq:maxpool}
 \bar{X}_{l,m} & = & {{\cal M}ax}  \left( X^{\prime}_{K(l-1)+j,K(m-1)+i} \bigg \rvert ^{K}_{i,j=1} \right)
\end{eqnarray}

\noindent Next, to standardise the activation in the preceding layers for a given set of input images, the {\it batch normalisation} is carried out. This results in more efficient learning and prevents over-fitting of the models. This is performed for a given batch of input images by re-scaling the output from each node in the preceding layer. So each neuron carries a zero mean and unit variance. The batch normalized output layer is given as

\begin{eqnarray}
\label{eq:batchnorm}
 {X}^{N}_{l,m} & = & \frac{\bar{X}_{l,m} - \mu_{l,m}}{\sigma_{l,m}}.
\end{eqnarray}
\noindent Here, $\mu_{l,m}$ and $\sigma_{l,m}$ are the mean and standard deviation of $\bar{X}_{l,m}$ for the entire batch.\\

\begin{figure*}
\centering
\resizebox{18 cm}{!}{\rotatebox{0}{\includegraphics{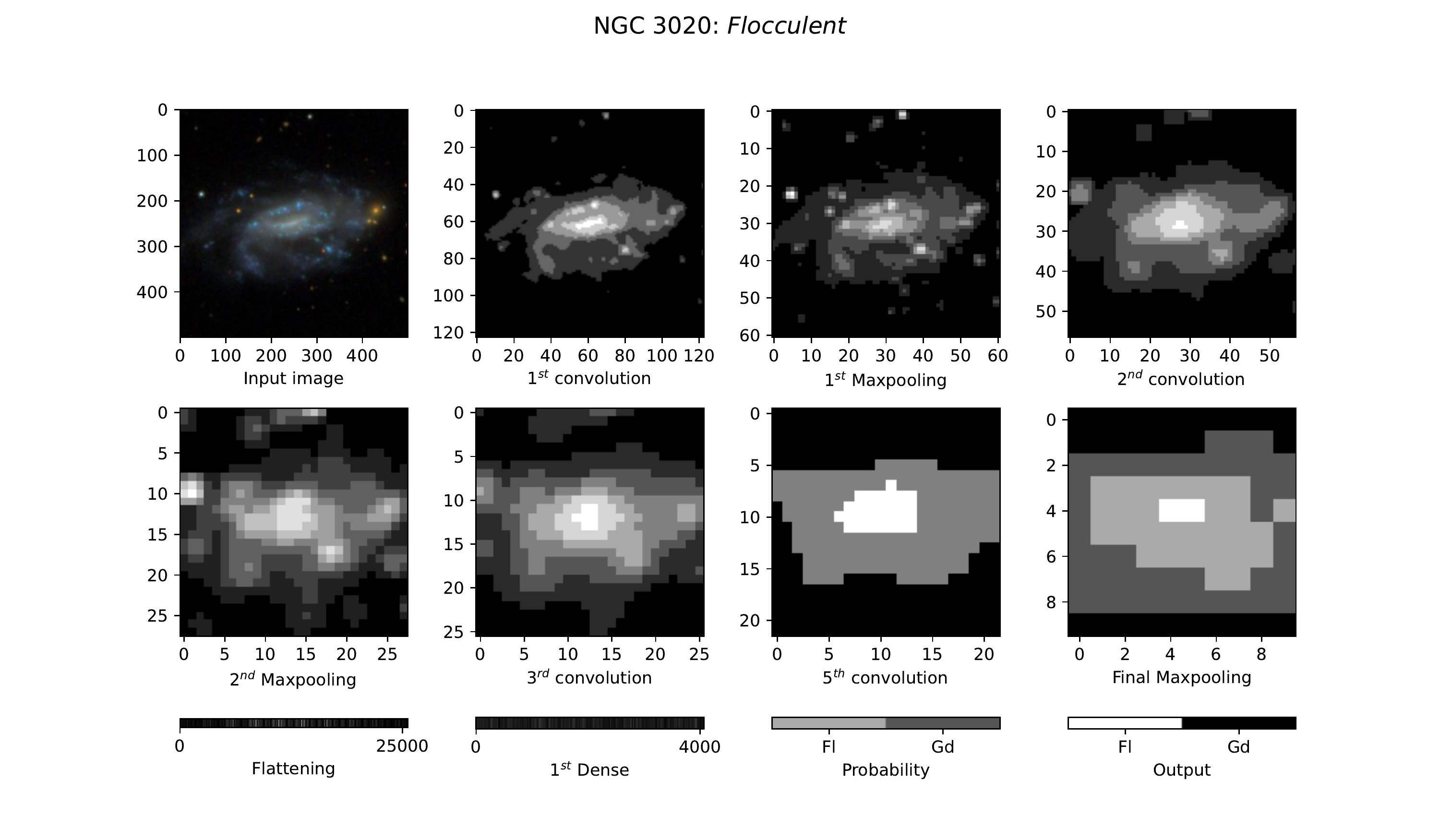}}}\\
\resizebox{18 cm}{!}{\rotatebox{0}{\includegraphics{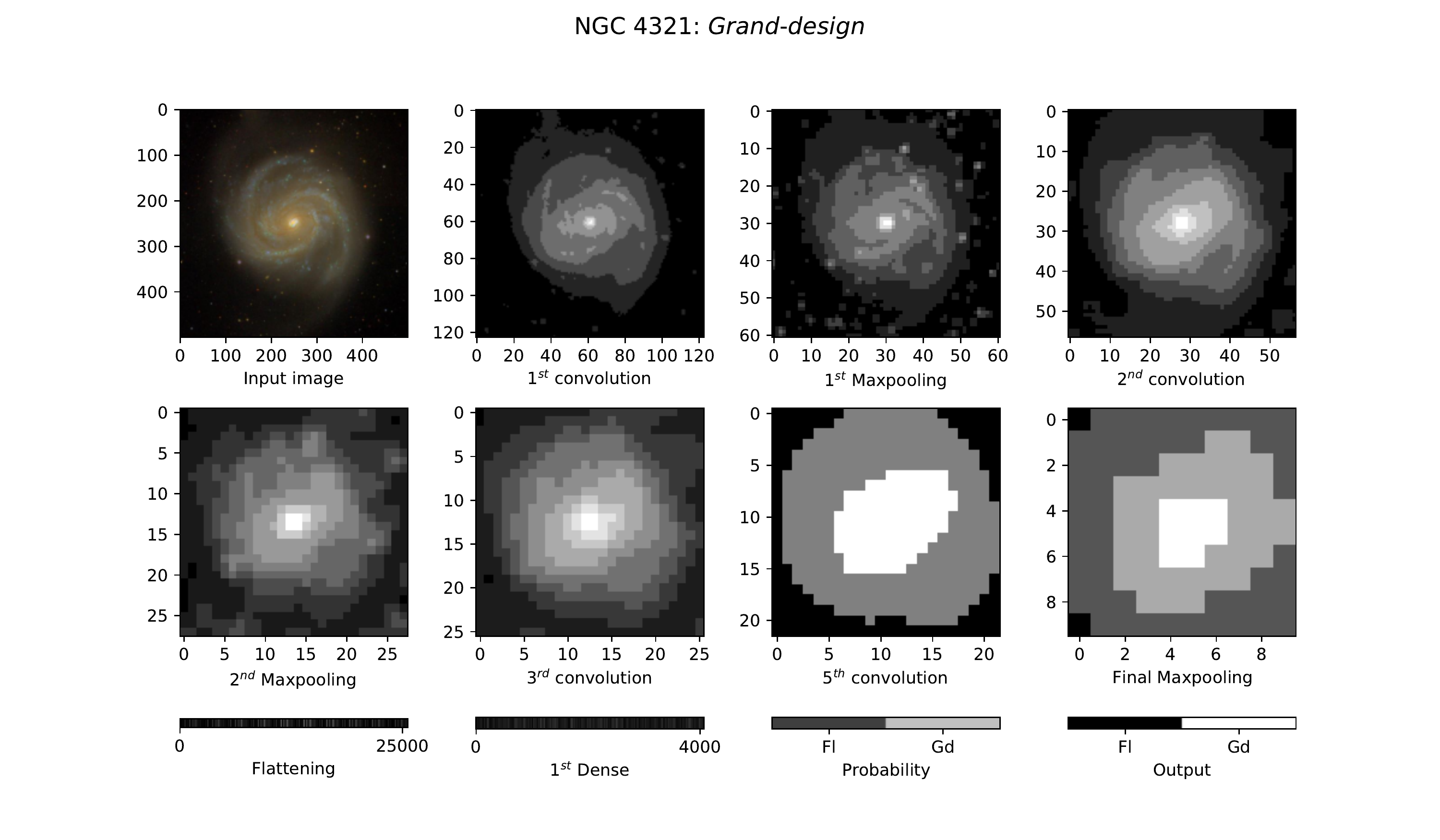}}}
\label{fig:feature_map}
\caption{[Top] Feature maps of the {\it Flocculent} galaxy {\it NGC 3020} at various stages of the DCNN. [Bottom] Feature maps of  {\it Grand-design} {\it NGC 4321}. Both the galaxies are from the training set. The cumulative projection of all filters is shown for the feature maps. The gray color scale is set to show values from 0 (black) to 1 (white).}
\end{figure*}

We use 5 convolution layers and 3 max-pooling layers, with 2 intermediate batch normalization operations. The output of the final pooling layer is then flattened into a 1-D array. This flattened matrix then passes through successive {\it dense} \citep{huang16,joseph21} and {\it dropout} \citep{krizhevsky12, srivastava14} layers  (\autoref{tab:model}). In a dense layer, neurons of the layer are connected to every neuron in the preceding layer. Dense is also activated using the ReLU activation function. Dropout layers reduce the chances of overfitting the model by randomly dropping neurons. The final dense layer is activated using the {\it softmax} \citep{bridle90,gold96,bishop06} activation function. The feature maps at different stages of our model building are shown in \autoref{fig:feature_map}.

\small{
\begin{table*}
 \centering
 \caption{Details of the architecture for our DCNN model}
 \label{tab:model}
\begin{tabular}{cccccccc}
\hline
\multirow{2}{*}{\textbf{ Sl. No.}} & \textbf{ Layer / }   &  \textbf { Kernel size /}  & \textbf{Size of feature map /}  & \textbf{Number of}   &  \textbf{Number of} \\
                                   & \textbf{ process } &  \textbf { pool size }    & \textbf{Flattened layer}      & \textbf{Filters}  &  \textbf{Parameters}\\
\hline
& & & & & &\\
1.  & 2D Convolution (1) & $11\times11$ & $123 \times 123$ & $96$ & $34944$\\
& & & & & &\\
2.  & Max-pooling (1) & $3\times3$ & $61 \times 61$ & $96$ & - \\
& & & & & &\\
3.  & Batch normalization (1) & - & $61 \times 61$ & $96$ & $384$ \\
& & & & & &\\
4.  & 2D convolution (2) & $5\times5$ & $57 \times 57$ & $256$ & $614656$\\
& & & & & &\\
5.  & Max-pooling (2) & $3\times3$ & $28 \times 28$ & $256$ & - \\
& & & & & &\\
6.  & Batch normalization (2) & - & $28 \times 28$ & $256$ & $1024$\\
& & & & & &\\
7.  & 2D convolution (3) & $3\times3$ & $26 \times26$ & $384$ & $885120$ \\
& & & & & &\\
8.  & 2D convolution (4) & $3\times3$ & $24 \times24$ & $384$ & $1327488$ \\
& & & & & &\\
9.  & 2D convolution (5) & $3\times3$ & $22\times22$ & $256$ & $884992$ \\
& & & & & &\\
10. & Max-pooling (3) & $3\times3$ &  $10\times10$ & $256$ & - \\
& & & & & &\\
\hline
& & & & & &\\
11. & Flattening & -  & $1\times 25600$ & - & - \\
& & & & & &\\
12. & Dense (1) & -  & $ 1\times 4096 $ & - & $4198400$\\
& & & & & &\\
13. & Dropout - 50 \%  (1) & - &  $ 1\times 4096 $ & - & - \\
& & & & & &\\
14. & Dense (2) & - & $ 1\times 4096 $ & - & $16781312$\\
& & & & & &\\
15  & Dropout - 50 \% (2) & - & $1\times 4096 $ & - & - \\
& & & & & &\\
16  & Dense (3) & - & $1\times 2$  & - & $8194$\\
& & & & & &\\
\hline
 \end{tabular}
\end{table*}
}

\subsection{Loss and Accuracy}
\begin{figure*}
\resizebox{8. cm}{!}{\rotatebox{0}{\includegraphics{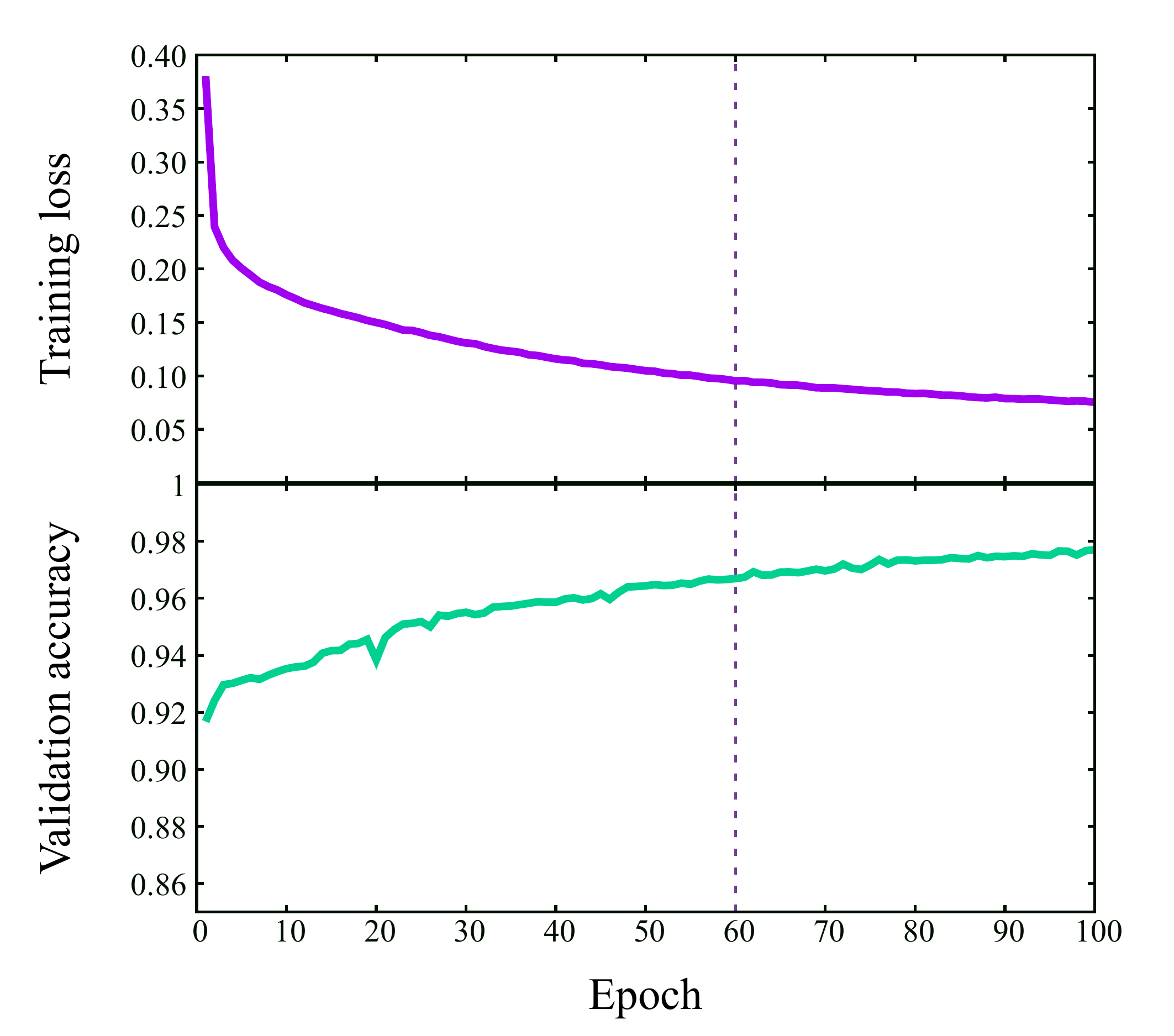}}}  \hspace{12px}
\resizebox{8.2 cm}{!}{\rotatebox{0}{\includegraphics{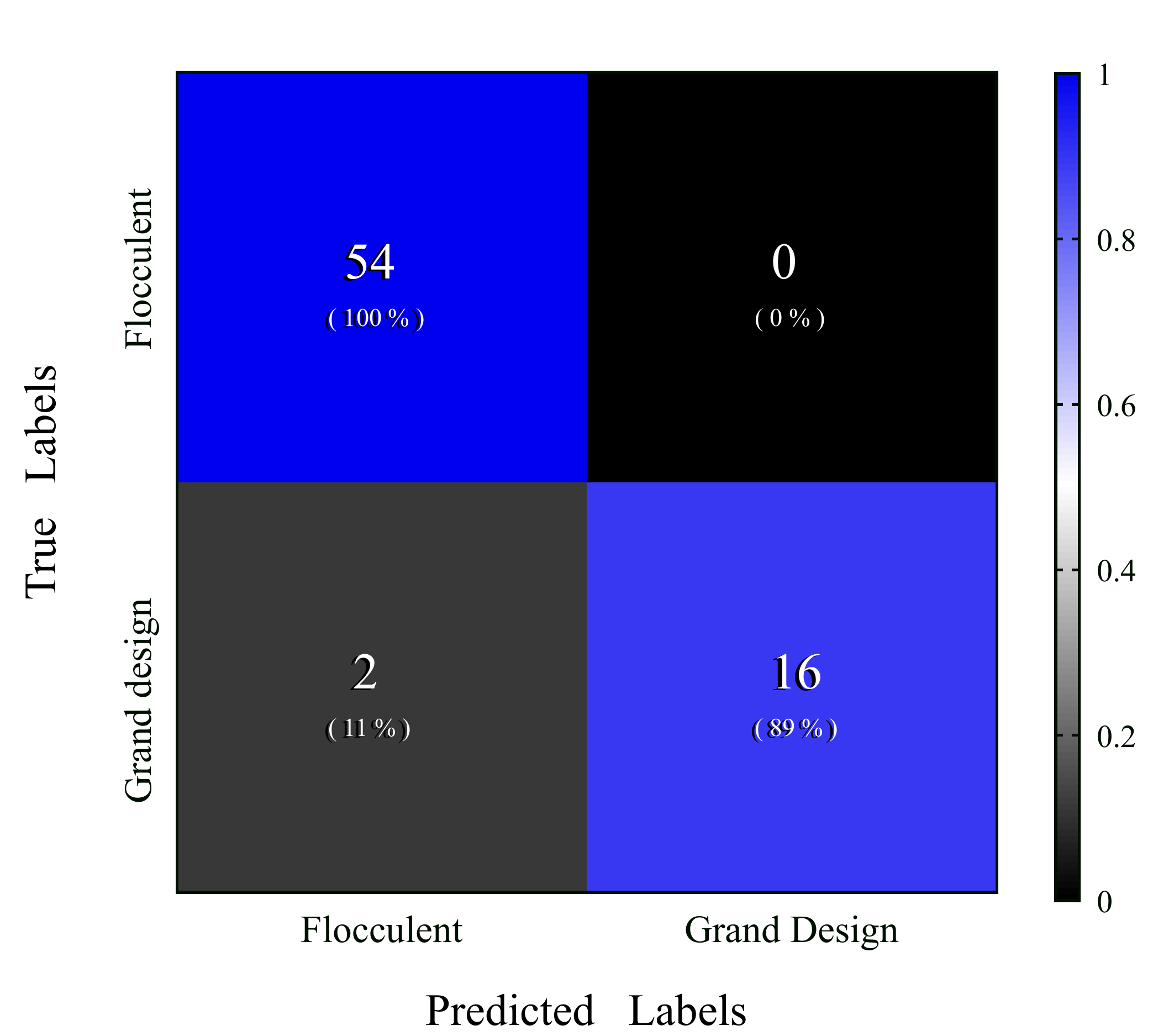}}}
\label{fig:val_acc}
\caption{[Left] The top panel shows the estimated loss during training, and the bottom panel gives the validation accuracy at different epochs.
The vertical dashed line indicates the epoch at which the model is found to have reached optimum testing accuracy. [Right] Confusion matrix
showing the number (percentage) of images correctly and falsely predicted in each class. }
\end{figure*}

While training, the model weights are updated by minimizing the loss function. The loss function is the metric that quantifies the dissimilarity between the actual values and the predictions made by the model. A drop in the loss value indicates the model is tending towards perfection. In this case, we chose \textit{binary cross-entropy} \citep{ruben04} as the loss function. Binary cross-entropy, averaged over support $N$, can be written as  
\begin{eqnarray}
\label{eq:loss}
L(W_i)  =  - \frac{1}{N} \sum\limits_{j=1}^{N}  { \left[ T_j \cdot \log{M_j (W_i)} + ( 1 - T_j ) \cdot \log{\left( 1 - M_j (W_i) \right)} \right]} .
\end{eqnarray}
Here $M_j$ is the modelled value of the $j^{th}$ scalar for the $i^{th}$ set of weights ($W_i$), and $T_j$ is the actual value we target to achieve. Meanwhile, accuracy is calculated as
the ratio of the correct predictions to the total predictions. In \autoref{fig:val_acc} [Left Panel], we plot the training loss and the validation accuracy as a function of number of epochs. \\

\subsection{Optimization of the model}
There are many optimization techniques available for updating the model weights to reach an optimum level. We opt for the {\it Stochastic Gradient Descent (SGD)} optimizer \citep{bottou98} while training our DCNN model. The updated model weights at the end of the i$^{th}$ epoch can be computed as
\begin{eqnarray}
\label{eq:opt1}
 W_{i+1} & = & W_i+V_{i+1}.
\end{eqnarray}
Here, $V$  is the incremental weight correction, which changes with each iteration as follows.
\begin{eqnarray}
\label{eq:opt2}
 V_{i+1} & = & \alpha V_i - \epsilon_i \bigg \langle \frac{\Delta L(W)}{\Delta W } \big \rvert _{W_i} \bigg \rangle_{B_i}.
\end{eqnarray}

\noindent $\frac{\Delta L}{\Delta W} \big \rvert _{W_i}$ is the gradient of the loss for a particular weight $W_{i}$. An ensemble average of the gradients are taken over the entire batch $B$ at the $i^{th}$ iteration. $\alpha$ is called the {\it momentum } coefficient and $\epsilon_i$ is the {\it learning rate} for the $i^{th}$ iteration. The learning rate is a parameter that controls the fractional change in model weights brought on by the differential change in the loss function. The learning rate is revised at every epoch using the {\it decay} parameter ($\delta$). The gradual reduction of the learning rate helps the model learn about complex patterns as it reaches the global minima. With the initial value being $\epsilon_0$, at any $i^{th}$ epoch the learning rate is found to be $\epsilon_{i} =\epsilon_0 (1 + i \delta )^{-1}$. On the other hand, the momentum coefficient ($\alpha$) determines how much of the previous learning will carry forward to the next epoch. Its primary goal is to ensure that a local minima do not halt learning and that the global minima of the loss function is quickly located. The choices of the hyper-parameters that led to our optimised model are $\alpha=0.9$, $\epsilon_0= 5 \times 10^{-5}$, $\delta = 2 \times 10^{-4}$. The batch size for training is set at $64$.

\subsection{Testing of the model}
The performance of the DCNN model in each class is assessed using the following parameters: {\it precision}, {\it recall}, and {\it $f_1$-score}.These parameters are calculated from the number of \textit{true positives} ($P_T$), \textit{true negatives} ($N_T$), \textit{false positives} ($P_F$) and \textit{false negatives} ($N_F$) identifications in each class.

\noindent \textit{Precision} is the positive predictive value, given by  
\begin{eqnarray}
\label{eq:pr}
\mathrm{precision} &=& \mathrm{\frac{P_T}{P_T+P_F}}. \\ \nn
\end{eqnarray}
\noindent while \textit{recall}  measures the hit rate or the sensitivity of the test and is given by
\begin{eqnarray}
\label{eq:rec}
\mathrm{recall} &=& \mathrm{\frac{P_T}{P_T+N_F}}. \\ \nn
\end{eqnarray}
\noindent  \textit{$f_1$-score} is the harmonic mean of precision and recall,
\begin{eqnarray}
\label{eq:f1}
f_1 &=& \mathrm{2 \cdot \, \frac{precision \times recall }{ precision + recall}}. \\ \nn
\end{eqnarray}
\noindent Finally, the overall testing accuracy can be estimated by the ratio of correct predictions to total predictions
\begin{eqnarray}
\label{eq:acc}
\mathrm{accuracy} = \mathrm{\frac{P_T + N_T}{P_T+N_T+P_F+N_F}}. \\ \nn
\end{eqnarray}

\noindent The performance of our DCNN model and the testing accuracy are presented in \autoref{tab:pred_res}. Although we go up to 100 epochs, the testing accuracy attains its maximum value of $97.22 \%$ after the $58^{th}$ epoch. Nevertheless, we check up to 100 epochs for any increment in testing accuracy. In \autoref{fig:val_acc} [Right Panel], we present the confusion matrix, indicating the number of true and false classifications in each category. To ensure that the model weights are optimum and free from over-fitting, we choose the model weights at epoch 60 to be the \textit{best weights}.

\begin{table*}
\centering
\caption{Prediction accuracy for the testing set from our DCNN model}
\label{tab:pred_res}
\begin{tabular}{lcccccc}
\hline
                    & True        &\multicolumn{2}{c}{Predicted as}   & Precision &  Recall   &  $f_1$-score  \\
                    & labels &   {\it Flocculent}  &   {\it Grand-design}   &           &           &            \\
\hline
{\it Flocculent}          &     54      &       54          &       0           &    0.96   &   1.00    &   0.98     \\
 {\it Grand-design}        &     18      &         2        &       16           &    1.00   &   0.89    &   0.94     \\
\hline
Accuracy    &\multicolumn{6}{c}{\textbf{ 97.22 \% }}  \\
\hline
\end{tabular}
\end{table*}

\begin{figure*}
\resizebox{8.6 cm}{!}{\rotatebox{0}{\includegraphics{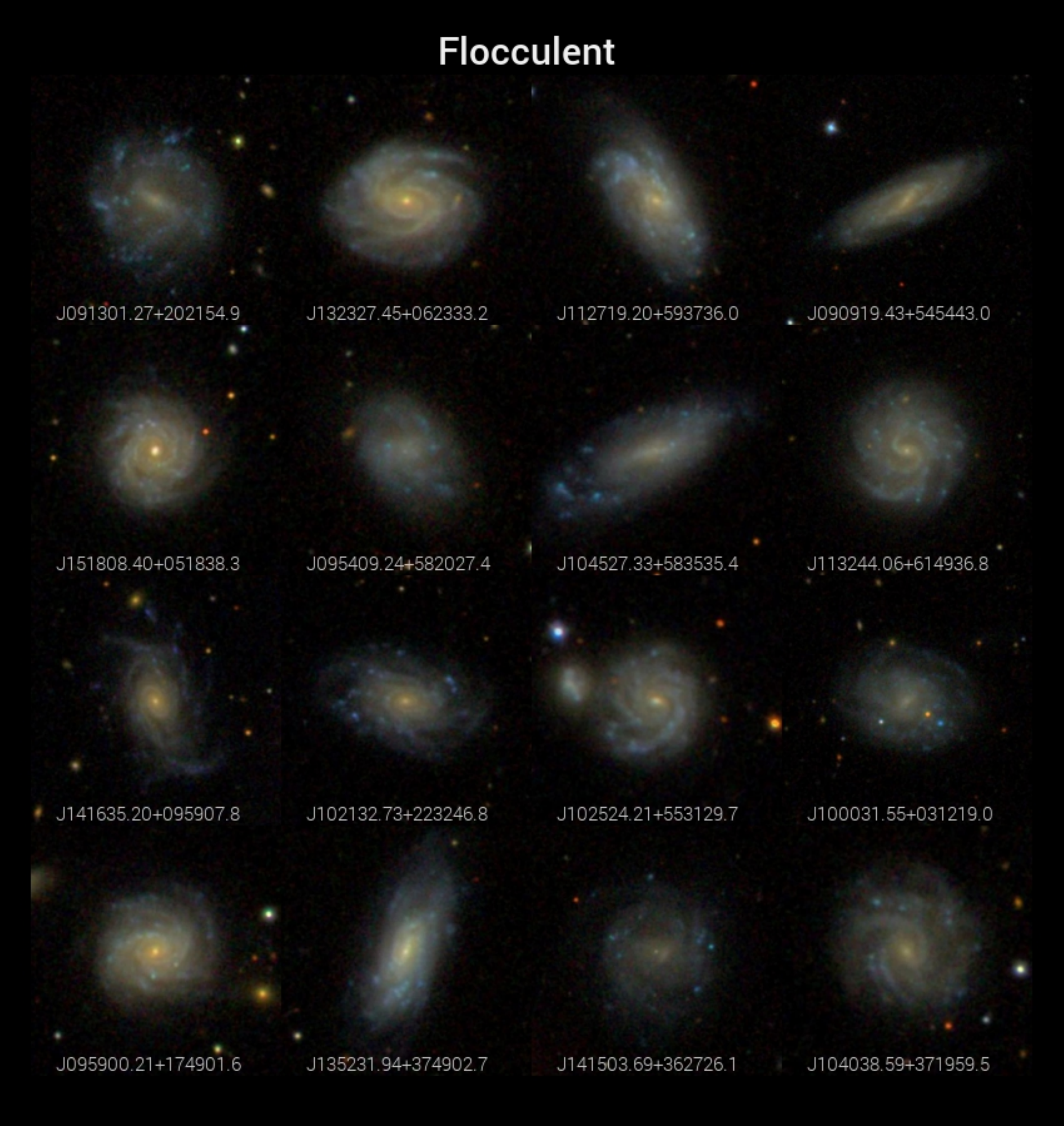}}} \hspace{0.3 cm}
\resizebox{8.6 cm}{!}{\rotatebox{0}{\includegraphics{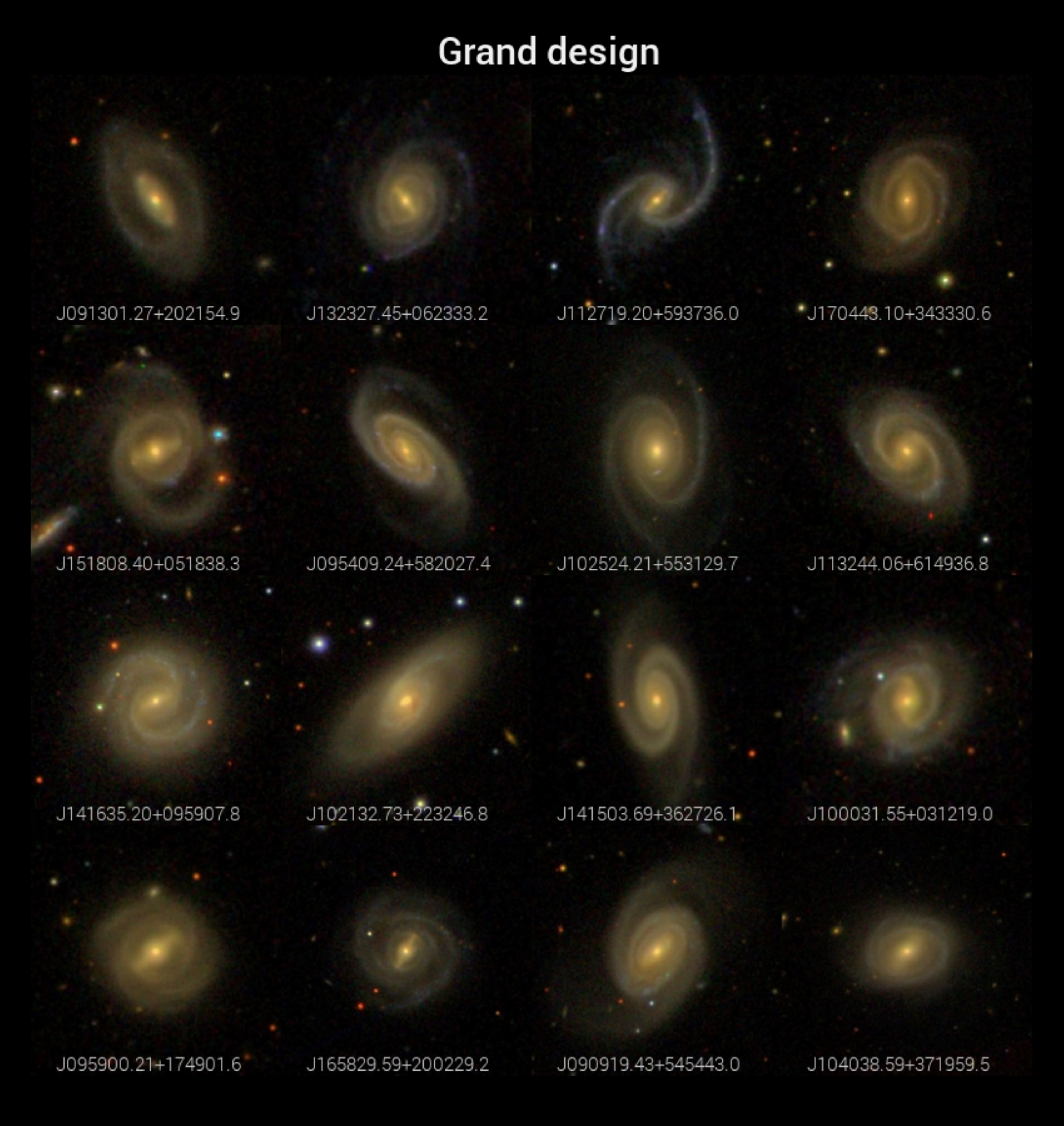}}}
\label{fig:sdssclass}
\caption{A subset of the {\it Flocculent} [Left] and  {\it Grand-design} [Right] galaxies from SDSS newly classified by our DCNN model }
\end{figure*}
\section{Classification of spiral galaxies using the trained DCNN model}

\noindent Finally, we use our trained DCNN model for classifications of a large set of spiral galaxies from SDSS DR17, which were hitherto unclassified. We begin by considering the $32059$ galaxies identified as spirals in the galaxy sample targeted for morphological classification in Galaxy zoo2 \citep{willett13}.  Galaxy zoo2, the advanced phase of Galaxy zoo \citep{lintott08,lintott11}, gives the morphological classifications for 3 million galaxies. However, the  {\it Grand-design} and {\it Flocculent} nature of the spiral galaxies are not inspected deeply therein. In the next step, we obtain the image cut-outs of these spiral galaxies from SDSS DR17 server. Then the {\it zoo2MainSpecz} table in the DR17 database of SDSS is joined with the {\it PhotoObjAll} table, in order to find the galaxies which have $PetroR90\_r > 20 \SI{}\arcsecond$ and $total\_{votes} > 20$; $PetroR90\_r$ is the radius that contains 90\% of the Petrosian flux. Setting a threshold for the $total_votes$ flag ensures the reliability of the classifications in Galaxy zoo2. The criteria used here ensures that the resolution of the image is good enough for classification. While extracting the data, we do not impose any redshift ($z$) cut on the sample. The high threshold value of $PetroR90\_r$  allows the galaxies at greater distances but of large physical sizes to be included in the sample with a workable image resolution. However, we find such galaxies of large physical sizes only up to $z \sim 0.12$. In order to eliminate the multiple-arm spirals from the set, we also confirm that our chosen galaxies have no more than two prominent arms. Further, edge-on galaxies, mergers, and spirals with {\it odd features} are carefully eliminated as well. Next, we exclude the images from our set that are already classified by \cite{buta15}; there are 298 such images.Finally, after discarding a number of low resolution and background noise dominated images, we are left with $1354$ galaxies. Intrinsically, the {\it Flocculents} have low arm-interarm contrast. Hence inclusion of low contrast images will bias our classification towards the {\it Flocculent} class. Therefore we discard images with low contrast by visual inspection to ensure that the classification is least affected by the image quality.

\noindent We then use our DCNN model to classify these $1354$ spiral galaxies into two classes:  {\it Grand-designs} and {\it Flocculents}. This time, we do not go up to the final softmax layer, which yields a binary classification of the images. Rather, we look at the predicted probabilities of them being  {\it Grand-design} ($p_G$) or {\it Flocculent} ($p_F$). $p_F$ and $p_G$ are complementary to each other and always add up to unity. Next, a five point classification scheme is adopted based on the values of $p_G$ and $p_F$. A galaxy is considered to be {\it Flocculent} with full certainty when $p_F =1$; we label these with a flag `$F$'. Similarly, galaxies with $p_G=1$ are labelled as `$G$'. {\it Grand-designs} and {\it Flocculents} classified with $1>p_F\geq 0.8$ and $1>p_G\geq0.8$ have lesser degree of certainty in their classification and are labelled as `$f$' and `$g$' respectively. Other galaxies with prediction probability $<0.8$ in both the classes are termed as {\it uncertain} (`$U$') as presented in \autoref{tab:sdss_class_num}. There are $134$ such galaxies out of the total $1354$, which we drop from the classified galaxy set. \emph{We identify 499 new  {\it Grand-designs} and 721 new {\it Flocculents} in our study.} The classifications of all the $1220$ galaxies, are tabulated in \autoref{tab:sdssclass}, in Appendix A. A subset of our classified  {\it Grand-designs} and {\it Flocculents} is presented in \autoref{fig:sdssclass}. The absence of {\it Grand-designs} or stationary density  waves in galaxies with low dynamical masses may be possibly explained due to the absence of an Inner Lindblad Resonance (ILR). A low dynamical mass indicates a low value of the rotational velocity, and hence the angular and epicyclic frequencies. This rules out the possibility of the angular velocity of the pattern relative to that of the disc being equal to half of the epicyclic frequency, and hence the presence of an ILR. This explains the preponderance of Flocculent or transient spirals in low mass galaxies instead of the stationary Grand-design patterns. \\

\begin{figure*}
\centering
\resizebox{80 mm}{!}{\rotatebox{0}{\includegraphics{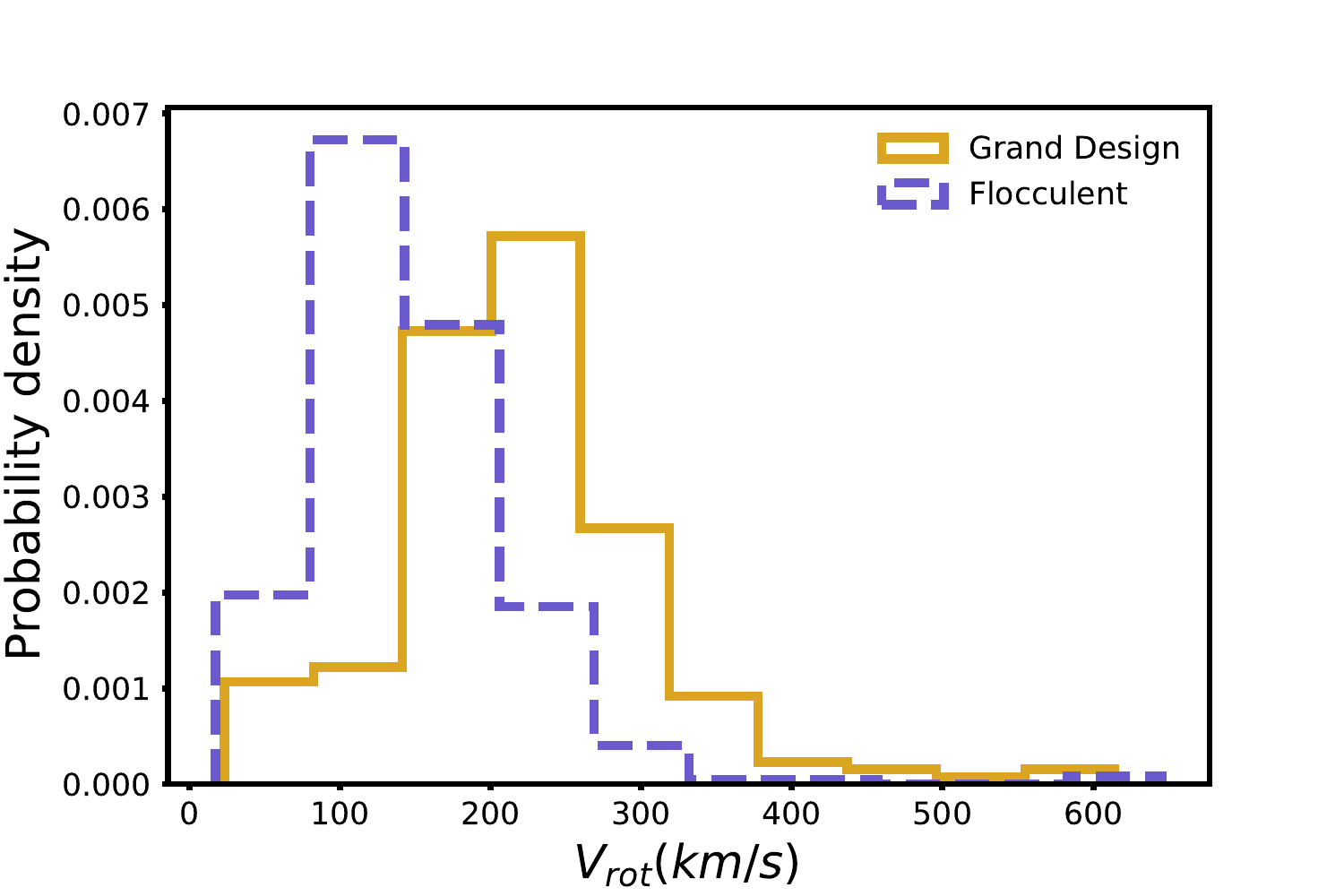}}} 
\resizebox{80 mm}{!}{\rotatebox{0}{\includegraphics{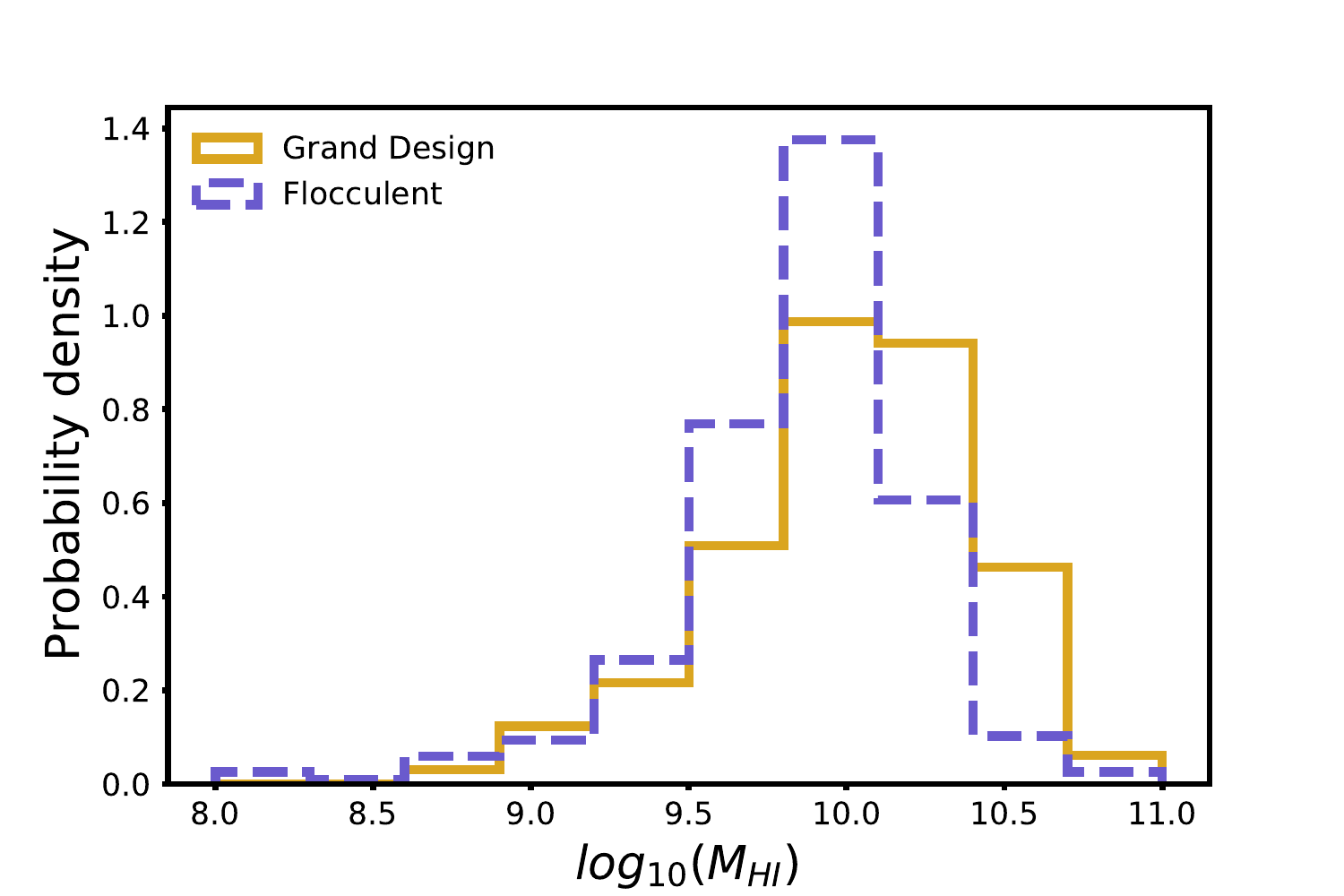}}} \\
\resizebox{80 mm}{!}{\rotatebox{0}{\includegraphics{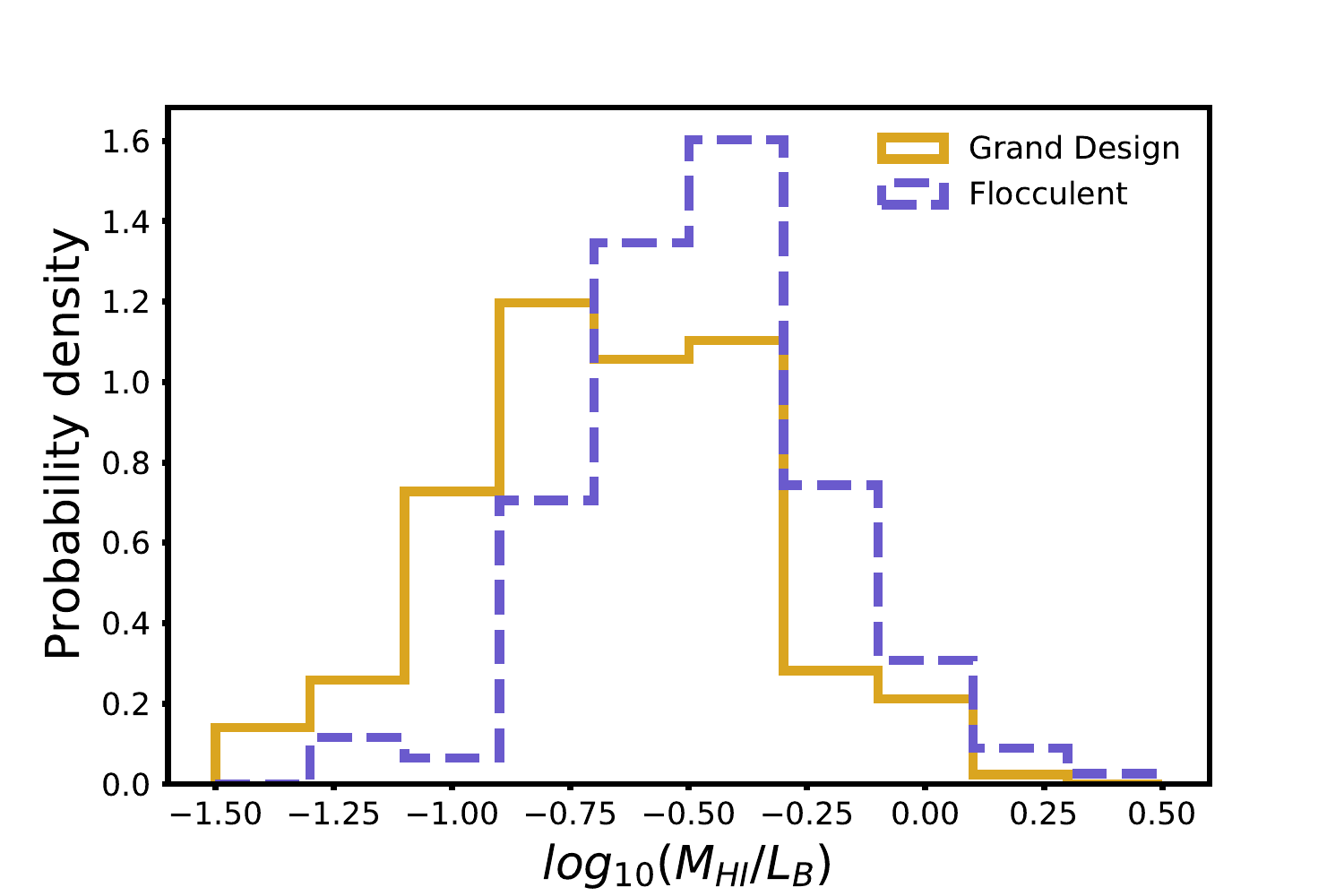}}}
\resizebox{80 mm}{!}{\rotatebox{0}{\includegraphics{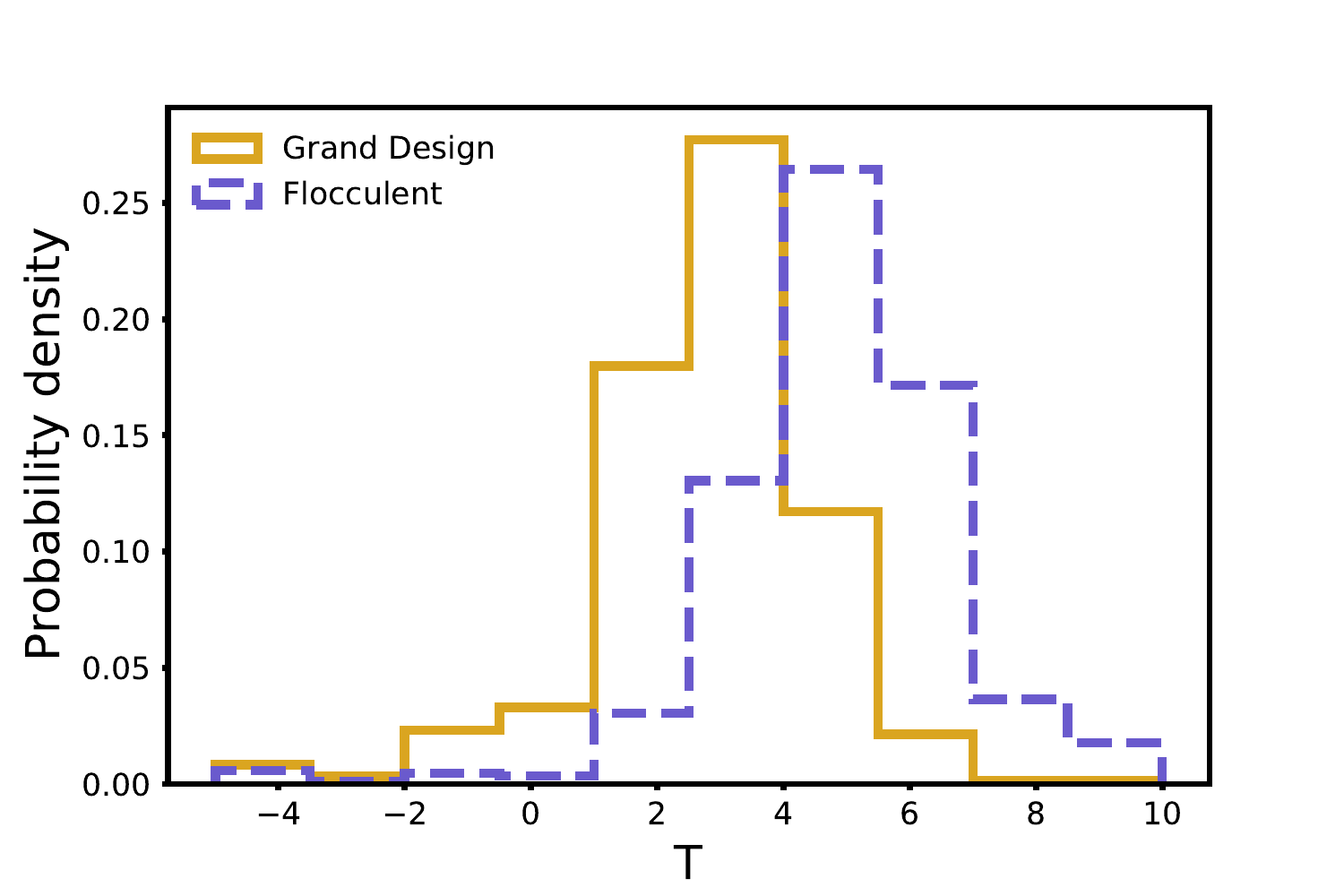}}}
\label{fig:propstudy}
\caption{Distribution of asymptotic rotational velocity $\rm{V_{rot}}$, total atomic hydrogen (HI) mass $M_{HI}$, total blue luminosity $L_B$ and ratio of HI mass-to-blue luminosity $M_{HI}/L_B$ from HyperLedA \citep{Markrov2014} for the newly classified  {\it Grand-design} and {\it Flocculent} galaxies.}
\end{figure*}

\begin{table}
	\centering
	\caption{Classified Grand-designs and Flocculents from SDSS}
	\label{tab:sdss_class_num}
	\begin{tabular}{lcc} 
	    \hline
    Confidence &  {\it Grand-designs}  & {\it Flocculents}  \\ 
		  \hline
            100 \%      &  252 & 604 \\
        80\% to 100\%      &  247 & 117 \\
            Total       &  499 & 721 \\
\hline
	\end{tabular}
\end{table}

\noindent \textbf{A comparison of physical properties:} In \autoref{fig:propstudy}, we present the distribution of physical properties of our newly classified  {\it Grand-designs} and {\it Flocculents} as collected from HyperLeda \citep{Markrov2014}. The top left panel shows the distribution of asymptotic rotational velocity ($V_{\rm{rot}}$), the mean values of which are $218\, \rm{km}\,\rm{s}^{-1}$ and $146 \,\rm{km}\, \rm{s}^{-1}$ for  {\it Grand-designs} and {\it Flocculents} galaxies respectively. This possibly implies that the  {\it Grand-designs} are more massive than {\it Flocculents} in general. The top right panel depicts the $log_{10}(M_{HI})$ distribution, with  {\it Grand-designs} having slightly higher gas mass than the {\it Flocculents}, with respective mean values in $log_{10}(M_{\odot})$ scale being 10. against 9.8. Also, the mean values of the $B$-band luminosity for the two classes are comparable, being $10.7$ and $10.3$ in $log_{10}(L_{\odot})$ respectively. The bottom left figure shows the $log_{10}(M_{HI}/L_B)$ distribution, with the  {\it Grand-designs} having a mean of $-0.7$  against $-0.5$ of {\it Flocculents}. This indicates that the  {\it Grand-designs} have a lower $(M_{HI}/L_B)$ ratio compared to {\it Flocculents}. In this context, it may be noted that the gaseous disk may be instrumental in regulating growth rate and lifetime of spirals in the galaxy disk \citep{ghosh15}.

The bottom right panel shows the distribution of de Vaucouleurs numerical stage indices $T$ \citep{deVaucouleurs1976RC2}. The mean values of $T$ for  {\it Grand-designs} and {\it Flocculents} are $2.6 \pm 1.8$ \& $4.7 \pm 1.9$ respectively, which again imply {\it Flocculents} are mostly late-types as found in \cite{Bittner2017}, and also consistent with the findings of \cite{romanishin85}. A higher mean value of T for {\it Flocculents} can also correspond to a lower ratio of bulge to disk \citep{Obreschkow_2009}. Morphology of the galaxies were obtained from HyperLeda extragalactic database where the recent morphological classification catalogue \citep{Ann2015} is used. We find that, 320 out of our 1220 newly classified spirals are barred-spirals, the rest being pure spiral galaxies. Out of the 320 barred-spiral galaxies, 142 are  {\it Grand-designs} and the rest are {\it Flocculents}. Out of the pure spirals, 357 are  {\it Grand-designs} and 543 are {\it Flocculents}. This observation is in compliance with earlier findings \citep{Bittner2017}. A similar fraction of galaxies with bars ($0.35$ and $0.32$) in both classes shows that presence of a bar do not play a significant role in modulating the arm class of the galaxy. Other observable differences between the {\it Grand-designs} and {\it Flocculents} in the final sample are shown in the Appendix B (\autoref{fig:oprops}). The above physical properties of our newly-classified spiral galaxies are presented in \autoref{tab:props}. Finally, out of our 1220 new classifications of spiral galaxies, $\sim$ 33 were low surface brightness galaxies; Out of these 3 were  {\it Grand-designs} and 30 were {\it Flocculents}. This is in line with earlier observations that LSBs have weak and fragmentary spiral features \citep{McGaugh1995,Schombert2011A}.

\noindent We estimate the approximate physical sizes of the galaxies in our sample using their $PetroR90\_r$ and $Redshift\;(z)$. The Physical radius ($r_p$) is computed from its angular diameter distance as, 
\begin{eqnarray}
\label{eq:redshift_dist}
r_p &=& \frac{\chi (z) \cdot PetroR90\_r}{1+z}. 
%r_p &=& \frac{ c \cdot PetroR90\_r}{H_0(1+z)}  \int_{0}^{z}\frac{dz^{\prime}}{(1-\Omega_{m})+\Omega_{m}(1+z^{\prime})^{3}}.
\end{eqnarray}

\begin{figure*}
\resizebox{18 cm}{!}{\rotatebox{0}{\includegraphics{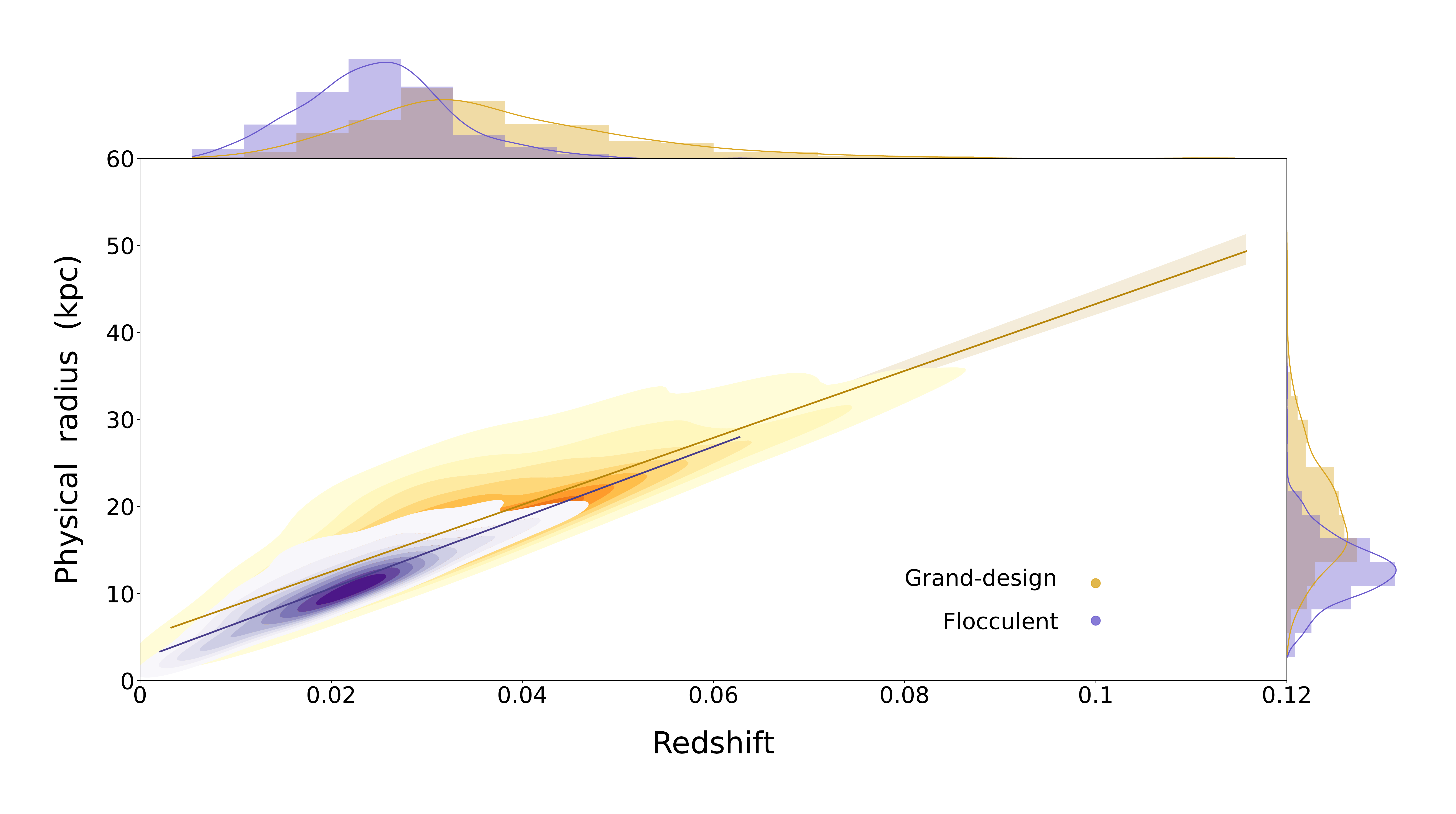}}}\vspace{-25px}
\label{fig:redshift_size}
\caption{This figure shows the distribution of \textit{ Physical radius} ($r_p$) \& \textit{Redshift} ($z$) for the newly classified Grand-designs and Flocculents.}
\end{figure*}

The comoving distance $\chi(z)$ is derived from the redshift by considering $\Lambda$CDM cosmology with $\Omega_m =0.315$ and $ H_0=67.4 \,\mathrm{ Km \; s^{-1}\; Mpc^{-1}}$ \citep{planck18}. The physical radius calculated here corresponds to the apparent observable size that contains 90\% of the flux. Hence, the actual radius of a galaxy that also contains the non-luminous matter would be larger than $r_p$. However, in this part of the study we try to make an overall comparison between the physical sizes of the {\it Grand-designs} and {\it Flocculents} in our sample which are classified with $100\%$ confidence ($p_F=1$ or $p_G=1$). For that purpose, $r_p$ works fairly well. In \autoref{fig:redshift_size} we show the distribution of the {\it Grand-designs} and {\it Flocculents} on the $ z - r_p $ plane, with the help of iso-probability contours and best fitted regression lines. It is found that the {\it Flocculents}, despite of having a larger set extend up to $z \sim 0.06$ with a median around $ z \sim 0.025$. Whereas, the {\it Grand-designs} with a higher median redshift ($z \sim 0.032$) extends up to $z \sim 0.12$. This is probably due to the comparatively lower surface brightness of the {\it Flocculents} which makes them undetectable after a certain depth. The {\it Grand-designs} are also found to be $\sim 1.5$ times larger compared to the {\it Flocculents}, in general. Most of the {\it Flocculents} are found to have physical radius below $ 20 \; \mathrm{kpc}$. Whereas, the largest {\it Grand design} that we found (J110116.68+405239.6) is around $47 \; \mathrm{kpc}$ in size. The median physical radius of the newly classified {\it Flocculents} and {\it Grand-designs} are $11 \; \mathrm{kpc}$ and $17 \; \mathrm{kpc}$ respectively.

\begin{table}
	\centering
	\caption{Properties of the newly classified galaxies}
	\label{tab:props}
	\begin{tabular}{lcc} % four columns, alignment for each
		\hline
		  &  {\it Grand-design} & {\it Flocculent} \\ 
		\hline
		%Total & 499 & 721 \\
		
		$\rm{V_{rot}(km/s)}$ & $218 \pm 86$  & $146 \pm 67$  \\
		$\rm{log_{10}(M_{HI})}$  & $10.0 \pm 0.4$ & $9.8 \pm 0.4$   \\
		$\rm{log_{10}(L_{B})}$ & $10.7 \pm 0.3 $ & $10.3 \pm 0.5$ \\
		$\rm{log_{10}(M_{HI}/L_{B})}$ & $-0.7 \pm 0.3 $ & $-0.5 \pm 0.3$ \\
		$T$\protect\footnotemark & $2.6 \pm 1.8$ & $4.7 \pm 1.9$ \\
		Barred-Spirals & 142 & 178 \\
		Pure Spirals & 357 & 543 \\
		\hline
	\end{tabular}
\end{table}

\section{Conclusion}
We train a Deep Convolutional Neural Network (DCNN) model based on the {\it AlexNet} architecture to distinguish between the {\it Grand-design} and {\it Flocculents} spiral galaxies. We obtain the SDSS DR17 images of the galaxies, which were identified as spiral galaxies in the Galaxyzoo2 project, but hitherto unclassified into  {\it Grand-designs} and {\it Flocculents}. Employing our trained DCNN model, we produce a new catalogue of $1220$ spiral galaxies classified into {\it Grand-designs} and {\it Flocculents}. Out of these, $499$ are {\it Grand-designs} and $721$ {\it Flocculents}. Among these, $252$ of the  {\it Grand-designs} and $604$ of the {\it Flocculents} are identified with prediction probability 1; the rest of the galaxies all are classified with prediction probability between $0.8$ and $1$, in their respective classes. We next study the physical properties of our newly classified spiral galaxies in the two classes and find that the mean asymptotic rotational velocities of the  {\it Grand-designs} and the {\it Flocculents} are $\sim$ 218 $\rm{km\;s^{-1}}$ and 145 $\rm{km\;s^{-1}}$, indicating that {\it Grand-designs} are hosted in galaxies with higher dynamical masses than the {\it Flocculents}. Besides, $M_{HI}/L_B$ for {\it Flocculents} is found to be higher than the  {\it Grand-designs} possibly implying that the {\it Flocculents} constitute a younger galaxy population compared to the {\it Grand-designs}. This is also confirmed by a higher morphological index $T$ $\sim$ 4.7 for the {\it Flocculents}  than the {\it Grand-designs} ($T$ $\sim$ 2.6), indicating that the {\it Flocculents} fundamentally comprise of a later-type galaxy population in contrast to the {\it Grand-designs}. Further, an almost equal fraction of $\sim$ 0.3 of bars in both the classes of spiral galaxies reveals that the presence of a bar component does not regulate the type of spiral arm hosted by a galaxy. Finally, out of the 33 LSBs among our newly classified spirals, only 3 were {\it Grand-designs} and the rest {\it Flocculents}, in compliance with earlier observations. To conclude, the characteristic features of {\it Grand-designs} and {\it Flocculents} are not fairly distinct and easy to identify. It is much more difficult to train the model with an intermediate multiple-arms class. We carry out this work as the first step and aim to include the {\it Multiple-arm} galaxies next. This opens up a possibility for the detection of more extensive sets of {\it Grand-designs}, {\it Mulpiple-arms} and {\it Flocculents} in deep sky surveys like DES \& Euclid.

% footnote for table {tab:props} 
\footnotetext{de Vaucoulers' Numerical Stage Index}

\section*{data Availability}
The image data utilized in this study can be found in SDSS databases, which are open to the public. Data produced through this work will be shared on 
reasonable request to the corresponding author.

\section*{Acknowledgement}
The authors sincerely thank the anonymous reviewer for the insightful comments and suggestions, 
which significantly improved the manuscript. \\

Funding for the Sloan Digital Sky 
Survey IV has been provided by the 
Alfred P. Sloan Foundation, the U.S. 
Department of Energy Office of 
Science, and the Participating 
Institutions. SDSS-IV acknowledges support and 
resources from the Center for High 
Performance Computing  at the 
University of Utah. The SDSS 
website is www.sdss.org.SDSS-IV is managed by the 
Astrophysical Research Consortium 
for the Participating Institutions 
of the SDSS Collaboration including 
the Brazilian Participation Group, 
the Carnegie Institution for Science, 
Carnegie Mellon University, Center for 
Astrophysics | Harvard \& 
Smithsonian, the Chilean Participation 
Group, the French Participation Group, 
Instituto de Astrof\'isica de 
Canarias, The Johns Hopkins 
University, Kavli Institute for the 
Physics and Mathematics of the 
Universe (IPMU) / University of 
Tokyo, the Korean Participation Group, 
Lawrence Berkeley National Laboratory, 
Leibniz Institut f\"ur Astrophysik 
Potsdam (AIP),  Max-Planck-Institut 
f\"ur Astronomie (MPIA Heidelberg), 
Max-Planck-Institut f\"ur 
Astrophysik (MPA Garching), 
Max-Planck-Institut f\"ur 
Extraterrestrische Physik (MPE), 
National Astronomical Observatories of 
China, New Mexico State University, 
New York University, University of 
Notre Dame, Observat\'ario 
Nacional / MCTI, The Ohio State 
University, Pennsylvania State 
University, Shanghai 
Astronomical Observatory, United 
Kingdom Participation Group, 
Universidad Nacional Aut\'onoma 
de M\'exico, University of Arizona, 
University of Colorado Boulder, 
University of Oxford, University of 
Portsmouth, University of Utah, 
University of Virginia, University 
of Washington, University of 
Wisconsin, Vanderbilt University, 
and Yale University.\\

\noindent The authors acknowledge the attempts made by Mr. P. S. Vishnuprasad and Dr. Pavan. K. Perupu to develop the DCNN model during the initial phase of the work.

%%%%%%%%%%%%%%%%%%%% REFERENCES %%%%%%%%%%%%%%%%%%
\bibliographystyle{mnras}
\bibliography{gdfl}

\begin{thebibliography}{}
\makeatletter
\relax
\def\mn@urlcharsother{\let\do\@makeother \do\$\do\&\do\#\do\^\do\_\do\%\do\~}
\def\mn@doi{\begingroup\mn@urlcharsother \@ifnextchar [ {\mn@doi@}
  {\mn@doi@[]}}
\def\mn@doi@[#1]#2{\def\@tempa{#1}\ifx\@tempa\@empty \href
  {http://dx.doi.org/#2} {doi:#2}\else \href {http://dx.doi.org/#2} {#1}\fi
  \endgroup}
\def\mn@eprint#1#2{\mn@eprint@#1:#2::\@nil}
\def\mn@eprint@arXiv#1{\href {http://arxiv.org/abs/#1} {{\tt arXiv:#1}}}
\def\mn@eprint@dblp#1{\href {http://dblp.uni-trier.de/rec/bibtex/#1.xml}
  {dblp:#1}}
\def\mn@eprint@#1:#2:#3:#4\@nil{\def\@tempa {#1}\def\@tempb {#2}\def\@tempc
  {#3}\ifx \@tempc \@empty \let \@tempc \@tempb \let \@tempb \@tempa \fi \ifx
  \@tempb \@empty \def\@tempb {arXiv}\fi \@ifundefined
  {mn@eprint@\@tempb}{\@tempb:\@tempc}{\expandafter \expandafter \csname
  mn@eprint@\@tempb\endcsname \expandafter{\@tempc}}}

\bibitem[\protect\citeauthoryear{{Abraham}, {Aniyan}, {Kembhavi}, {Philip}  \&
  {Vaghmare}}{{Abraham} et~al.}{2018}]{abraham18}
{Abraham} S.,  {Aniyan} A.~K.,  {Kembhavi} A.~K.,  {Philip} N.~S.,   {Vaghmare}
  K.,  2018, \mn@doi [\mnras] {10.1093/mnras/sty627}, \href
  {https://ui.adsabs.harvard.edu/abs/2018MNRAS.477..894A} {477, 894}

\bibitem[\protect\citeauthoryear{{Ackermann}, {Schawinski}, {Zhang}, {Weigel}
  \& {Turp}}{{Ackermann} et~al.}{2018}]{ackermann18}
{Ackermann} S.,  {Schawinski} K.,  {Zhang} C.,  {Weigel} A.~K.,   {Turp} M.~D.,
   2018, \mn@doi [\mnras] {10.1093/mnras/sty1398}, \href
  {https://ui.adsabs.harvard.edu/abs/2018MNRAS.479..415A} {479, 415}

\bibitem[\protect\citeauthoryear{{Aniyan} \& {Thorat}}{{Aniyan} \&
  {Thorat}}{2017}]{aniyan17}
{Aniyan} A.~K.,  {Thorat} K.,  2017, \mn@doi [\apjs]
  {10.3847/1538-4365/aa7333}, \href
  {https://ui.adsabs.harvard.edu/abs/2017ApJS..230...20A} {230, 20}

\bibitem[\protect\citeauthoryear{{Ann}, {Seo}  \& {Ha}}{{Ann}
  et~al.}{2015}]{Ann2015}
{Ann} H.~B.,  {Seo} M.,   {Ha} D.~K.,  2015, \mn@doi [\apjs]
  {10.1088/0067-0049/217/2/27}, \href
  {https://ui.adsabs.harvard.edu/abs/2015ApJS..217...27A} {217, 27}

\bibitem[\protect\citeauthoryear{{Athanassoula}, {Romero-G{\'o}mez}  \&
  {Masdemont}}{{Athanassoula} et~al.}{2009a}]{athan09}
{Athanassoula} E.,  {Romero-G{\'o}mez} M.,   {Masdemont} J.~J.,  2009a, \mn@doi
  [\mnras] {10.1111/j.1365-2966.2008.14273.x}, \href
  {https://ui.adsabs.harvard.edu/abs/2009MNRAS.394...67A} {394, 67}

\bibitem[\protect\citeauthoryear{{Athanassoula}, {Romero-G{\'o}mez}, {Bosma}
  \& {Masdemont}}{{Athanassoula} et~al.}{2009b}]{athan09b}
{Athanassoula} E.,  {Romero-G{\'o}mez} M.,  {Bosma} A.,   {Masdemont} J.~J.,
  2009b, \mn@doi [\mnras] {10.1111/j.1365-2966.2009.15583.x}, \href
  {https://ui.adsabs.harvard.edu/abs/2009MNRAS.400.1706A} {400, 1706}

\bibitem[\protect\citeauthoryear{{Bertin}}{{Bertin}}{1988}]{Bertin1988}
{Bertin} G.,  1988, in {Benney} D.~J.,  {Shu} F.~H.,   {Chi} Y.,  eds, Applied
  Mathematics, Fluid Mechanics, Astrophysics. p.~319

\bibitem[\protect\citeauthoryear{Bertin}{Bertin}{2014}]{bertin14}
Bertin G.,  2014, Dynamics of Galaxies, 2 edn.
Cambridge University Press, \mn@doi{10.1017/CBO9780511731990}

\bibitem[\protect\citeauthoryear{{Bickley} et~al.,}{{Bickley}
  et~al.}{2021}]{bickley21}
{Bickley} R.~W.,  et~al., 2021, \mn@doi [\mnras] {10.1093/mnras/stab806}, \href
  {https://ui.adsabs.harvard.edu/abs/2021MNRAS.504..372B} {504, 372}

\bibitem[\protect\citeauthoryear{{Binney} \& {Tremaine}}{{Binney} \&
  {Tremaine}}{2008}]{Binney2008}
{Binney} J.,  {Tremaine} S.,  2008, {Galactic Dynamics: Second Edition}.
Princeton University Press

\bibitem[\protect\citeauthoryear{Bishop}{Bishop}{2006}]{bishop06}
Bishop C.~M.,  2006, Pattern Recognition and Machine Learning.
Springer

\bibitem[\protect\citeauthoryear{{Bittner}, {Gadotti}, {Elmegreen},
  {Athanassoula}, {Elmegreen}, {Bosma}  \& {Mu{\~n}oz-Mateos}}{{Bittner}
  et~al.}{2017}]{Bittner2017}
{Bittner} A.,  {Gadotti} D.~A.,  {Elmegreen} B.~G.,  {Athanassoula} E.,
  {Elmegreen} D.~M.,  {Bosma} A.,   {Mu{\~n}oz-Mateos} J.-C.,  2017, \mn@doi
  [\mnras] {10.1093/mnras/stx1646}, \href
  {https://ui.adsabs.harvard.edu/abs/2017MNRAS.471.1070B} {471, 1070}

\bibitem[\protect\citeauthoryear{Bottou}{Bottou}{1998}]{bottou98}
Bottou L.,  1998, in Saad D.,  ed., , Online Learning and Neural Networks.
Cambridge University Press, Cambridge, UK, \url
  {http://leon.bottou.org/papers/bottou-98x}

\bibitem[\protect\citeauthoryear{{Bottrell} et~al.,}{{Bottrell}
  et~al.}{2019}]{bottrell19}
{Bottrell} C.,  et~al., 2019, \mn@doi [\mnras] {10.1093/mnras/stz2934}, \href
  {https://ui.adsabs.harvard.edu/abs/2019MNRAS.490.5390B} {490, 5390}

\bibitem[\protect\citeauthoryear{Bridle}{Bridle}{1990}]{bridle90}
Bridle J.~S.,  1990, in Touretzky D.~S.,  ed., , Vol.~2, {Advances in Neural
  Information Processing Systems}.
{Morgan Kaufmann, California}, {San Mateo CA}, pp 211--217

\bibitem[\protect\citeauthoryear{Buta, Knapen, Elmegreen, Salo, Laurikainen,
  Elmegreen, Puerari  \& Block}{Buta et~al.}{2009}]{buta09}
Buta R.~J.,  Knapen J.~H.,  Elmegreen B.~G.,  Salo H.,  Laurikainen E.,
  Elmegreen D.~M.,  Puerari I.,   Block D.~L.,  2009, The Astronomical Journal,
  137, 4487

\bibitem[\protect\citeauthoryear{{Buta} et~al.,}{{Buta} et~al.}{2015}]{buta15}
{Buta} R.~J.,  et~al., 2015, \mn@doi [\apjs] {10.1088/0067-0049/217/2/32},
  \href {https://ui.adsabs.harvard.edu/abs/2015ApJS..217...32B} {217, 32}

\bibitem[\protect\citeauthoryear{{Cabrera-Vives}, {Reyes}, {F{\"o}rster},
  {Est{\'e}vez}  \& {Maureira}}{{Cabrera-Vives} et~al.}{2017}]{cabrera17}
{Cabrera-Vives} G.,  {Reyes} I.,  {F{\"o}rster} F.,  {Est{\'e}vez} P.~A.,
  {Maureira} J.-C.,  2017, \mn@doi [\apj] {10.3847/1538-4357/836/1/97}, \href
  {https://ui.adsabs.harvard.edu/abs/2017ApJ...836...97C} {836, 97}

\bibitem[\protect\citeauthoryear{{Cheng} et~al.,}{{Cheng}
  et~al.}{2021}]{cheng21}
{Cheng} T.-Y.,  et~al., 2021, \mn@doi [\mnras] {10.1093/mnras/stab2142}, \href
  {https://ui.adsabs.harvard.edu/abs/2021MNRAS.507.4425C} {507, 4425}

\bibitem[\protect\citeauthoryear{{Conroy} \& {Gunn}}{{Conroy} \&
  {Gunn}}{2010}]{conroy10}
{Conroy} C.,  {Gunn} J.~E.,  2010, {FSPS: Flexible Stellar Population
  Synthesis}, Astrophysics Source Code Library, record ascl:1010.043
  (\mn@eprint {ascl} {1010.043})

\bibitem[\protect\citeauthoryear{Cun, Boser, Denker, Henderson, Howard, Hubbard
   \& Jackel}{Cun et~al.}{1990}]{lecun90}
Cun L.,  Boser B.,  Denker J.~S.,  Henderson D.,  Howard R.~E.,  Hubbard W.,
  Jackel L.~D.,  1990, in Advances in Neural Information Processing Systems.
  Morgan Kaufmann, pp 396--404

\bibitem[\protect\citeauthoryear{{De Vaucouleurs}, {de Vaucouleurs}  \&
  {Corwin}}{{De Vaucouleurs} et~al.}{1976}]{deVaucouleurs1976RC2}
{De Vaucouleurs} G.,  {de Vaucouleurs} A.,   {Corwin} J.~R.,  1976, Second
  reference catalogue of bright galaxies, \href
  {https://ui.adsabs.harvard.edu/abs/1976RC2...C......0D} {1976, 0}

\bibitem[\protect\citeauthoryear{D{\'{\i}}az-Garc{\'{\i}}a, Salo, Knapen  \&
  Herrera-Endoqui}{D{\'{\i}}az-Garc{\'{\i}}a et~al.}{2019}]{diaz19}
D{\'{\i}}az-Garc{\'{\i}}a S.,  Salo H.,  Knapen J.~H.,   Herrera-Endoqui M.,
  2019, Astronomy {\&} Astrophysics, 631, A94

\bibitem[\protect\citeauthoryear{{Dieleman}, {Willett}  \& {Dambre}}{{Dieleman}
  et~al.}{2015}]{dieleman15}
{Dieleman} S.,  {Willett} K.~W.,   {Dambre} J.,  2015, \mn@doi [\mnras]
  {10.1093/mnras/stv632}, \href
  {https://ui.adsabs.harvard.edu/abs/2015MNRAS.450.1441D} {450, 1441}

\bibitem[\protect\citeauthoryear{{Elmegreen} \& {Elmegreen}}{{Elmegreen} \&
  {Elmegreen}}{1982}]{elmegreen82}
{Elmegreen} D.~M.,  {Elmegreen} B.~G.,  1982, \mn@doi [\mnras]
  {10.1093/mnras/201.4.1021}, \href
  {https://ui.adsabs.harvard.edu/abs/1982MNRAS.201.1021E} {201, 1021}

\bibitem[\protect\citeauthoryear{{Elmegreen} \& {Elmegreen}}{{Elmegreen} \&
  {Elmegreen}}{1987}]{elmegreen87}
{Elmegreen} D.~M.,  {Elmegreen} B.~G.,  1987, \mn@doi [\apj] {10.1086/165034},
  \href {https://ui.adsabs.harvard.edu/abs/1987ApJ...314....3E} {314, 3}

\bibitem[\protect\citeauthoryear{{Elmegreen} et~al.,}{{Elmegreen}
  et~al.}{2011}]{elmegreen11}
{Elmegreen} D.~M.,  et~al., 2011, \mn@doi [\apj] {10.1088/0004-637X/737/1/32},
  \href {https://ui.adsabs.harvard.edu/abs/2011ApJ...737...32E} {737, 32}

\bibitem[\protect\citeauthoryear{{Filistov}}{{Filistov}}{2012}]{filistov12}
{Filistov} E.,  2012, Astronomical and Astrophysical Transactions

\bibitem[\protect\citeauthoryear{{Fukugita}, {Ichikawa}, {Gunn}, {Doi},
  {Shimasaku}  \& {Schneider}}{{Fukugita} et~al.}{1996}]{fukugita96}
{Fukugita} M.,  {Ichikawa} T.,  {Gunn} J.~E.,  {Doi} M.,  {Shimasaku} K.,
  {Schneider} D.~P.,  1996, \mn@doi [\aj] {10.1086/117915}, \href
  {https://ui.adsabs.harvard.edu/abs/1996AJ....111.1748F} {111, 1748}

\bibitem[\protect\citeauthoryear{Fukushima}{Fukushima}{1975}]{fukushima75}
Fukushima K.,  1975, \mn@doi [Biological Cybernetics] {10.1007/BF00342633}, 20,
  121

\bibitem[\protect\citeauthoryear{{Garma-Oehmichen}, {Martinez-Medina},
  {Hern{\'a}ndez-Toledo}  \& {Puerari}}{{Garma-Oehmichen}
  et~al.}{2021}]{garma21}
{Garma-Oehmichen} L.,  {Martinez-Medina} L.,  {Hern{\'a}ndez-Toledo} H.,
  {Puerari} I.,  2021, \mn@doi [\mnras] {10.1093/mnras/stab333}, \href
  {https://ui.adsabs.harvard.edu/abs/2021MNRAS.502.4708G} {502, 4708}

\bibitem[\protect\citeauthoryear{{Gerola} \& {Seiden}}{{Gerola} \&
  {Seiden}}{1978}]{Gerola1978}
{Gerola} H.,  {Seiden} P.~E.,  1978, \mn@doi [\apj] {10.1086/156243}, \href
  {https://ui.adsabs.harvard.edu/abs/1978ApJ...223..129G} {223, 129}

\bibitem[\protect\citeauthoryear{{Ghosh} \& {Jog}}{{Ghosh} \&
  {Jog}}{2015}]{ghosh15}
{Ghosh} S.,  {Jog} C.~J.,  2015, \mn@doi [\mnras] {10.1093/mnras/stv1040},
  \href {https://ui.adsabs.harvard.edu/abs/2015MNRAS.451.1350G} {451, 1350}

\bibitem[\protect\citeauthoryear{{Goddard} \& {Shamir}}{{Goddard} \&
  {Shamir}}{2020}]{goddard20}
{Goddard} H.,  {Shamir} L.,  2020, \mn@doi [\apjs] {10.3847/1538-4365/abc0ed},
  \href {https://ui.adsabs.harvard.edu/abs/2020ApJS..251...28G} {251, 28}

\bibitem[\protect\citeauthoryear{Gold, Rangarajan  et~al.}{Gold
  et~al.}{1996}]{gold96}
Gold S.,  Rangarajan A.,   et~al., 1996, Journal of Artificial Neural Networks,
  2, 381

\bibitem[\protect\citeauthoryear{{Goldreich} \& {Lynden-Bell}}{{Goldreich} \&
  {Lynden-Bell}}{1965}]{Goldreich1965}
{Goldreich} P.,  {Lynden-Bell} D.,  1965, \mn@doi [\mnras]
  {10.1093/mnras/130.2.125}, \href
  {https://ui.adsabs.harvard.edu/abs/1965MNRAS.130..125G} {130, 125}

\bibitem[\protect\citeauthoryear{{Gunn} et~al.,}{{Gunn} et~al.}{1998}]{gunn98}
{Gunn} J.~E.,  et~al., 1998, \mn@doi [\aj] {10.1086/300645}, \href
  {https://ui.adsabs.harvard.edu/abs/1998AJ....116.3040G} {116, 3040}

\bibitem[\protect\citeauthoryear{{Hosseinzadeh} et~al.,}{{Hosseinzadeh}
  et~al.}{2020}]{hossein20}
{Hosseinzadeh} G.,  et~al., 2020, \mn@doi [\apj] {10.3847/1538-4357/abc42b},
  \href {https://ui.adsabs.harvard.edu/abs/2020ApJ...905...93H} {905, 93}

\bibitem[\protect\citeauthoryear{{Huang}, {Liu}, {van der Maaten}  \&
  {Weinberger}}{{Huang} et~al.}{2016}]{huang16}
{Huang} G.,  {Liu} Z.,  {van der Maaten} L.,   {Weinberger} K.~Q.,  2016, arXiv
  e-prints, \href {https://ui.adsabs.harvard.edu/abs/2016arXiv160806993H} {p.
  arXiv:1608.06993}

\bibitem[\protect\citeauthoryear{{Huertas-Company} et~al.,}{{Huertas-Company}
  et~al.}{2015}]{huertas15}
{Huertas-Company} M.,  et~al., 2015, \mn@doi [\apjs]
  {10.1088/0067-0049/221/1/8}, \href
  {https://ui.adsabs.harvard.edu/abs/2015ApJS..221....8H} {221, 8}

\bibitem[\protect\citeauthoryear{{Jacobs}, {Glazebrook}, {Collett}, {More}  \&
  {McCarthy}}{{Jacobs} et~al.}{2017}]{jacobs17}
{Jacobs} C.,  {Glazebrook} K.,  {Collett} T.,  {More} A.,   {McCarthy} C.,
  2017, \mn@doi [\mnras] {10.1093/mnras/stx1492}, \href
  {https://ui.adsabs.harvard.edu/abs/2017MNRAS.471..167J} {471, 167}

\bibitem[\protect\citeauthoryear{Jarrett, Kavukcuoglu, Ranzato  \&
  LeCun}{Jarrett et~al.}{2009}]{jarrett09}
Jarrett K.,  Kavukcuoglu K.,  Ranzato M.,   LeCun Y.,  2009, in 2009 IEEE 12th
  International Conference on Computer Vision. pp 2146--2153,
  \mn@doi{10.1109/ICCV.2009.5459469}

\bibitem[\protect\citeauthoryear{{Jog}}{{Jog}}{1992}]{Jog1992}
{Jog} C.~J.,  1992, \mn@doi [\apj] {10.1086/171289}, \href
  {https://ui.adsabs.harvard.edu/abs/1992ApJ...390..378J} {390, 378}

\bibitem[\protect\citeauthoryear{{Jog}}{{Jog}}{1996}]{Jog1996}
{Jog} C.~J.,  1996, \mn@doi [\mnras] {10.1093/mnras/278.1.209}, \href
  {https://ui.adsabs.harvard.edu/abs/1996MNRAS.278..209J} {278, 209}

\bibitem[\protect\citeauthoryear{Josephine, Nirmala  \& Alluri}{Josephine
  et~al.}{2021}]{joseph21}
Josephine V.,  Nirmala A.,   Alluri V.,  2021, \mn@doi [IOP Conference Series:
  Materials Science and Engineering] {10.1088/1757-899X/1131/1/012007}, 1131,
  012007

\bibitem[\protect\citeauthoryear{{Julian} \& {Toomre}}{{Julian} \&
  {Toomre}}{1966}]{Julian1966}
{Julian} W.~H.,  {Toomre} A.,  1966, \mn@doi [\apj] {10.1086/148957}, \href
  {https://ui.adsabs.harvard.edu/abs/1966ApJ...146..810J} {146, 810}

\bibitem[\protect\citeauthoryear{{Kim} \& {Brunner}}{{Kim} \&
  {Brunner}}{2017}]{kim17}
{Kim} E.~J.,  {Brunner} R.~J.,  2017, \mn@doi [\mnras] {10.1093/mnras/stw2672},
  \href {https://ui.adsabs.harvard.edu/abs/2017MNRAS.464.4463K} {464, 4463}

\bibitem[\protect\citeauthoryear{Krizhevsky, Sutskever  \& Hinton}{Krizhevsky
  et~al.}{2012}]{krizhevsky12}
Krizhevsky A.,  Sutskever I.,   Hinton G.~E.,  2012, Advances in neural
  information processing systems, 25, 1097

\bibitem[\protect\citeauthoryear{{Lanusse}, {Ma}, {Li}, {Collett}, {Li},
  {Ravanbakhsh}, {Mandelbaum}  \& {P{\'o}czos}}{{Lanusse}
  et~al.}{2018}]{lanusse18}
{Lanusse} F.,  {Ma} Q.,  {Li} N.,  {Collett} T.~E.,  {Li} C.-L.,  {Ravanbakhsh}
  S.,  {Mandelbaum} R.,   {P{\'o}czos} B.,  2018, \mn@doi [\mnras]
  {10.1093/mnras/stx1665}, \href
  {https://ui.adsabs.harvard.edu/abs/2018MNRAS.473.3895L} {473, 3895}

\bibitem[\protect\citeauthoryear{Lecun, Bottou, Bengio  \& Haffner}{Lecun
  et~al.}{1998}]{lecun98}
Lecun Y.,  Bottou L.,  Bengio Y.,   Haffner P.,  1998, \mn@doi [Proceedings of
  the IEEE] {10.1109/5.726791}, 86, 2278

\bibitem[\protect\citeauthoryear{{Lin} \& {Shu}}{{Lin} \& {Shu}}{1966}]{lin66}
{Lin} C.~C.,  {Shu} F.~H.,  1966, \mn@doi [Proceedings of the National Academy
  of Science] {10.1073/pnas.55.2.229}, \href
  {https://ui.adsabs.harvard.edu/abs/1966PNAS...55..229L} {55, 229}

\bibitem[\protect\citeauthoryear{{Lintott} et~al.,}{{Lintott}
  et~al.}{2008}]{lintott08}
{Lintott} C.~J.,  et~al., 2008, \mn@doi [\mnras]
  {10.1111/j.1365-2966.2008.13689.x}, \href
  {https://ui.adsabs.harvard.edu/abs/2008MNRAS.389.1179L} {389, 1179}

\bibitem[\protect\citeauthoryear{{Lintott} et~al.,}{{Lintott}
  et~al.}{2011}]{lintott11}
{Lintott} C.,  et~al., 2011, \mn@doi [\mnras]
  {10.1111/j.1365-2966.2010.17432.x}, \href
  {https://ui.adsabs.harvard.edu/abs/2011MNRAS.410..166L} {410, 166}

\bibitem[\protect\citeauthoryear{{Makarov}, {Prugniel}, {Terekhova}, {Courtois}
   \& {Vauglin}}{{Makarov} et~al.}{2014}]{Markrov2014}
{Makarov} D.,  {Prugniel} P.,  {Terekhova} N.,  {Courtois} H.,   {Vauglin} I.,
  2014, \mn@doi [\aap] {10.1051/0004-6361/201423496}, \href
  {http://adsabs.harvard.edu/abs/2014A%26A...570A..13M} {570, A13}

\bibitem[\protect\citeauthoryear{{McGaugh}, {Schombert}  \& {Bothun}}{{McGaugh}
  et~al.}{1995}]{McGaugh1995}
{McGaugh} S.~S.,  {Schombert} J.~M.,   {Bothun} G.~D.,  1995, \mn@doi [\aj]
  {10.1086/117427}, \href
  {https://ui.adsabs.harvard.edu/abs/1995AJ....109.2019M} {109, 2019}

\bibitem[\protect\citeauthoryear{Mcculloch \& Pitts}{Mcculloch \&
  Pitts}{1943}]{mcculloch43a}
Mcculloch W.,  Pitts W.,  1943, Bulletin of Mathematical Biophysics, 5, 127

\bibitem[\protect\citeauthoryear{{M{\"o}ller} et~al.,}{{M{\"o}ller}
  et~al.}{2016}]{moller16}
{M{\"o}ller} A.,  et~al., 2016, \mn@doi [\jcap]
  {10.1088/1475-7516/2016/12/008}, \href
  {https://ui.adsabs.harvard.edu/abs/2016JCAP...12..008M} {2016, 008}

\bibitem[\protect\citeauthoryear{Mondal \& Chattopadhyay}{Mondal \&
  Chattopadhyay}{2021}]{mondal21}
Mondal D.,  Chattopadhyay T.,  2021, Celestial Mechanics and Dynamical
  Astronomy

\bibitem[\protect\citeauthoryear{Nair \& Hinton}{Nair \& Hinton}{2010}]{nair10}
Nair V.,  Hinton G.~E.,  2010, in ICML 2010. pp 807--814

\bibitem[\protect\citeauthoryear{Obreschkow, Croton, Lucia, Khochfar  \&
  Rawlings}{Obreschkow et~al.}{2009}]{Obreschkow_2009}
Obreschkow D.,  Croton D.,  Lucia G.~D.,  Khochfar S.,   Rawlings S.,  2009,
  \mn@doi [The Astrophysical Journal] {10.1088/0004-637x/698/2/1467}, 698, 1467

\bibitem[\protect\citeauthoryear{{Odewahn}, {Stockwell}, {Pennington},
  {Humphreys}  \& {Zumach}}{{Odewahn} et~al.}{1992}]{odewahn92}
{Odewahn} S.~C.,  {Stockwell} E.~B.,  {Pennington} R.~L.,  {Humphreys} R.~M.,
  {Zumach} W.~A.,  1992, \mn@doi [\aj] {10.1086/116063}, \href
  {https://ui.adsabs.harvard.edu/abs/1992AJ....103..318O} {103, 318}

\bibitem[\protect\citeauthoryear{{Planck Collaboration} et~al.,}{{Planck
  Collaboration} et~al.}{2020}]{planck18}
{Planck Collaboration} et~al., 2020, \mn@doi [\aap]
  {10.1051/0004-6361/201833910}, \href
  {https://ui.adsabs.harvard.edu/abs/2020A&A...641A...6P} {641, A6}

\bibitem[\protect\citeauthoryear{{Prakash}, {Banerjee}  \& {Perepu}}{{Prakash}
  et~al.}{2020}]{prakash20}
{Prakash} P.,  {Banerjee} A.,   {Perepu} P.~K.,  2020, \mn@doi [\mnras]
  {10.1093/mnras/staa2109}, \href
  {https://ui.adsabs.harvard.edu/abs/2020MNRAS.497.3323P} {497, 3323}

\bibitem[\protect\citeauthoryear{{Romanishin}}{{Romanishin}}{1985}]{romanishin85}
{Romanishin} W.,  1985, \mn@doi [\apj] {10.1086/162917}, \href
  {https://ui.adsabs.harvard.edu/abs/1985ApJ...289..570R} {289, 570}

\bibitem[\protect\citeauthoryear{{Romero-G{\'o}mez}, {Athanassoula}, {Bosma}
  \& {Masdemont}}{{Romero-G{\'o}mez} et~al.}{2011}]{Romero2011}
{Romero-G{\'o}mez} M.,  {Athanassoula} E.,  {Bosma} A.,   {Masdemont} J.~J.,
  2011, in {Zapatero Osorio} M.~R.,  {Gorgas} J.,  {Ma{\'\i}z Apell{\'a}niz}
  J.,  {Pardo} J.~R.,   {Gil de Paz} A.,  eds, Highlights of Spanish
  Astrophysics VI. pp 314--319

\bibitem[\protect\citeauthoryear{{Ross} et~al.,}{{Ross} et~al.}{2011}]{ross11}
{Ross} A.~J.,  et~al., 2011, \mn@doi [\mnras]
  {10.1111/j.1365-2966.2011.19351.x}, \href
  {https://ui.adsabs.harvard.edu/abs/2011MNRAS.417.1350R} {417, 1350}

\bibitem[\protect\citeauthoryear{Rubinstein \& Kroese}{Rubinstein \&
  Kroese}{2004}]{ruben04}
Rubinstein R.,  Kroese D.~P.,  2004, in The Cross-Entropy Method: A Unified
  Approach to Combinatorial Optimization, Monte-Carlo Simulation and Machine
  Learning. Springer

\bibitem[\protect\citeauthoryear{{Rumelhart}, {Hinton}  \&
  {Williams}}{{Rumelhart} et~al.}{1986}]{rumel86}
{Rumelhart} D.~E.,  {Hinton} G.~E.,   {Williams} R.~J.,  1986, \mn@doi [\nat]
  {10.1038/323533a0}, \href
  {https://ui.adsabs.harvard.edu/abs/1986Natur.323..533R} {323, 533}

\bibitem[\protect\citeauthoryear{{Salo}, {Laurikainen}, {Buta}  \&
  {Knapen}}{{Salo} et~al.}{2010}]{salo10}
{Salo} H.,  {Laurikainen} E.,  {Buta} R.,   {Knapen} J.~H.,  2010, \mn@doi
  [\apjl] {10.1088/2041-8205/715/1/L56}, \href
  {https://ui.adsabs.harvard.edu/abs/2010ApJ...715L..56S} {715, L56}

\bibitem[\protect\citeauthoryear{Samuel}{Samuel}{1959}]{samuel59}
Samuel A.~L.,  1959, \mn@doi [IBM Journal of Research and Development]
  {10.1147/rd.33.0210}, 3, 210

\bibitem[\protect\citeauthoryear{{Schaefer}, {Geiger}, {Kuntzer}  \&
  {Kneib}}{{Schaefer} et~al.}{2018}]{schaefer18}
{Schaefer} C.,  {Geiger} M.,  {Kuntzer} T.,   {Kneib} J.~P.,  2018, \mn@doi
  [\aap] {10.1051/0004-6361/201731201}, \href
  {https://ui.adsabs.harvard.edu/abs/2018A&A...611A...2S} {611, A2}

\bibitem[\protect\citeauthoryear{{Schombert}, {Maciel}  \&
  {McGaugh}}{{Schombert} et~al.}{2011}]{Schombert2011A}
{Schombert} J.,  {Maciel} T.,   {McGaugh} S.,  2011, \mn@doi [Advances in
  Astronomy] {10.1155/2011/143698}, \href
  {https://ui.adsabs.harvard.edu/abs/2011AdAst2011E..12S} {2011, 143698}

\bibitem[\protect\citeauthoryear{{Schwarz}}{{Schwarz}}{1984}]{schwarz84}
{Schwarz} M.~P.,  1984, \mn@doi [\mnras] {10.1093/mnras/209.1.93}, \href
  {https://ui.adsabs.harvard.edu/abs/1984MNRAS.209...93S} {209, 93}

\bibitem[\protect\citeauthoryear{{Seigar} \& {James}}{{Seigar} \&
  {James}}{1998}]{1998Seigar}
{Seigar} M.~S.,  {James} P.~A.,  1998, \mn@doi [\mnras]
  {10.1046/j.1365-8711.1998.01778.x}, \href
  {https://ui.adsabs.harvard.edu/abs/1998MNRAS.299..672S} {299, 672}

\bibitem[\protect\citeauthoryear{{Sellwood}}{{Sellwood}}{2011}]{Sellwood2011}
{Sellwood} J.~A.,  2011, \mn@doi [\mnras] {10.1111/j.1365-2966.2010.17545.x},
  \href {https://ui.adsabs.harvard.edu/abs/2011MNRAS.410.1637S} {410, 1637}

\bibitem[\protect\citeauthoryear{{Sellwood} \& {Carlberg}}{{Sellwood} \&
  {Carlberg}}{1984}]{1984Sellwood}
{Sellwood} J.~A.,  {Carlberg} R.~G.,  1984, \mn@doi [\apj] {10.1086/162176},
  \href {https://ui.adsabs.harvard.edu/abs/1984ApJ...282...61S} {282, 61}

\bibitem[\protect\citeauthoryear{Sellwood \& Carlberg}{Sellwood \&
  Carlberg}{2014a}]{Sellwood_2014}
Sellwood J.~A.,  Carlberg R.~G.,  2014a, \mn@doi [The Astrophysical Journal]
  {10.1088/0004-637x/785/2/137}, 785, 137

\bibitem[\protect\citeauthoryear{{Sellwood} \& {Carlberg}}{{Sellwood} \&
  {Carlberg}}{2014b}]{SellwoodCarlberg2014}
{Sellwood} J.~A.,  {Carlberg} R.~G.,  2014b, \mn@doi [\apj]
  {10.1088/0004-637X/785/2/137}, \href
  {https://ui.adsabs.harvard.edu/abs/2014ApJ...785..137S} {785, 137}

\bibitem[\protect\citeauthoryear{Shu}{Shu}{1982}]{shu82}
Shu F.,  1982, The Physical Universe: An Introduction to Astronomy.
Series of books in astronomy, University Science Books, \url
  {https://books.google.co.in/books?id=GUyQQAAACAAJ}

\bibitem[\protect\citeauthoryear{Srivastava, Hinton, Krizhevsky, Sutskever  \&
  Salakhutdinov}{Srivastava et~al.}{2014}]{srivastava14}
Srivastava N.,  Hinton G.,  Krizhevsky A.,  Sutskever I.,   Salakhutdinov R.,
  2014, J. Mach. Learn. Res., 15, 1929–1958

\bibitem[\protect\citeauthoryear{{Thornley}}{{Thornley}}{1996}]{thornley96}
{Thornley} M.~D.,  1996, \mn@doi [\apjl] {10.1086/310250}, \href
  {https://ui.adsabs.harvard.edu/abs/1996ApJ...469L..45T} {469, L45}

\bibitem[\protect\citeauthoryear{{Thornley} \& {Mundy}}{{Thornley} \&
  {Mundy}}{1997}]{Thornley1997}
{Thornley} M.~D.,  {Mundy} L.~G.,  1997, \mn@doi [\apj] {10.1086/304907}, \href
  {https://ui.adsabs.harvard.edu/abs/1997ApJ...490..682T} {490, 682}

\bibitem[\protect\citeauthoryear{{Toomre}}{{Toomre}}{1964}]{toomre1964gravitational}
{Toomre} A.,  1964, \mn@doi [\apj] {10.1086/147861}, \href
  {https://ui.adsabs.harvard.edu/abs/1964ApJ...139.1217T} {139, 1217}

\bibitem[\protect\citeauthoryear{{Toomre}}{{Toomre}}{1981}]{Toomre1981}
{Toomre} A.,  1981, in {Fall} S.~M.,  {Lynden-Bell} D.,  eds, Structure and
  Evolution of Normal Galaxies. pp 111--136

\bibitem[\protect\citeauthoryear{{Visser}, {Bosma}  \& {Postma}}{{Visser}
  et~al.}{2021}]{visser21}
{Visser} K.,  {Bosma} B.,   {Postma} E.,  2021, arXiv e-prints, \href
  {https://ui.adsabs.harvard.edu/abs/2021arXiv210506292V} {p. arXiv:2105.06292}

\bibitem[\protect\citeauthoryear{{Voglis}, {Tsoutsis}  \&
  {Efthymiopoulos}}{{Voglis} et~al.}{2006}]{voglis06}
{Voglis} N.,  {Tsoutsis} P.,   {Efthymiopoulos} C.,  2006, \mn@doi [\mnras]
  {10.1111/j.1365-2966.2006.11021.x}, \href
  {https://ui.adsabs.harvard.edu/abs/2006MNRAS.373..280V} {373, 280}

\bibitem[\protect\citeauthoryear{{Weir}, {Fayyad}  \& {Djorgovski}}{{Weir}
  et~al.}{1995}]{weir95}
{Weir} N.,  {Fayyad} U.~M.,   {Djorgovski} S.,  1995, \mn@doi [\aj]
  {10.1086/117459}, \href
  {https://ui.adsabs.harvard.edu/abs/1995AJ....109.2401W} {109, 2401}

\bibitem[\protect\citeauthoryear{Werbos}{Werbos}{1975}]{werbos75}
Werbos P.,  1975, Beyond Regression: New Tools for Prediction and Analysis in
  the Behavioral Sciences.
Harvard University, \url {https://books.google.co.in/books?id=z81XmgEACAAJ}

\bibitem[\protect\citeauthoryear{{Willett} et~al.,}{{Willett}
  et~al.}{2013}]{willett13}
{Willett} K.~W.,  et~al., 2013, \mn@doi [\mnras] {10.1093/mnras/stt1458}, \href
  {https://ui.adsabs.harvard.edu/abs/2013MNRAS.435.2835W} {435, 2835}

\bibitem[\protect\citeauthoryear{{York} et~al.,}{{York} et~al.}{2000}]{york00}
{York} D.~G.,  et~al., 2000, \mn@doi [\aj] {10.1086/301513}, \href
  {https://ui.adsabs.harvard.edu/abs/2000AJ....120.1579Y} {120, 1579}

\makeatother
\end{thebibliography}

%%%%%%%%%%%%%%%%%%%% APPENDIX %%%%%%%%%%%%%%%%%%
\appendix

\section{Catalogue of Newly Classified spirals}
\autoref{tab:sdssclass} presents the catalogue of the newly classified $1,220$ spiral galaxies from SDSS. {\it Grand-designs} and {\it Flocculents} 
classified with 100 \% prediction probability are flagged ass $G$ and, $F$ respectively. The {\it Flocculents} and {\it Grand-designs} with a prediction 
probability between 80 - 100 \% are flagged as $f$ and $g$.

\label{sec:new}
%% longtable 1st page
\small{
\begin{table*}
\label{tab:sdssclass} 
\caption{  {\it Grand-design} and {\it Flocculent} galaxies identified from SDSS. \\ \hspace*{30px}  \textbf{F:} {\it Flocculent} ( $p_F =1$); \hspace{10px} \textbf{f:} {\it Flocculent} ($0.8 \leq p_F < 1$ ); \hspace*{10px} \textbf{G:}  {\it Grand-design} ( $p_G =1 $); \hspace{10px} \textbf{g:}  {\it Grand-design} ( $ 0.8 \leq p_G  < 1$) } 
\begin{tabular}{cccccccc}
\hline
\textbf{Galaxy} & \textbf{Class} & \textbf{Galaxy} & \textbf{Class} & \textbf{Galaxy}  & \textbf{Class} & \textbf{Galaxy} & \textbf{Class} \\ 
\hline 
J002624.97+010112.0	&	F	&	J080845.83+181139.1	&	g	&	J084614.11+413447.2	&	g	&	J091633.43+395222.0	&	F	\\
J002820.53-001304.2	&	G	&	J080959.95+400611.4	&	F	&	J084645.21+065735.8	&	G	&	J091639.79+071558.9	&	F	\\
J003514.53+004145.5	&	g	&	J081200.33+192147.4	&	g	&	J084706.91+281411.6	&	G	&	J091644.18+195608.5	&	F	\\
J003934.83+005135.8	&	F	&	J081246.43+262142.4	&	F	&	J084806.88+415133.2	&	F	&	J091814.20+453906.0	&	F	\\
J004037.85-001950.4	&	F	&	J081314.20+522731.4	&	G	&	J084824.64+181952.0	&	F	&	J091859.85+511924.2	&	F	\\
J004215.87+005043.7	&	G	&	J081321.11+575108.0	&	G	&	J084825.19+234337.7	&	g	&	J091933.11+055257.9	&	G	\\
J004351.87+004807.0	&	F	&	J081416.48+182626.4	&	F	&	J084840.22+010219.6	&	G	&	J091937.94+272728.0	&	G	\\
J005751.70-010809.6	&	G	&	J081422.03+391504.8	&	G	&	J084922.55+364237.2	&	G	&	J091953.44+480558.2	&	F	\\
J005848.86+003514.1	&	F	&	J081602.15+283729.0	&	F	&	J084927.19+293111.8	&	F	&	J092032.71+174207.8	&	F	\\
J005942.12+005459.7	&	F	&	J081618.48+522518.9	&	F	&	J085241.66+404104.0	&	F	&	J092040.65+150603.4	&	G	\\
J011233.68+001736.2	&	F	&	J081625.36+255928.8	&	G	&	J085311.32+090853.3	&	g	&	J092041.03+181422.4	&	G	\\
J011832.25-011150.4	&	G	&	J081759.80+463414.7	&	F	&	J085335.85+452007.8	&	F	&	J092054.48+280913.6	&	F	\\
J012223.77-005230.7	&	g	&	J081915.52+244733.5	&	F	&	J085421.60+324051.0	&	f	&	J092101.68+390923.6	&	F	\\
J015148.51+001549.7	&	G	&	J081932.11+212339.4	&	f	&	J085552.29+023127.4	&	F	&	J092206.21+035350.1	&	F	\\
J015840.93+003145.1	&	g	&	J082016.56+205230.2	&	F	&	J085640.68+002230.0	&	F	&	J092238.18+605155.6	&	F	\\
J020212.33-000602.2	&	G	&	J082040.80+255419.2	&	f	&	J085658.00+520357.4	&	G	&	J092253.06+215830.6	&	G	\\
J021219.68-004841.4	&	G	&	J082046.07+163844.8	&	G	&	J085701.08+131156.7	&	f	&	J092253.83+291944.4	&	F	\\
J022854.70+002213.1	&	F	&	J082049.32+223927.9	&	F	&	J085850.47+061734.7	&	f	&	J092308.11+022909.8	&	F	\\
J024927.95-005222.7	&	g	&	J082059.90+551630.4	&	g	&	J090007.89+165526.4	&	F	&	J092313.36+023604.6	&	F	\\
J025627.12+005232.7	&	g	&	J082202.66+571830.3	&	F	&	J090015.62+401748.3	&	G	&	J092317.35+401200.2	&	F	\\
J072333.23+412605.6	&	F	&	J082210.67+031604.9	&	g	&	J090020.25+522939.2	&	g	&	J092319.53+221625.2	&	g	\\
J072916.63+421646.5	&	f	&	J082224.79+253032.6	&	g	&	J090022.63+645451.7	&	G	&	J092335.52+244542.1	&	G	\\
J072954.31+372706.3	&	g	&	J082258.58+274227.7	&	F	&	J090125.79+070149.7	&	G	&	J092401.10+051233.8	&	F	\\
J073353.32+370133.3	&	g	&	J082356.76+040904.9	&	F	&	J090131.19+491931.4	&	G	&	J092455.63+555348.1	&	F	\\
J073737.12+415649.4	&	F	&	J082430.93+183549.5	&	F	&	J090140.94+110458.7	&	F	&	J092518.91+340644.9	&	F	\\
J073749.41+462351.5	&	g	&	J082509.36+421801.0	&	F	&	J090206.16+232313.5	&	F	&	J092528.24+233630.9	&	G	\\
J073836.50+373800.6	&	F	&	J082512.07+202005.0	&	G	&	J090215.69+031011.7	&	g	&	J092547.99+341637.5	&	F	\\
J073906.00+290936.2	&	g	&	J082545.40+192657.4	&	g	&	J090215.81+165017.8	&	F	&	J092609.43+491836.7	&	G	\\
J073945.26+484431.1	&	F	&	J082607.43+212724.1	&	F	&	J090243.84+252516.1	&	F	&	J092651.02+233037.9	&	f	\\
J074438.73+402158.8	&	F	&	J082656.13+040503.5	&	G	&	J090617.35+500521.4	&	G	&	J092723.49+302626.9	&	g	\\
J074618.81+390400.6	&	F	&	J082726.52+171702.6	&	g	&	J090649.87+484620.9	&	G	&	J092752.72+483124.0	&	F	\\
J074637.70+444725.8	&	G	&	J082727.67+254327.9	&	F	&	J090658.46+175749.5	&	g	&	J092810.03+443953.0	&	f	\\
J074951.23+184944.3	&	F	&	J082814.37+280326.3	&	g	&	J090711.08+504245.8	&	g	&	J093011.76+555108.7	&	G	\\
J075000.02+300129.7	&	g	&	J082844.42+580905.8	&	F	&	J090719.58+461321.2	&	G	&	J093015.19+040839.1	&	F	\\
J075135.63+425248.3	&	f	&	J082959.75+310207.8	&	G	&	J090720.92+402706.0	&	F	&	J093107.87+462303.3	&	F	\\
J075208.95+502225.1	&	g	&	J083002.52+212919.1	&	G	&	J090758.29+414232.1	&	F	&	J093123.37+051609.7	&	g	\\
J075318.07+445157.5	&	g	&	J083041.28+521802.4	&	F	&	J090813.24+513906.4	&	g	&	J093136.57+411903.1	&	F	\\
J075329.20+143641.8	&	F	&	J083054.11+201453.0	&	F	&	J090842.62+444838.3	&	F	&	J093225.05+572858.4	&	F	\\
J075341.04+525238.2	&	g	&	J083127.04+605942.6	&	F	&	J090919.43+545443.0	&	F	&	J093249.24+622012.3	&	F	\\
J075546.72+465456.4	&	g	&	J083203.50+240039.2	&	F	&	J090920.30+204150.3	&	F	&	J093303.64+295542.0	&	g	\\
J075616.63+113943.8	&	g	&	J083322.68+523156.1	&	F	&	J090933.47+323022.8	&	F	&	J093316.39+230805.6	&	F	\\
J075625.07+270045.2	&	g	&	J083356.66+265821.5	&	F	&	J090941.78+373605.6	&	g	&	J093341.87+395902.7	&	F	\\
J075719.70+111221.8	&	G	&	J083517.47+554943.8	&	G	&	J090958.07+621450.4	&	G	&	J093352.12+465149.2	&	F	\\
J075756.49+250939.1	&	g	&	J083524.74+233130.5	&	F	&	J091005.47+543449.1	&	F	&	J093356.87+513144.5	&	f	\\
J075913.05+325452.8	&	G	&	J083534.43+524720.0	&	G	&	J091046.43+332239.0	&	G	&	J093402.71+092845.1	&	F	\\
J075953.80+331725.8	&	G	&	J083742.96+203013.8	&	F	&	J091135.56+325055.6	&	F	&	J093438.64+055029.1	&	g	\\
J075957.14+354851.5	&	g	&	J083834.00+304755.2	&	F	&	J091301.27+202154.9	&	F	&	J093526.27+294845.4	&	G	\\
J080006.28+130907.6	&	F	&	J083844.47+433251.1	&	G	&	J091302.23+493822.0	&	F	&	J093630.83+482810.4	&	F	\\
J080236.52+272614.9	&	f	&	J083909.28+450747.7	&	f	&	J091313.39+031349.7	&	F	&	J093637.36+342847.2	&	G	\\
J080249.63+154829.8	&	F	&	J083939.93+605808.0	&	f	&	J091316.34+312130.4	&	f	&	J093652.46+374141.6	&	F	\\
J080315.45+304747.4	&	g	&	J084009.47+522721.5	&	G	&	J091331.36+285706.3	&	g	&	J093709.24+195009.4	&	g	\\
J080328.94+332744.6	&	G	&	J084015.50+560254.6	&	F	&	J091335.59+122626.9	&	G	&	J093710.07+165837.9	&	G	\\
J080415.02+084355.4	&	F	&	J084022.70+233222.8	&	f	&	J091346.29+621954.4	&	G	&	J093715.19+233526.1	&	g	\\
J080534.10+102336.2	&	f	&	J084052.60+425013.7	&	g	&	J091402.54+400647.1	&	F	&	J093731.00+193521.7	&	f	\\
J080550.85+122847.9	&	f	&	J084132.11+511446.6	&	F	&	J091412.74+164432.3	&	g	&	J093832.18+371129.5	&	F	\\
J080611.20+123236.6	&	g	&	J084141.20+403926.8	&	G	&	J091426.23+360644.1	&	g	&	J093833.50+203922.4	&	F	\\
J080627.40+505717.3	&	F	&	J084153.06+325205.3	&	g	&	J091503.52+291611.9	&	G	&	J093907.96+340022.7	&	F	\\
J080642.79+390524.7	&	G	&	J084158.53+573530.9	&	g	&	J091504.79+415948.9	&	G	&	J094032.90+252925.7	&	G	\\
J080724.84+391140.1	&	F	&	J084252.68+250413.4	&	F	&	J091506.31-004306.2	&	G	&	J094112.93+610340.7	&	F	\\
J080756.40+174831.2	&	F	&	J084408.28+344302.0	&	F	&	J091555.00+100757.0	&	F	&	J094146.39+465825.9	&	F	\\
J080844.71+563532.4	&	g	&	J084455.20+474444.8	&	F	&	J091601.77+173523.3	&	G	&	J094255.15+285850.7	&	g	\\
\hline
\end{tabular}
\end{table*}

%%% longtable 2nd page

\begin{table*}
\caption{continued from previous page}   
%\contcaption{ from previous page}
\label{tab:sdss_class_c2}
\begin{tabular}{cccccccc}
\hline 
\textbf{Galaxy} & \textbf{Class} & \textbf{Galaxy} & \textbf{Class} & \textbf{Galaxy}  & \textbf{Class} & \textbf{Galaxy} & \textbf{Class} \\ 
\hline
J094302.18+374923.3	&	g	&	J100750.54+341855.0	&	F	&	J103521.00+033330.2	&	g	&	J105609.59+472332.8	&	f	\\
J094308.54+211056.4	&	G	&	J100804.27+144814.9	&	F	&	J103625.68+583322.3	&	g	&	J105615.48+151324.8	&	F	\\
J094356.95+424020.9	&	g	&	J100836.40+181352.3	&	G	&	J103638.66+350310.4	&	F	&	J105632.99+095602.1	&	g	\\
J094357.45+414114.3	&	G	&	J100901.53+322930.1	&	F	&	J103657.36+001347.1	&	f	&	J105659.30+181330.2	&	F	\\
J094423.49+111352.6	&	F	&	J100920.11+104635.9	&	g	&	J103737.91+372720.3	&	G	&	J105804.08+060247.3	&	F	\\
J094434.96+554546.2	&	F	&	J101000.23+115459.6	&	F	&	J103805.16+014442.3	&	g	&	J105805.35+093006.8	&	f	\\
J094453.63+225306.3	&	G	&	J101027.86+021341.6	&	F	&	J103837.87+111303.3	&	g	&	J105828.10+242231.5	&	F	\\
J094456.56+310552.2	&	f	&	J101027.91+275721.9	&	F	&	J103846.77+054148.8	&	F	&	J105833.33+461604.8	&	G	\\
J094457.09+164226.7	&	g	&	J101117.85+002632.6	&	F	&	J103911.78-002434.1	&	f	&	J105907.08+050022.4	&	G	\\
J094508.97+683540.4	&	G	&	J101352.58+003303.3	&	F	&	J103952.48+205049.3	&	g	&	J105909.00+613150.4	&	f	\\
J094513.39+392618.6	&	F	&	J101421.98+301025.5	&	F	&	J103957.93+240528.4	&	G	&	J110004.53+121404.7	&	F	\\
J094514.39+090637.2	&	f	&	J101439.55-004951.2	&	g	&	J104011.30+234121.2	&	G	&	J110032.51+020657.8	&	G	\\
J094522.17+230416.7	&	F	&	J101500.43+650823.7	&	F	&	J104038.59+371959.5	&	F	&	J110047.95+104341.3	&	G	\\
J094530.26+062236.5	&	F	&	J101511.42+564019.5	&	F	&	J104108.78+362222.1	&	f	&	J110111.30+122829.1	&	F	\\
J094549.96+282822.8	&	F	&	J101542.26+554002.9	&	g	&	J104152.94+211509.1	&	G	&	J110116.68+405239.6	&	G	\\
J094700.09+254045.8	&	g	&	J101550.49+203902.7	&	g	&	J104153.40+004735.5	&	F	&	J110117.20+670622.9	&	f	\\
J094704.29+423116.2	&	F	&	J101556.01+485737.0	&	G	&	J104221.91+152134.5	&	g	&	J110131.48+014527.2	&	g	\\
J094720.16+463636.2	&	F	&	J101610.24+582537.2	&	g	&	J104238.11+235706.8	&	G	&	J110137.99+504955.9	&	F	\\
J094722.94+344701.3	&	F	&	J101620.49+044919.2	&	g	&	J104312.47+213925.3	&	F	&	J110234.92+503456.2	&	F	\\
J094741.23+350336.1	&	g	&	J101725.41+644259.4	&	G	&	J104325.44+404627.1	&	f	&	J110238.87+511031.5	&	F	\\
J094749.00+234322.1	&	F	&	J101942.81+572524.4	&	F	&	J104351.09+212806.0	&	F	&	J110325.27+391518.2	&	g	\\
J094843.63+440453.1	&	F	&	J102009.62+131946.8	&	F	&	J104358.39+015608.9	&	F	&	J110338.42+451047.8	&	G	\\
J094854.50+245229.0	&	f	&	J102016.67+243550.9	&	F	&	J104436.43+213958.9	&	F	&	J110439.62+274325.9	&	F	\\
J094925.72+214042.4	&	F	&	J102031.72+430117.7	&	G	&	J104509.23+185822.7	&	F	&	J110441.90+041750.4	&	G	\\
J095011.13+441742.1	&	F	&	J102132.73+223246.8	&	F	&	J104509.45+220442.5	&	g	&	J110602.35+042545.7	&	G	\\
J095011.30+161711.9	&	F	&	J102138.25+123433.9	&	F	&	J104509.98+045640.1	&	g	&	J110606.88+295558.0	&	G	\\
J095042.50+302934.1	&	g	&	J102212.93+164806.1	&	F	&	J104527.33+583535.4	&	F	&	J110613.05+242119.4	&	g	\\
J095045.55+224515.9	&	f	&	J102224.40+035950.4	&	f	&	J104542.67+112039.0	&	F	&	J110649.56+574107.6	&	F	\\
J095106.02+090030.9	&	F	&	J102238.11+272121.7	&	G	&	J104546.94+371240.7	&	f	&	J110658.99+230022.7	&	F	\\
J095123.28+354520.3	&	F	&	J102240.72+461419.2	&	F	&	J104554.26+371103.3	&	F	&	J110704.29+074814.4	&	g	\\
J095208.28+041508.2	&	F	&	J102246.44+483813.6	&	g	&	J104559.59+224914.4	&	F	&	J110731.60+004659.1	&	g	\\
J095229.66+020916.0	&	G	&	J102257.38+555449.3	&	F	&	J104609.50+493236.8	&	G	&	J110753.59+663616.9	&	f	\\
J095318.33+360508.8	&	F	&	J102307.41+415031.1	&	F	&	J104615.88+510442.7	&	F	&	J110759.56+365220.2	&	F	\\
J095409.24+582027.4	&	F	&	J102318.79+222326.8	&	g	&	J104622.05+131317.1	&	F	&	J110808.30+455904.5	&	F	\\
J095410.67+021713.8	&	F	&	J102341.80+334626.3	&	f	&	J104636.45+520858.5	&	F	&	J110817.59+534927.1	&	F	\\
J095417.78+231714.9	&	F	&	J102408.56+164430.8	&	f	&	J104702.59+263234.4	&	F	&	J111044.88+043039.0	&	G	\\
J095524.67+410916.2	&	F	&	J102524.21+553129.7	&	F	&	J104710.75+025949.2	&	G	&	J111106.36+433758.8	&	G	\\
J095533.59+162558.2	&	G	&	J102625.53+173037.3	&	F	&	J104717.63+472521.0	&	g	&	J111152.20+453229.0	&	F	\\
J095552.53+355756.1	&	g	&	J102631.65+152024.0	&	F	&	J104739.35+261741.3	&	g	&	J111221.91+540138.2	&	F	\\
J095557.04+133315.1	&	G	&	J102649.63+345513.4	&	f	&	J104801.03+431111.5	&	F	&	J111224.71+312244.3	&	F	\\
J095703.23+153441.7	&	F	&	J102707.58+012934.4	&	F	&	J104801.05+281446.9	&	g	&	J111242.31+231828.5	&	F	\\
J095758.41+152801.7	&	F	&	J102720.23+221424.3	&	F	&	J104826.52+065801.4	&	F	&	J111313.29+553320.9	&	g	\\
J095900.21+174901.6	&	F	&	J102733.57+042800.3	&	F	&	J104827.28+263501.6	&	F	&	J111316.56+590050.0	&	F	\\
J095926.18-001514.1	&	G	&	J102810.12+495837.2	&	F	&	J104827.91+382349.8	&	g	&	J111456.83+334932.3	&	F	\\
J100031.55+031219.0	&	F	&	J102829.39+535159.3	&	F	&	J104839.19+214420.5	&	F	&	J111628.05+291936.1	&	G	\\
J100105.40+165619.8	&	F	&	J102841.90+034039.8	&	F	&	J104847.15+041856.6	&	G	&	J111629.13+410441.9	&	F	\\
J100140.48+104523.1	&	F	&	J102910.22+050822.8	&	F	&	J104916.99+001945.5	&	g	&	J111733.59+360350.5	&	f	\\
J100242.47+674349.7	&	F	&	J102929.32+193722.0	&	F	&	J105010.53+380952.5	&	F	&	J111814.71+263713.9	&	F	\\
J100351.02+502052.9	&	F	&	J102951.60+194959.3	&	g	&	J105023.99+644834.0	&	G	&	J111816.67+281537.5	&	g	\\
J100355.24+544254.8	&	g	&	J103129.97+245209.9	&	g	&	J105034.05+484136.4	&	F	&	J111856.18+001033.8	&	F	\\
J100431.82+380021.7	&	f	&	J103135.15+002831.3	&	f	&	J105039.84+654337.9	&	F	&	J111952.53+542746.3	&	G	\\
J100445.91+444220.7	&	F	&	J103145.04+464018.0	&	F	&	J105045.14+465032.4	&	F	&	J112000.62+360603.1	&	F	\\
J100447.30+573605.8	&	G	&	J103228.17+271014.2	&	G	&	J105100.31+121714.5	&	G	&	J112104.80+311508.1	&	F	\\
J100450.64+350657.2	&	F	&	J103238.49+435929.6	&	F	&	J105129.49+091646.2	&	f	&	J112135.03+182722.8	&	F	\\
J100513.48+212721.4	&	F	&	J103238.99+121040.3	&	g	&	J105259.06+373648.2	&	G	&	J112306.93+651506.6	&	G	\\
J100535.78+041645.8	&	g	&	J103247.56+155138.9	&	F	&	J105300.09+173429.4	&	F	&	J112309.72+354108.1	&	F	\\
J100547.32+142019.4	&	G	&	J103322.84+643005.8	&	F	&	J105310.05+372528.7	&	G	&	J112359.64+024130.7	&	F	\\
J100628.77+125214.0	&	F	&	J103349.99+125242.0	&	f	&	J105341.80+384612.8	&	G	&	J112405.87+454839.9	&	F	\\
J100640.36+381007.9	&	F	&	J103351.36-003340.9	&	g	&	J105514.13+484327.5	&	F	&	J112424.62+511405.7	&	g	\\
J100718.96+470023.4	&	F	&	J103353.37+111225.3	&	f	&	J105533.12+421759.8	&	G	&	J112425.79+272723.1	&	F	\\
J100727.69+121628.1	&	F	&	J103408.66+194215.4	&	g	&	J105547.39+145156.0	&	g	&	J112511.42+321531.7	&	G	\\
\hline
\end{tabular}
\end{table*}

%%% longtable 3rd page

\begin{table*}
\caption{continued from previous page}   
%\contcaption{ from previous page}
\label{tab:sdss_class_c1}
\begin{tabular}{cccccccc}
\hline 
\textbf{Galaxy} & \textbf{Class} & \textbf{Galaxy} & \textbf{Class} & \textbf{Galaxy}  & \textbf{Class} & \textbf{Galaxy} & \textbf{Class} \\ 
\hline
J112519.05+634345.2	&	F	&	J114429.06+553908.0	&	F	&	J120514.73+381408.1	&	G	&	J122947.56+271435.9	&	F	\\
J112529.20+570113.6	&	g	&	J114450.47-013604.7	&	f	&	J120520.97+630923.8	&	F	&	J122956.37+103655.2	&	g	\\
J112530.45+632646.6	&	g	&	J114503.88+195825.2	&	g	&	J120527.91+201831.4	&	g	&	J123051.38+082135.9	&	G	\\
J112533.36+224908.4	&	F	&	J114505.90+202616.4	&	g	&	J120550.06+202837.1	&	f	&	J123106.76+522451.5	&	G	\\
J112545.04+144035.6	&	G	&	J114517.56+264602.6	&	G	&	J120644.47+174255.2	&	G	&	J123250.27+054753.2	&	F	\\
J112545.33+240823.9	&	F	&	J114529.61+192400.6	&	G	&	J120649.46+250011.3	&	F	&	J123256.44+320832.4	&	F	\\
J112546.29+185623.8	&	F	&	J114651.69+652153.7	&	g	&	J120818.93+553833.6	&	g	&	J123301.79+565209.6	&	F	\\
J112615.74+275201.6	&	F	&	J114800.21+042917.3	&	g	&	J120825.58+100100.0	&	F	&	J123511.78+340719.9	&	F	\\
J112634.03+112624.2	&	F	&	J114803.36+302133.5	&	f	&	J120854.19+321339.7	&	g	&	J123523.97+292931.4	&	F	\\
J112719.20+593736.0	&	F	&	J114812.79+131233.7	&	F	&	J120910.00+291036.8	&	F	&	J123541.68+261319.8	&	F	\\
J112721.91+400046.9	&	F	&	J114850.40-020156.0	&	F	&	J120917.52+440524.6	&	F	&	J123713.53+492653.6	&	f	\\
J112839.91+090555.3	&	F	&	J114918.76+260717.7	&	F	&	J120928.84+261334.5	&	F	&	J123733.55+042203.6	&	F	\\
J112911.08+423133.9	&	G	&	J114922.08+245618.4	&	F	&	J121008.52+390308.6	&	F	&	J123807.68+224154.8	&	F	\\
J112924.07+345215.9	&	F	&	J115020.54-024839.9	&	g	&	J121010.75+491709.6	&	g	&	J123815.76+074913.3	&	F	\\
J112938.61+353052.7	&	G	&	J115044.20+210632.5	&	F	&	J121036.76+345723.6	&	F	&	J123845.43+064605.3	&	F	\\
J112945.19+220735.5	&	F	&	J115101.10+202357.5	&	f	&	J121041.85+131952.5	&	F	&	J123848.02+320529.3	&	F	\\
J113017.27+580802.0	&	F	&	J115122.68+532639.0	&	G	&	J121049.60+392822.1	&	G	&	J123951.67+345829.8	&	G	\\
J113050.49+603008.0	&	F	&	J115159.83+210630.3	&	F	&	J121058.75+431457.8	&	F	&	J124056.23+292756.0	&	g	\\
J113052.15+013250.6	&	g	&	J115202.61+254504.0	&	F	&	J121155.12+161353.7	&	F	&	J124114.11+264410.0	&	F	\\
J113111.66+514153.0	&	F	&	J115212.38+344738.7	&	G	&	J121227.95+390639.3	&	F	&	J124144.54+350345.9	&	F	\\
J113112.50+341210.9	&	F	&	J115244.11+183651.1	&	F	&	J121322.20+283039.4	&	F	&	J124155.00+114626.9	&	F	\\
J113144.59+342000.2	&	F	&	J115258.32+611257.8	&	F	&	J121418.09+593655.5	&	F	&	J124205.40+630645.4	&	G	\\
J113213.46+004909.4	&	F	&	J115311.13+252615.3	&	f	&	J121422.05+560041.1	&	G	&	J124231.46+531216.3	&	g	\\
J113220.35+535416.0	&	F	&	J115320.32+204506.1	&	g	&	J121426.30+241055.5	&	F	&	J124240.72+012044.8	&	F	\\
J113241.54+202619.1	&	G	&	J115346.36+102411.3	&	f	&	J121505.37+133541.0	&	F	&	J124302.99-011657.8	&	F	\\
J113244.06+614936.8	&	F	&	J115350.32+194501.7	&	G	&	J121544.80+545127.3	&	F	&	J124313.89+310505.8	&	g	\\
J113245.43+405033.2	&	F	&	J115359.66+203421.0	&	F	&	J121552.51+002402.5	&	g	&	J124428.12-030018.8	&	F	\\
J113315.79+242648.8	&	g	&	J115400.95+062035.3	&	F	&	J121608.52-033412.5	&	F	&	J124531.87+544415.3	&	g	\\
J113329.95+341858.4	&	F	&	J115433.67+582201.4	&	G	&	J121609.04+280746.7	&	f	&	J124611.08+464338.3	&	F	\\
J113342.04+232445.8	&	f	&	J115451.28+025733.3	&	F	&	J121623.92+470135.0	&	G	&	J124701.00-013441.0	&	F	\\
J113412.74+341845.0	&	G	&	J115458.53+261209.0	&	F	&	J121648.86+472726.6	&	F	&	J124845.86+351957.7	&	F	\\
J113423.32-023145.5	&	g	&	J115458.72+582936.9	&	F	&	J121746.92+105040.6	&	F	&	J124903.69+305535.1	&	F	\\
J113453.54+202917.0	&	F	&	J115516.72+172915.2	&	G	&	J121808.54-010350.8	&	f	&	J124936.86+305043.8	&	F	\\
J113537.58+250519.0	&	G	&	J115524.71+391324.3	&	F	&	J121821.43+251300.3	&	F	&	J125059.51+474017.4	&	f	\\
J113606.64+621456.9	&	G	&	J115535.13+255321.9	&	F	&	J121835.71+420050.1	&	g	&	J125117.92+270622.1	&	g	\\
J113612.28+595835.4	&	f	&	J115628.20+394430.8	&	F	&	J121912.21+492116.7	&	F	&	J125137.96+312109.9	&	F	\\
J113630.48+265138.8	&	F	&	J115719.70+362456.9	&	F	&	J121913.46+222553.2	&	F	&	J125236.28+264459.6	&	g	\\
J113658.03+550943.4	&	F	&	J115746.05+141750.7	&	g	&	J121943.87+285146.0	&	F	&	J125348.59+293518.1	&	f	\\
J113701.89+153414.1	&	F	&	J115823.78-021638.5	&	f	&	J121952.58+273715.5	&	F	&	J125448.59+191037.8	&	F	\\
J113704.17+240546.5	&	G	&	J115827.21+573546.9	&	g	&	J122004.77+275831.1	&	G	&	J125448.96+440920.1	&	G	\\
J113730.43+222358.9	&	F	&	J115836.16+161030.0	&	g	&	J122202.61+320545.4	&	F	&	J125524.83+521603.7	&	G	\\
J113914.88+170837.1	&	G	&	J115844.63+281722.4	&	G	&	J122203.59+124427.3	&	g	&	J125749.94+293915.1	&	F	\\
J113926.83+032816.0	&	F	&	J115845.67+272708.7	&	F	&	J122207.31+085926.0	&	F	&	J125813.56+262536.7	&	G	\\
J114013.94+244149.4	&	G	&	J115905.47+245920.2	&	F	&	J122238.99+064037.3	&	F	&	J125822.80+135221.7	&	F	\\
J114039.28+285139.0	&	F	&	J115910.17+374736.2	&	F	&	J122246.75+655037.6	&	G	&	J125822.89+451622.0	&	F	\\
J114046.67+174745.2	&	F	&	J115926.88+174525.1	&	F	&	J122311.63+060419.6	&	g	&	J125842.04+103659.4	&	F	\\
J114103.67+101330.3	&	F	&	J115944.25-032356.4	&	F	&	J122322.27+060226.6	&	g	&	J125850.87+093914.2	&	F	\\
J114111.13+432635.3	&	F	&	J120044.28+281943.0	&	g	&	J122358.84+484646.0	&	g	&	J125945.33+320242.0	&	F	\\
J114223.73+101550.9	&	G	&	J120050.49+315242.3	&	F	&	J122413.99+445615.4	&	F	&	J130016.12+361514.8	&	F	\\
J114224.50+200709.4	&	f	&	J120144.30+175355.9	&	f	&	J122426.73+131400.3	&	F	&	J130055.56+471319.5	&	g	\\
J114239.45+244921.0	&	F	&	J120203.14+295052.7	&	F	&	J122500.48+283330.9	&	g	&	J130117.11+535138.9	&	G	\\
J114245.16+200156.3	&	g	&	J120250.68+483810.3	&	F	&	J122540.89-025702.6	&	G	&	J130137.79+542325.6	&	F	\\
J114302.15+193859.0	&	G	&	J120405.42+140355.9	&	F	&	J122546.70+043035.5	&	F	&	J130204.91+402430.0	&	F	\\
J114327.02+604034.8	&	F	&	J120409.74+014933.4	&	F	&	J122547.25+455434.4	&	F	&	J130213.92+670454.8	&	F	\\
J114327.28+524240.0	&	F	&	J120428.94+300530.8	&	G	&	J122549.39+500545.3	&	F	&	J130316.10+503714.7	&	F	\\
J114328.96+100942.7	&	f	&	J120432.52+201217.7	&	f	&	J122553.80+451653.3	&	F	&	J130316.58+324317.2	&	F	\\
J114331.44+224331.5	&	F	&	J120443.31+311038.2	&	g	&	J122655.46+375432.2	&	g	&	J130321.21+142239.0	&	f	\\
J114358.96+200437.2	&	F	&	J120447.20-024312.2	&	G	&	J122708.76+161931.4	&	F	&	J130355.17+161014.9	&	f	\\
J114403.93-023330.1	&	F	&	J120456.25+211425.7	&	F	&	J122727.62+031807.6	&	g	&	J130411.85+611140.9	&	G	\\
J114404.89+600711.2	&	G	&	J120456.49+431938.2	&	F	&	J122806.79+135442.3	&	F	&	J130415.04+091324.4	&	F	\\
J114425.77+332118.2	&	G	&	J120512.76+284654.7	&	F	&	J122903.86+274643.8	&	f	&	J130421.98+434834.2	&	g	\\
\hline
\end{tabular}
\end{table*}

%%% longtable 4th page

\begin{table*}
\caption{continued from previous page}   
%\contcaption{ from previous page}
\label{tab:sdss_class_c3}
\begin{tabular}{cccccccc}
\hline 
\textbf{Galaxy} & \textbf{Class} & \textbf{Galaxy} & \textbf{Class} & \textbf{Galaxy}  & \textbf{Class} & \textbf{Galaxy} & \textbf{Class} \\ 
\hline
J130455.77+473013.1	&	f	&	J132840.22+124240.3	&	F	&	J134923.66+083027.4	&	g	&	J141311.61+130013.2	&	F	\\
J130524.86+561921.9	&	F	&	J132916.19+530455.1	&	G	&	J135021.91+180916.4	&	F	&	J141316.08+270029.0	&	f	\\
J130603.55+355849.1	&	F	&	J132925.22-002356.5	&	F	&	J135032.73+350753.9	&	F	&	J141340.03+435159.6	&	F	\\
J130612.74+252033.1	&	G	&	J132955.10+485049.7	&	F	&	J135048.19+294606.4	&	g	&	J141420.85+073057.2	&	g	\\
J130727.96+522418.0	&	F	&	J132956.71+110419.5	&	F	&	J135216.75+223117.6	&	G	&	J141427.21+362416.1	&	g	\\
J130817.52+520028.6	&	F	&	J132956.76+413733.7	&	F	&	J135222.75+213221.6	&	G	&	J141452.03+140733.1	&	F	\\
J130837.55+540427.7	&	F	&	J133002.25+132457.8	&	G	&	J135231.94+374902.7	&	F	&	J141503.69+362726.1	&	F	\\
J130841.73+524627.3	&	g	&	J133004.20-014303.6	&	F	&	J135232.73+075127.6	&	G	&	J141512.04+045219.4	&	F	\\
J130910.84+222758.8	&	F	&	J133006.14-014314.1	&	F	&	J135256.56+024851.3	&	g	&	J141516.22+342054.1	&	F	\\
J130949.99+243439.2	&	f	&	J133011.49-013947.2	&	g	&	J135302.44+243336.7	&	g	&	J141623.85+393007.8	&	G	\\
J130952.10+282256.7	&	F	&	J133015.67+072910.3	&	F	&	J135341.92+032238.6	&	f	&	J141635.20+095907.8	&	F	\\
J131012.91+523103.3	&	F	&	J133023.78-025415.6	&	F	&	J135403.81+035704.8	&	F	&	J141814.90+005327.9	&	F	\\
J131042.02+421704.6	&	g	&	J133025.96+313714.8	&	G	&	J135406.26+052122.8	&	F	&	J141818.50+285805.9	&	g	\\
J131047.03+500534.2	&	g	&	J133026.37+300144.4	&	F	&	J135428.53+305445.3	&	g	&	J141823.95+262309.1	&	G	\\
J131056.52+112838.5	&	F	&	J133036.95+345502.6	&	G	&	J135428.94+234524.1	&	F	&	J141848.50+105037.7	&	G	\\
J131118.07+453938.0	&	f	&	J133037.05+411015.7	&	F	&	J135507.96+401003.3	&	F	&	J141912.79+244755.4	&	G	\\
J131126.35-001459.0	&	f	&	J133117.44+292205.4	&	G	&	J135531.29+264749.8	&	G	&	J141916.58+261754.9	&	g	\\
J131127.45+473258.3	&	g	&	J133130.19+333223.4	&	G	&	J135559.49-011542.2	&	F	&	J141936.98+381403.3	&	F	\\
J131156.35+452614.7	&	G	&	J133220.37+013400.7	&	F	&	J135603.40+133021.0	&	G	&	J142125.10+351614.2	&	g	\\
J131206.26+461146.1	&	g	&	J133231.22+044809.0	&	F	&	J135747.63+072346.5	&	g	&	J142131.32+235653.9	&	G	\\
J131211.30+212418.6	&	F	&	J133248.69+415218.5	&	F	&	J135804.70+151853.4	&	F	&	J142152.58+395844.7	&	f	\\
J131258.27+311530.9	&	F	&	J133310.22+324107.2	&	G	&	J135805.66+214750.9	&	G	&	J142209.21+560005.1	&	g	\\
J131313.46+335903.9	&	g	&	J133328.05+052849.1	&	F	&	J135814.18+363900.5	&	F	&	J142246.68+375942.8	&	G	\\
J131320.97+060340.5	&	f	&	J133329.01+330232.5	&	G	&	J135820.15+071330.3	&	F	&	J142259.11+234739.6	&	F	\\
J131325.70+274548.4	&	G	&	J133329.18+172814.1	&	g	&	J140004.44+120641.4	&	G	&	J142342.38+340032.4	&	F	\\
J131326.95+274808.5	&	g	&	J133430.38+601550.2	&	F	&	J140109.47+074208.1	&	F	&	J142422.94+243650.8	&	g	\\
J131427.60-020751.5	&	g	&	J133439.26+040747.8	&	F	&	J140121.81+102851.3	&	F	&	J142433.16+011038.3	&	g	\\
J131432.44+304220.8	&	F	&	J133455.34+312336.5	&	F	&	J140204.80-012128.4	&	g	&	J142449.72+351705.7	&	F	\\
J131432.68+443027.0	&	G	&	J133457.26+340238.6	&	F	&	J140211.30+092624.3	&	F	&	J142452.99+275300.5	&	g	\\
J131700.02+340605.9	&	g	&	J133531.94+133943.7	&	F	&	J140212.19+043508.6	&	g	&	J142456.61+250129.2	&	f	\\
J131730.95+310547.3	&	F	&	J133548.23+025956.1	&	G	&	J140217.59+074102.9	&	G	&	J142502.97+274526.7	&	F	\\
J131745.19+273411.5	&	g	&	J133558.58+014348.2	&	F	&	J140225.36+511231.1	&	g	&	J142533.21+335053.2	&	F	\\
J131810.07-011437.2	&	F	&	J133602.56+661809.1	&	F	&	J140232.92+090446.9	&	F	&	J142538.08+013601.5	&	F	\\
J131834.58+042910.7	&	g	&	J133626.20+072210.5	&	f	&	J140248.93+575828.6	&	F	&	J142553.88+025226.8	&	F	\\
J131903.09+393521.5	&	F	&	J133649.31+445253.2	&	F	&	J140307.15+092146.9	&	g	&	J142602.63+493118.2	&	G	\\
J131905.03-023055.1	&	F	&	J133659.25+333412.8	&	F	&	J140308.75+592600.3	&	F	&	J142612.19+483350.3	&	F	\\
J131908.23+283024.9	&	G	&	J133700.11+383526.3	&	g	&	J140323.13+092653.1	&	F	&	J142630.62+055834.9	&	f	\\
J131933.40+165911.6	&	F	&	J133813.05+324922.3	&	F	&	J140339.04+112241.6	&	F	&	J142702.32+395726.2	&	F	\\
J132056.68+140921.8	&	F	&	J133823.47+065315.5	&	G	&	J140348.59+152408.2	&	F	&	J142709.50+305653.5	&	F	\\
J132113.05+311318.6	&	F	&	J133826.06-001339.4	&	G	&	J140436.76+291159.8	&	F	&	J142714.56+112020.6	&	G	\\
J132132.18+121115.8	&	f	&	J133910.84+285733.1	&	F	&	J140501.82+110042.7	&	F	&	J142718.23+044810.2	&	G	\\
J132140.53+312103.5	&	F	&	J133913.27+043724.0	&	f	&	J140520.32+304841.6	&	F	&	J142720.37+010133.1	&	G	\\
J132226.13-022505.7	&	f	&	J133937.80+060912.9	&	F	&	J140658.94+123338.8	&	g	&	J142728.53+110226.6	&	g	\\
J132300.38+135702.3	&	F	&	J133955.41+282344.3	&	F	&	J140708.54+563224.6	&	f	&	J142821.11+352420.5	&	g	\\
J132300.91+231823.1	&	F	&	J134001.34+072202.7	&	F	&	J140814.40+354412.5	&	F	&	J142852.05+164322.1	&	g	\\
J132310.53+430510.5	&	F	&	J134009.09+365139.9	&	F	&	J140829.16+070328.4	&	F	&	J142856.99+253312.2	&	F	\\
J132324.62+263236.8	&	F	&	J134105.20+050620.9	&	F	&	J140830.45+362943.3	&	F	&	J142917.13+203916.3	&	g	\\
J132327.45+062333.2	&	F	&	J134129.66+231001.0	&	g	&	J140832.97+114858.4	&	g	&	J142923.42+450419.3	&	g	\\
J132336.93+135824.7	&	F	&	J134142.23+405228.0	&	f	&	J140902.32+141901.1	&	F	&	J142932.44+384011.2	&	F	\\
J132341.56+313846.7	&	F	&	J134142.40+300731.5	&	g	&	J140919.79+551656.4	&	f	&	J142934.00+352743.0	&	f	\\
J132412.57+414910.8	&	f	&	J134205.95+370228.3	&	F	&	J140932.37+144954.5	&	G	&	J142938.52+103502.6	&	F	\\
J132531.22-032225.7	&	g	&	J134215.02+015126.8	&	G	&	J140951.93+145220.6	&	F	&	J143020.15+230342.4	&	F	\\
J132538.66+334049.5	&	F	&	J134223.88+144420.8	&	g	&	J141006.91+173656.7	&	F	&	J143117.97+075640.6	&	g	\\
J132609.14+355603.9	&	f	&	J134241.92+073057.9	&	F	&	J141045.45+151233.8	&	F	&	J143205.83+281121.2	&	F	\\
J132628.53+360037.0	&	F	&	J134308.83+302015.8	&	f	&	J141057.23+252950.0	&	G	&	J143245.14+025454.0	&	F	\\
J132645.33+202038.8	&	g	&	J134459.54+292535.8	&	F	&	J141208.01+551347.9	&	g	&	J143317.52+035412.2	&	F	\\
J132654.23+113341.8	&	g	&	J134510.34+351308.6	&	F	&	J141212.71+083942.6	&	f	&	J143320.56+261228.5	&	F	\\
J132711.90+552913.8	&	F	&	J134516.17+231308.2	&	g	&	J141229.04+243807.8	&	g	&	J143348.33+035724.7	&	G	\\
J132716.84+320150.5	&	g	&	J134609.00+215339.4	&	F	&	J141235.85+461218.8	&	F	&	J143358.63+401439.7	&	F	\\
J132746.96+174643.7	&	f	&	J134744.90+174306.4	&	F	&	J141238.13+391836.6	&	g	&	J143409.98+540444.1	&	g	\\
J132816.17+025857.5	&	F	&	J134755.75+374934.7	&	g	&	J141251.24+354239.2	&	g	&	J143521.40+501711.3	&	F	\\
\hline
\end{tabular}
\end{table*}

%%% longtable 5th page

\begin{table*}
\caption{continued from previous page}
%\contcaption{ from previous page}
\label{tab:sdss_class_c4}
\begin{tabular}{cccccccc}
\hline 
\textbf{Galaxy} & \textbf{Class} & \textbf{Galaxy} & \textbf{Class} & \textbf{Galaxy}  & \textbf{Class} & \textbf{Galaxy} & \textbf{Class} \\ 
\hline 
J143545.74+244332.8	&	g	&	J150734.75+193456.7	&	G	&	J154521.60+252637.3	&	g	&	J162122.05+404837.8	&	F	\\
J143644.71+210421.9	&	f	&	J151006.40+562221.3	&	g	&	J154601.32+024251.2	&	F	&	J162133.88+380030.8	&	f	\\
J143657.07+414941.1	&	f	&	J151018.40+074219.6	&	G	&	J154601.82+050813.2	&	g	&	J162139.74+190000.9	&	F	\\
J143713.89+083845.5	&	F	&	J151024.21+450958.1	&	F	&	J154630.00+320702.1	&	G	&	J162254.36+413114.0	&	g	\\
J143722.15+363404.2	&	G	&	J151036.09+410317.0	&	g	&	J154645.21+310040.4	&	G	&	J162321.07+371530.2	&	G	\\
J143808.92+201024.8	&	F	&	J151207.19+554706.3	&	F	&	J154745.95+283830.0	&	F	&	J162412.19+300944.1	&	G	\\
J143838.28+523452.1	&	G	&	J151220.64+092059.7	&	F	&	J154812.64+111646.2	&	G	&	J162415.16+201100.7	&	G	\\
J143917.95+032206.0	&	G	&	J151231.48+190928.9	&	g	&	J154928.80+115609.2	&	g	&	J162900.40+411703.3	&	g	\\
J144019.51+164106.4	&	f	&	J151307.36+411552.6	&	G	&	J154934.29+134942.8	&	G	&	J162951.04+394559.5	&	G	\\
J144132.03+443045.9	&	F	&	J151320.80+023049.5	&	G	&	J155022.72+185621.3	&	f	&	J163038.87+350323.0	&	G	\\
J144142.96+055709.9	&	F	&	J151415.14+202843.1	&	g	&	J155025.91+180822.4	&	g	&	J163103.88+410921.7	&	F	\\
J144232.64+042549.6	&	g	&	J151431.41+585035.2	&	f	&	J155050.64+221419.2	&	F	&	J163121.57+224149.4	&	f	\\
J144233.24+284335.2	&	F	&	J151542.84+012720.7	&	F	&	J155100.98+514713.9	&	F	&	J163134.53+403356.1	&	G	\\
J144252.58+183712.9	&	F	&	J151600.93+542856.2	&	f	&	J155106.91+201209.0	&	G	&	J163425.48+213227.2	&	G	\\
J144317.61-024514.4	&	f	&	J151619.12+070944.4	&	g	&	J155111.85+215633.4	&	F	&	J163446.00+451927.5	&	F	\\
J144349.20+532402.3	&	F	&	J151639.57+425713.9	&	F	&	J155218.55+232035.9	&	F	&	J163456.87+254133.3	&	G	\\
J144352.05+411404.2	&	F	&	J151747.35+071305.8	&	G	&	J155220.73+243735.7	&	F	&	J163632.56+390140.3	&	g	\\
J144406.31+091649.0	&	F	&	J151800.09+133135.3	&	g	&	J155243.12+201359.3	&	F	&	J163654.31+362524.5	&	G	\\
J144429.23+044441.0	&	G	&	J151808.40+051838.3	&	F	&	J155311.25+391133.9	&	G	&	J163743.75+251514.9	&	G	\\
J144503.28+003137.1	&	F	&	J151815.11+122919.1	&	F	&	J155323.18+115733.0	&	f	&	J163809.04+395504.1	&	G	\\
J144545.28+323748.3	&	g	&	J151858.17+204855.3	&	G	&	J155325.41+392310.2	&	g	&	J163849.53+172112.0	&	F	\\
J144549.05+502338.5	&	F	&	J151925.27+455248.9	&	f	&	J155554.86+265759.4	&	F	&	J164014.64+230118.3	&	f	\\
J144621.38+130115.5	&	g	&	J152041.32+490623.5	&	g	&	J155555.36+170949.4	&	F	&	J164022.68+334047.0	&	G	\\
J144640.70+123618.3	&	f	&	J152114.23+414333.3	&	F	&	J155601.60+302502.6	&	g	&	J164208.18+401636.8	&	F	\\
J144648.64-015731.1	&	G	&	J152136.81+111526.6	&	F	&	J155801.84+145748.8	&	G	&	J164641.59+360525.4	&	G	\\
J144710.44+632527.0	&	g	&	J152145.69+230909.4	&	F	&	J155802.78+162051.3	&	f	&	J164714.78+401442.3	&	f	\\
J144742.26+183007.2	&	g	&	J152305.03+111441.8	&	g	&	J155807.96+120413.0	&	F	&	J165422.27+412007.6	&	F	\\
J144842.57+122725.9	&	F	&	J152352.72+233251.2	&	F	&	J155818.83+365312.1	&	g	&	J165728.60+223557.5	&	G	\\
J144843.39+181721.5	&	g	&	J152356.52+380719.7	&	F	&	J155945.79+184803.9	&	F	&	J165829.59+200229.2	&	F	\\
J144911.08+110654.8	&	G	&	J152408.13+133537.0	&	F	&	J155946.58+370214.2	&	G	&	J170128.22+634128.0	&	F	\\
J145026.66+164442.7	&	G	&	J152449.27+421021.2	&	F	&	J160142.96+484940.5	&	G	&	J170443.10+343330.6	&	F	\\
J145058.17+160622.8	&	f	&	J152552.58+074910.6	&	G	&	J160149.70+080853.8	&	F	&	J170628.51+382148.6	&	f	\\
J145109.91+171120.0	&	f	&	J152602.66+161923.6	&	F	&	J160151.95+154732.1	&	G	&	J170727.45+311451.7	&	g	\\
J145346.72+200657.4	&	f	&	J152639.07+501908.7	&	g	&	J160216.34+491211.4	&	F	&	J170732.78+362703.7	&	G	\\
J145449.24+180621.4	&	F	&	J152658.39+285109.3	&	g	&	J160240.55+372134.1	&	g	&	J171140.20+595944.3	&	F	\\
J145459.83+621633.3	&	G	&	J152854.93+074443.0	&	G	&	J160324.16+205328.4	&	f	&	J171159.71+232247.8	&	g	\\
J145608.25+121848.1	&	G	&	J152925.77+050506.5	&	g	&	J160600.16+181145.4	&	g	&	J171512.79+574538.0	&	F	\\
J145629.01+092116.3	&	f	&	J152945.00+425507.1	&	G	&	J160604.39+204805.7	&	f	&	J171637.46+582442.8	&	g	\\
J145653.06+091617.8	&	g	&	J153000.84+125921.5	&	F	&	J160619.53+162553.1	&	G	&	J171647.73+615512.4	&	F	\\
J145711.20+194153.6	&	F	&	J153036.59+424301.7	&	F	&	J160622.82+194640.6	&	F	&	J171843.72+580806.4	&	G	\\
J145756.35+534705.8	&	F	&	J153144.08-012241.7	&	g	&	J160642.59+315339.8	&	G	&	J172241.11+600038.5	&	F	\\
J145859.66+535523.8	&	f	&	J153221.50+482420.2	&	f	&	J160707.03+220338.0	&	G	&	J172243.82+620957.8	&	F	\\
J145934.31+270658.5	&	f	&	J153232.68+415849.6	&	F	&	J160725.10+102532.8	&	F	&	J172408.08+585942.2	&	F	\\
J145934.75+325028.8	&	F	&	J153350.64+051625.3	&	G	&	J160758.36+104646.4	&	F	&	J172959.32+602100.9	&	g	\\
J150027.60+062717.0	&	F	&	J153407.82+142344.3	&	F	&	J160915.71+254244.8	&	G	&	J174215.28+555910.6	&	F	\\
J150056.68+113102.6	&	g	&	J153433.38+410755.8	&	F	&	J160948.31+422002.4	&	f	&	J205222.36+000432.6	&	G	\\
J150150.90+475231.4	&	g	&	J153510.22+163258.4	&	F	&	J161010.84+164158.5	&	g	&	J205310.36+001356.8	&	g	\\
J150241.99+232001.0	&	F	&	J153642.16+433221.5	&	G	&	J161026.54+121820.4	&	G	&	J205404.36+004638.7	&	g	\\
J150411.78+072723.1	&	G	&	J153726.08+214437.7	&	g	&	J161126.44+292110.8	&	G	&	J210107.75-001142.6	&	g	\\
J150429.73+021958.9	&	F	&	J153840.43+410018.8	&	F	&	J161145.91+372740.3	&	F	&	J211627.64-004935.3	&	G	\\
J150437.84+402219.4	&	g	&	J153854.16+170134.3	&	g	&	J161235.23+292154.2	&	g	&	J213241.75-000740.2	&	G	\\
J150440.58+123800.4	&	F	&	J153939.45+031156.5	&	g	&	J161403.28+141655.5	&	G	&	J213702.37+002541.7	&	g	\\
J150447.23+081016.2	&	F	&	J154006.76+204050.1	&	F	&	J161536.91+280930.9	&	F	&	J214907.29+002650.3	&	F	\\
J150501.03+054748.3	&	G	&	J154008.83+213056.1	&	G	&	J161711.61+495751.9	&	G	&	J215824.95-004422.3	&	F	\\
J150530.55+083127.6	&	f	&	J154028.65+533042.6	&	g	&	J161725.63+350808.7	&	g	&	J220316.00+003415.9	&	g	\\
J150553.27+393119.8	&	g	&	J154056.06+324700.4	&	g	&	J161749.43+204128.6	&	F	&	J220756.59+002212.5	&	G	\\
J150639.21+371707.1	&	g	&	J154109.86-014222.2	&	G	&	J161811.83+400541.1	&	f	&	J221152.92+000631.4	&	G	\\
J150642.04+124035.1	&	F	&	J154122.58+281347.1	&	F	&	J161854.59+350914.2	&	G	&	J223207.24+000213.4	&	g	\\
J150653.40+125131.2	&	G	&	J154235.27+373614.6	&	g	&	J161918.14+370542.7	&	F	&	J233025.65+000924.1	&	F	\\
J150654.64+563032.4	&	f	&	J154355.10+441821.4	&	G	&	J161941.51+360517.6	&	G	&	J235106.26+010324.1	&	f	\\
J150709.50+093808.0	&	F	&	J154401.51+281637.2	&	f	&	J161953.49+513247.4	&	G	&	J235627.28-005040.8	&	F	\\
\hline
\end{tabular}
\end{table*}
}

\section{Other observable differences between Flocculents and Grand-designs}
\begin{figure}
  \label{fig:oprops}
  \centering%
    \includegraphics[width=1.1\linewidth]{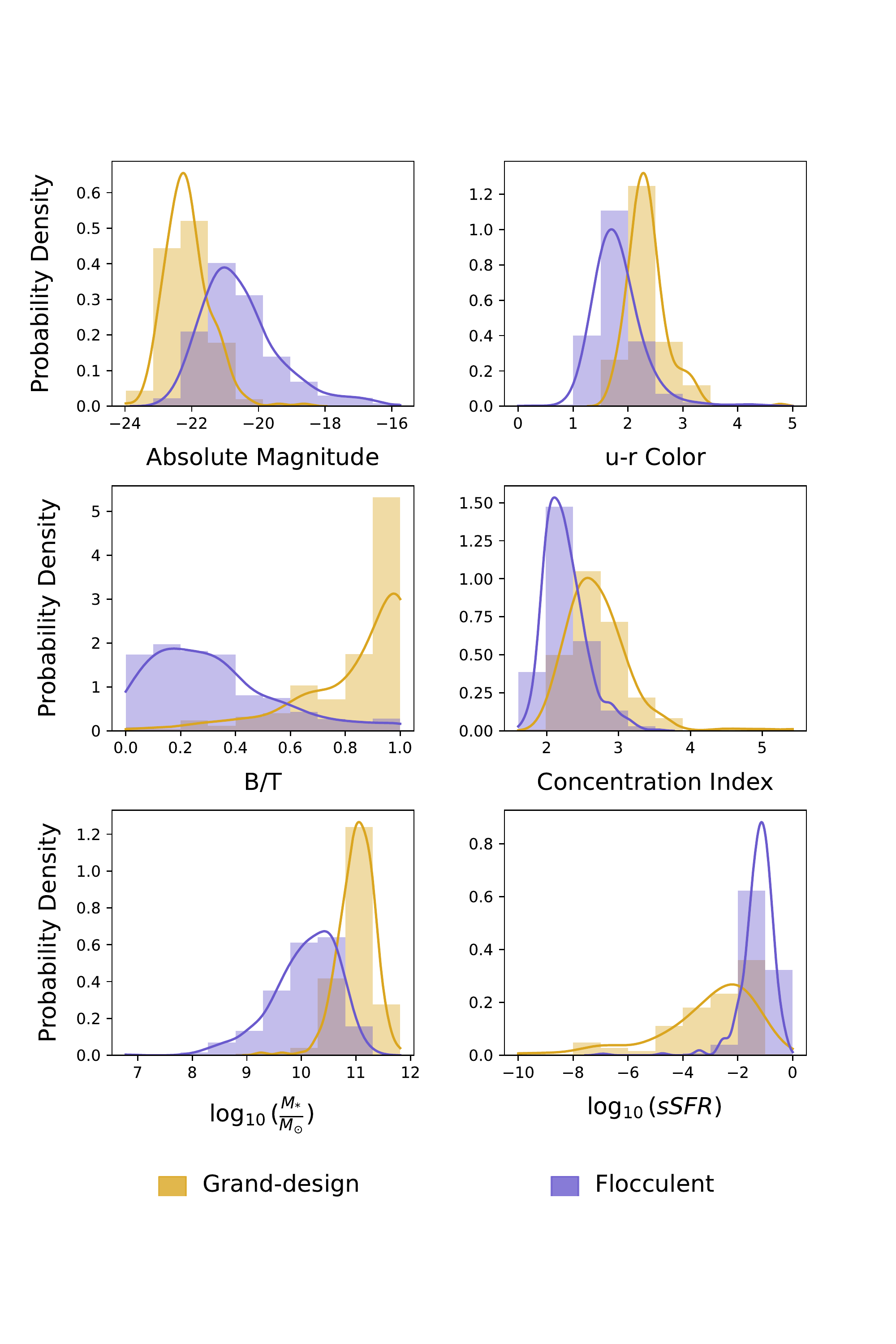}
    \vspace{-30px}
    \caption{This figure shows the PDFs of six diffrent properties for the {\it Grand-design} and {\it Flocculent} spirals in our sample. The slateblue colour 
            represents the {\it Flocculents}, whereas, the {\it Grand-designs} are shown in goldenrod.}
\end{figure}

\autoref{fig:oprops} shows the comparison between the {\it Flocculents} and {\it Grand-designs} in our final sample using the Probability 
Density Functions of a number of observable quantities derived from the SDSS spectroscopic and photometric data. The panels in the figure present: 
\textbf{[top left]} r-band absolute magnitude; 
\textbf{[top right]} (u-r) colour; 
\textbf{[middle left]}{\it bulge-to-total} ratio of flux for the disc; 
\textbf{[middle right]} Concentration index ( ratio of radii containing 90 \% \& 50\% Petrosian flux); 
\textbf{[bottom left]} stellar mass;
\textbf{[bottom right]} specific star formation rate.
The estimation of stellar-mass and sSFR is done by \cite{conroy10} using Granada Flexible Stellar Population Synthesis (FSPS) model.
This figure supports the claim of the {\it Flocculents} being gas rich younger galaxies compared to the {\it Grand-designs}.

%%%%%%%%%%%%%%%%%%%%%%%%%%%%%%%%%%%%%%%%%%%%%%%%%%
% Don't change these lines
\bsp	% typesetting comment
\label{lastpage}
\end{document}